\documentclass[reprint,superscriptaddress,amsmath,amssymb,aps,pre, nofootinbib]{revtex4-2}
\usepackage{graphicx}
\usepackage{dcolumn}
\usepackage{bm}
\usepackage{color}
\usepackage{units}
\usepackage{wasysym}
\usepackage{MnSymbol}
\usepackage{comment}
\usepackage{times}
\usepackage{amsmath}
\usepackage{footnote}
\usepackage{mathtools}
\usepackage[unicode=true,pdfusetitle,bookmarks=false,colorlinks=true,citecolor=blue,urlcolor=blue,linkcolor=blue]{hyperref}
\DeclareMathAlphabet{\mathpzc}{OT1}{pzc}{m}{it}

\begin{document}

\title{Multicomponent Linear Transport in the Absence of Local Equilibrium}

\author{Yu-Jen Chiu}
\thanks{These authors contributed equally to this work}
\affiliation{Department of Materials Science and Engineering, University of California, Berkeley, California 94720, USA}

\author{Eric M. Weiner}
\thanks{These authors contributed equally to this work}
\affiliation{Department of Materials Science and Engineering, University of California, Berkeley, California 94720, USA}

\author{Ahmad K. Omar}
\email{aomar@berkeley.edu}
\affiliation{Department of Materials Science and Engineering, University of California, Berkeley, California 94720, USA}
\affiliation{Materials Sciences Division, Lawrence Berkeley National Laboratory, Berkeley, California 94720, USA}

\begin{abstract}
The linear laws of transport phenomena are central in our description of irreversible processes in systems across the physical sciences.
Linear irreversible thermodynamics allows for the identification of the underlying forces driving transport and the structure of the relevant transport coefficients for systems that are locally in equilibrium. 
Increasingly, linear relations are found to describe transport in systems in which a local equilibrium hypothesis is unlikely to hold. 
Here, we derive a mechanical theory of multicomponent transport without appealing to equilibrium notions. 
Our theory for the Onsager transport tensor highlights the general breakdown of the familiar Onsager reciprocal relations and Einstein relations when a local equilibrium is absent. 
The procedure outlined is applied to a variety of systems, including passive systems, mixtures with nonreciprocal interactions, electrolytes under an electric field, and active systems, and can be straightforwardly used to understand other transport processes. 
The framework further provides a basis to extend numerical approaches for computing the transport coefficients of nonequilibrium systems, as is demonstrated for a system with nonreciprocal interactions.
\end{abstract}

\maketitle

\section{Introduction}
The canonical constitutive laws describing heat, mass, and momentum transfer were initially established empirically to explain universal macroscopic transport phenomena observed across diverse systems~\cite{Bird1960, DeGroot2013}. 
Their predictive success, broad applicability, and analytical tractability have made linear constitutive laws foundational to our understanding of irreversible processes~\cite{DeGroot2013}. 
The ubiquity of these linear relationships has stimulated numerous theoretical investigations aimed at understanding their origins. 
Onsager’s pioneering work~\cite{Onsager1931_1, Onsager1931_2} identified a central property of transport phenomena, the Onsager transport tensor $\mathbf{L}$, which couples flux responses to identified thermodynamic driving forces. 
His work further clarified fundamental properties of these transport coefficients through the formulation of the Onsager reciprocal relations~\cite{Onsager1931_1, Onsager1931_2}, demonstrating that $\mathbf{L}$ is symmetric when the underlying microscopic dynamics satisfy time-reversal symmetry.
These reciprocal relations can be extended in a controlled context beyond equilibrium, such as for systems in the presence of magnetic fields (the resulting relations are sometimes referred to as Onsager-Casimir relations~\cite{Casimir1945, DeGroot2013}).
This framework of linear irreversible thermodynamics is fundamentally rooted in the ``local equilibrium hypothesis''~\cite{Prigogine1968, DeGroot2013}, which conceptualizes nonequilibrium systems of interest as being composed of local thermodynamic subsystems in equilibrium. 
While this theoretical perspective provides a deeper understanding of some systems, there is recent interest in transport phenomena in systems in which the local equilibrium hypothesis does not hold, including systems comprised of active matter~\cite{Howse2007, Marchetti2013, Reichhardt2014, Bechinger2016, Ruben2017, Hargus2020, Hargus2021, Reichert2021, Solon2022, Fruchart2023, Hargus2025}. 
In these systems, the driving force for motion is inherently not thermodynamic in origin, but linear laws (e.g., a diffusive response to concentration gradients) often appear to hold.

It is convenient to frame our discussion of irreversible processes by distinguishing two classes of transport phenomena: transport driven by external forces that explicitly appear in the microscopic equations of motion and transport in response to spatial gradients in field variables. 
These two different forms of transport are referred to as ``mechanical'' and ``thermal'' transport by Kubo et al.~\cite{Kubo1957II}.
The importance of distinguishing these forms of transport was recently emphasized by Hargus et al. in the context of active systems~\cite{Hargus2021, Hargus2025}.
Here, we refer to Kubo's thermal transport as ``gradient'' transport to emphasize that such processes can also occur in athermal systems.
Linear mechanical and gradient transport are characterized by coefficients that, in equilibrium, are related through thermodynamic factors in the form of Einstein relations. 
The breakdown of these Einstein relations for nonequilibrium systems has motivated the pursuit of independent theories of mechanical and gradient transport coefficients. 
Mechanical transport coefficients can be understood in and out of equilibrium through the use of path ensemble frameworks, which generalize linear response theory~\cite{Baiesi2009,Baesi2013,Lesnicki2020, Singh2025}. 
Similarly, Green-Kubo relations~\cite{Green1954, Kubo1957I} for the determination of gradient coefficients are found to hold even out of equilibrium by postulating that fluctuations generate a flux consistent with a linear constitutive relation~\cite{Hargus2025}. 
The absence of a single framework that can describe both classes of transport processes for systems that do not admit states of local equilibrium is notable.

The local equilibrium hypothesis allows for the identification of the thermodynamic driving forces for transport, provides fundamental limits on the structure of transport coefficients, and results in the Einstein relations connecting mechanical and gradient transport coefficients. 
Absent local equilibrium, what fundamentally drives the motion of systems, what is the structure of the linear transport coefficients, and is there a connection between gradient and mechanical transport?  
These questions motivate our development of an entirely \emph{mechanical} framework of transport phenomena that no longer relies on the local equilibrium hypothesis.
Our approach is rooted in deriving the \textit{exact} dynamics for the desired fluxes and the systematic expansion and coarse-graining of these equations of motion to simultaneously identify the linear driving force and Onsager tensor. 
We demonstrate our approach by deriving the transport relations governing \textit{multicomponent} species transport  -- allowing us to identify a mechanical definition of both the multicomponent Onsager mobility tensor (a mechanical transport coefficient) and the mutual diffusion tensor (a gradient transport coefficient).
Our analysis is sufficiently general to recover equilibrium properties of these coefficients and explicitly identify conditions under which Einstein relations break down in nonequilibrium settings.
We validate the mechanical transport framework by applying it to passive multicomponent systems, confirm the recovery of equilibrium transport properties, and extend the framework to explore transport phenomena in systems with nonreciprocal interactions, chiral active dynamics, and the nonlinear electric field response of electrolytes.

The proposed framework can be readily applied to rigorously formulate constitutive relations for other forms of transport, including heat and momentum transfer. 
The mechanical approach further allows us to identify useful computational techniques for determining $\mathbf{L}$ numerically through what is traditionally called nonequilibrium molecular dynamics (NEMD), which we refer to here as a ``color field'' approach.
While NEMD was originally developed for systems with microscopic time-reversal symmetry~\cite{Ciccotti1975,Ciccotti2005,Evans1982,Evans1984,Evans2008}, we demonstrate that the color field approach is applicable to intrinsically nonequilibrium systems.

\section{The Mechanics of Multicomponent Transport}
\label{sec:mechanics_of_multicomponent_transport}
We consider a system consisting of $n_c$ distinct species in $d$ spatial dimensions.
The species could be atomic/molecular in nature or could represent larger coarse-grained entities (e.g., colloids, bacteria, supramolecules). 
For each species, the equation of motion for the underlying microscopic dynamics, which need not satisfy time-reversal symmetry, will eventually be required for our approach.
We consider the local number density field of species $i$, $\rho_i(\mathbf{x}; t)$, which is suitably averaged (over space, time, and/or noise) such that each field is smooth.
Here, we do not consider local sources or sinks of particles such that the density fields satisfy the continuity equation:
\begin{equation}
\label{eq:density_dynamic}
    \frac{\partial \rho_i}{\partial t} = - \boldsymbol{\nabla} \cdot \mathbf{J}_i,
\end{equation}
where $\mathbf{J}_i(\mathbf{x}; t) \equiv \rho_i(\mathbf{x}; t)\mathbf{v}_i(\mathbf{x}; t)$ is the absolute flux of species $i$, $\mathbf{v}_i$ is the local velocity field of the species, and $\boldsymbol{\nabla}\equiv \partial / \partial \mathbf{x}$. 
We consider the following constitutive equation for the species flux: 
\begin{subequations}
\begin{align}
\label{eq:flux_onsager}
    &\mathbf{J}_i = \sum_j^{n_c} \mathbf{L}_{ij} \cdot \mathbf{f}_j,
\end{align}
where $\mathbf{f}_i$ is a generalized ``direct force'' driving the flux of species $i$ and $\mathbf{L}_{ij}$ is an Onsager transport coefficient tensor coupling the fluxes and forces.
The direct force $\mathbf{f}_i$ includes externally applied forces (e.g., gravity, electric fields, etc.) as well as ``internal'' forces (e.g., those that may arise from density gradients).
We note that while we express the above equation for \textit{absolute} flux, we will find for truly Galilean invariant systems that we can only describe \textit{relative} fluxes using this kind of constitutive equation.
Crucially, $\mathbf{L}_{ij}$ and $\mathbf{f}_j$ are \textit{independent} of $\mathbf{J}_i$, such that Eq.~\eqref{eq:flux_onsager} is truly a linear constitutive equation describing ``small'' fluxes. 
Moreover, this constitutive relation is \textit{local} in space and time: the Onsager transport coefficients and the direct forces depend on the same location and time at which we seek to evaluate the fluxes. 
The linear form of this equation suggests a more compact expression:
\begin{align}
    \mathbf{J} = \mathbf{L}\cdot\mathbf{f},\label{eq:flux_onsager_mat}
\end{align}
\end{subequations}
where $\mathbf{J}_i$ is now a sub-vector of a larger vector $\mathbf{J}$ with $dn_c$ components of all of the species fluxes. 
Similarly, tensors of any rank with Roman indices (e.g.,  $\mathbf{L}_{ij}$) are understood as sub-tensors of a larger tensor (denoted without Roman indices e.g., $\mathbf{L}$) for all species.

Our aim is to formulate a procedure for determining the form of the generalized forces and transport coefficients that is applicable to systems that do not admit a local equilibrium. 
The formulation must recover the results of linear irreversible thermodynamics~\cite{DeGroot2013, Prigogine1968} for \textit{passive} systems, which results in thermodynamic contributions to $\mathbf{f}_i$ such as the spatial gradient of the chemical potential and the celebrated Onsager reciprocal relations of $\mathbf{L}$. 
Our starting point is to examine the \textit{exact} dynamics of the species fluxes by noting their intimate connection to the species \textit{momentum densities}. 
The momentum density, $m_i\rho_i\mathbf{v}_i = m_i\mathbf{J}_i$ (where $m_i$ is the mass of a particle of species $i$), is rarely considered in continuum mechanics, as even for passive systems, it is not conserved. 
Nevertheless, the connection between the species absolute flux and momentum density leads us to express the exact species momentum balance (see 
Appendix~\ref{sec:appendix_species_momentum_balance}):
\begin{equation}
\label{eq:species_momentum_balance}
    m_i \frac{\partial}{\partial t}\mathbf{J}_i + m_i \boldsymbol{\nabla}\cdot(\mathbf{J}_i\mathbf{J}_i /\rho_i) = \rho_i\mathbf{f}_i^{\rm eff},
\end{equation}
where the effective force $\mathbf{f}_i^{\rm eff}(\mathbf{x};t)$ acting on particles of species $i$ has an explicit microscopic form given in Appendix~\ref{sec:appendix_species_momentum_balance}.
These effective forces include external forces, nonconservative forces, ideal forces, and forces arising from interparticle interactions (reciprocal or otherwise).
The precise form of these forces thus intimately depends on the microscopic degrees of freedom and can be obtained through a systematic Irving-Kirkwood procedure~\cite{Irving1950}. 

From the species momentum balance [Eq.~\eqref{eq:species_momentum_balance}], we can immediately identify that the flux generally depends on the force and flux history with flux dynamics that are nonlinear and inhomogeneous. 
The inhomogeneous driving forces generating the flux are the effective forces, which themselves generally depend on the species fluxes (in addition to the density fields and any other system-dependent fields). 
As we ultimately restrict our analysis to linear constitutive equations, we consider fluxes that are near a steady state. 
For notational simplicity, we expand around zero flux here, but we will later examine expansions about finite flux (see Sec.~\ref{sec:nonlinear_transport}). 
We can perform a functional expansion\footnote{If this equation involves stochastic variables additional care is required depending on the interpretation and character of the noise.} of $\mathbf{f}^{\rm eff}$ with respect to the flux field:
\begin{subequations}
\label{eq:f_eff_nonlocal}
\begin{align}
    &\mathbf{f}^{\rm eff}(\mathbf{x}, t) = \mathbf{f}^{\rm static}(\mathbf{x}, t) \nonumber\\
    &\hspace{8pt}- \int_{-\infty}^t dt'\int_V d\mathbf{x}' \mathbf{R}(\mathbf{x}, \mathbf{x}',t,t')\cdot\mathbf{J}(\mathbf{x}',t') + \mathcal{O}(\|\mathbf{J}\|^2), \\
    &\mathbf{f}^{\rm static}(\mathbf{x},t) \equiv \left.\mathbf{f}^{\rm eff}(\mathbf{x}, t)\right|_{\mathbf{J}=\mathbf{0}},\\
    &\mathbf{R}(\mathbf{x}, \mathbf{x}',t,t') \equiv -\left.\frac{\delta \mathbf{f}^{\rm eff}(\mathbf{x}, t)}{\delta \mathbf{J}(\mathbf{x}', t')}\right|_{\mathbf{J}=\mathbf{0}},
\end{align}
\end{subequations}
where we have defined the resistance kernel $\mathbf{R}(\mathbf{x},\mathbf{x}',t,t')$ which determines how $\mathbf{f}^{\rm eff}(\mathbf{x}, t)$ varies with $\mathbf{J}(\mathbf{x}', t')$, and $V$ is the system volume.
We also introduce the effective force in the flux-free limit as the \textit{static forces} acting on each species, $\mathbf{f}^{\rm static}$. 
These forces include external forces independent of flux and internally generated forces described by the relevant field variables.
Furthermore, $\mathbf{f}^{\rm static}$ may also depend on the flux associated with other irreversible processes (e.g., heat flux in the case of heat transfer) present in the system.
We can understand $-\mathbf{f}^{\rm static}$ as the required force to exert such that the system remains flux-free in its current configuration.
With this perspective, we see $\mathbf{f}^{\rm static}$ generally represents a nonequilibrium force and can only reduce to thermodynamic forces in the absence of these other irreversible processes not explicitly considered. 
Ultimately, a dynamical description for each of the fields appearing in $\mathbf{f}^{\rm static}$ will be required to obtain a complete system of equations. 

In order to eventually recover a transport equation that is local in time and space like Eq.~\eqref{eq:flux_onsager_mat}, we can identify characteristic time and length scales that provide the resolution of $\mathbf{R}$ in time and space domains. 
We define the largest of these time and length scales demarcating the non-local dynamics as $\tau^{\rm NL}$ and $\lambda^{\rm NL}$ respectively (see Appendix~\ref{sec:app_locality}).
In this analysis we identify $\tau^{\rm NL}$ as a measure of microscopic reorganization time which determines how long $\mathbf{f}^{\rm eff}$ takes to respond to changes in $\mathbf{J}$.
For the validity of these linearized relations, we then self-consistently require that $\mathbf{J}$ varies over timescales much longer than $\tau^{\rm NL}$.
We find similar restrictions on gradients of $\mathbf{J}$ from $\lambda^{\rm NL}$, and these timescales and lengthscales help define the regime of applicability for the theory.
If we consider scenarios in which the system flux varies slowly in time  (slower than $\tau^{\rm NL}$) with small spatial variations over distances of $\lambda^{\rm NL}$, the resistance kernel can be well approximated as ${\mathbf{R}(\mathbf{x},\mathbf{x}',t,t')=\boldsymbol{\mathcal{R}}(\mathbf{x},t)\delta(\mathbf{x}- \mathbf{x}')\delta(t-t')}$ where  $\boldsymbol{\mathcal{R}}$ is a local resistance tensor.
In these limits, we arrive at an \textit{entirely local} expansion for the effective force with respect to the flux:
\begin{subequations}
\label{eq:effective_force_expansion}
\begin{align}
    & \mathbf{f}^{\rm eff}(\mathbf{x},t) = \mathbf{f}^{\rm static}(\mathbf{x},t) -  \boldsymbol{\mathcal{R}}(\mathbf{x},t) \cdot \mathbf{J}(\mathbf{x},t) + \mathcal{O}(\|\mathbf{J}\|^2),\\
    & \boldsymbol{\mathcal{R}}(\mathbf{x},t) \equiv - \left.\frac{\partial \mathbf{f}^{\rm eff}(\mathbf{x},t)}{\partial\mathbf{J}(\mathbf{x},t)} \right|_{\ \mathbf{J} = \mathbf{0}},\label{eq:resistance_tensor} \\
    &\boldsymbol{\mathcal{R}}(\mathbf{x},t) = \int_{-\infty}^t dt' \int_V d\mathbf{x}' \mathbf{R}(\mathbf{x},\mathbf{x}',t,t'),
\end{align}
\end{subequations}
where we find two compatible definitions for $\boldsymbol{\mathcal{R}}$ as a derivative of the \textit{steady state} effective force response with respect to flux evaluated at the current configuration and in the absence of any flux, and as an integral over space and time of the resistance kernel $\mathbf{R}$.
We emphasize here that $\boldsymbol{\mathcal{R}}$ itself is independent of flux variations but will generally depend on precisely the same fields as the static forces. 
This expanded force expression, while strictly linear in the forces and fluxes, can be nonlinear in the relevant fields that describe the system (e.g., both $\mathbf{f}^{\rm static}$ and $\boldsymbol{\mathcal{R}}$ can be nonlinear in the species density fields, density gradients, external fields, etc.). 

Substitution of our small flux force balance and the absence of large spatial gradients in the flux allows us to express the species momentum balance as:
\begin{equation}
\label{eq:species_momentum_balance_linear}
    m_i \frac{\partial}{\partial t}\mathbf{J}_i = \rho_i
    \left(\mathbf{f}_i^{\rm static} - \boldsymbol{\mathcal{R}}_{ij} \cdot \mathbf{J}_j \right) + \mathcal{O}(\|\mathbf{J}\|^2 + \|\nabla \cdot (\mathbf{J}\mathbf{J})\|),
\end{equation}
which importantly now results in \textit{linear} flux dynamics.
The linearization of the species momentum balance allows for the straightforward determination of the time-dependence of the species flux (see Appendix~\ref{sec:timescale} for a timescale analysis).
We identify the flux memory kernel and its associated momentum relaxation timescales, which are related to the time it takes the flux $\mathbf{J}$ to reach a steady state for a given value of force. 
These timescales are fundamentally connected to particle inertia, in contrast to those directly appearing in the resistance kernel (e.g., $\tau^{\rm NL}$).
The momentum relaxation timescales for $\mathbf{J}$ may vary spatially and directly follow from the eigenvalues of the tensor $\mathbf{A}$ with components:
\begin{align}
\label{eq:A_time_matrix}
    \mathbf{A}_{ij}(\mathbf{x}) \equiv \boldsymbol{\mathcal{R}}_{ij}(\mathbf{x})\rho_i(\mathbf{x})/m_i,
\end{align}
with each eigenvalue representing an (inverse) inertial relaxation time.
While the equations of motion for the species flux are local and Markovian for appropriately large timescales and lengthscale variations, we consider the overdamped limit -- the limit in which the time variation of the flux in Eq.~\eqref{eq:species_momentum_balance_linear} can be discarded -- such that the flux is independent of its history.
This limit is dictated by considering dynamics on timescales much longer than the longest inertial timescale in the system $\tau$:
\begin{align}
\frac{1}{\tau}=\inf_\mathbf{x}\{\Re(\lambda_{\rm min}(\mathbf{A}(\mathbf{x}))\},
\end{align}
where $\lambda_{\rm min}$ is a function that returns the eigenvalue with the smallest real part of its tensor argument, and $\Re(\lambda)$ returns the real part of the eigenvalue argument. 

If the real part of any of the eigenvalues of $\mathbf{A}$ is less than zero, this indicates the presence of non-dissipative forces.
That is, relative motion between particles of different species produces forces that promote rather than dampen this motion. 
This acceleration of the flux associated with these modes would eventually push the system away from a given steady state.
This kind of motion, absent for passive systems, cannot be described by linear constitutive relations of the form given in Eq.~\eqref{eq:flux_onsager_mat} and is therefore not considered in this work. 
Similarly, a zero eigenvalue implies that there are combinations of fluxes that induce no dampening forces in the system, meaning the values of the flux are dependent on their own history.
We anticipate zero eigenvalue modes (one for each of the $d$ spatial dimensions) associated with the collective motion of systems that are Galilean invariant. 
As we will discuss in detail later, coupling the system dynamics to the environment through dissipative forces breaks Galilean invariance and eliminates these expected zero eigenvalue modes.
To ensure the existence of the overdamped limit, we model fluxes and forces for systems such that $\mathbf{A}$ has only non-negative eigenvalues with potential zero eigenvalues corresponding to the undamped collective motion of the system. 

We first analyze the case of full-rank $\boldsymbol{\mathcal{R}}$.
For timescales $t \gg \tau$, our species momentum balance reduces to:
\begin{equation}
    \mathbf{0} \stackrel{\vphantom{\big|}{t \gg \tau}}{=\joinrel=}
    \mathbf{f}^{\rm static} - \boldsymbol{\mathcal{R}} \cdot \mathbf{J}  + \mathcal{O}(\|\mathbf{J}\|^2 + \|\nabla \cdot (\mathbf{J}\mathbf{J})\|).
\end{equation}
We see that at these timescales up to linear order in the fluxes, we recover a dynamical force balance, with a linear relationship between $\mathbf{f}^{\rm static}$ and $\mathbf{J}$.
Rearranging this force balance, we arrive at:
\begin{subequations}
\begin{equation}
\label{eq:flux_beff}
    \mathbf{J} =   \boldsymbol{\mathcal{R}}^{-1} \cdot \mathbf{f}^{\rm static},
\end{equation}
where the inversion of the tensor is an inversion of both the spatial dimension and the species dimension.
Equation~\eqref{eq:flux_beff} is precisely of the form of the postulated linear transport relation [Eq.~\eqref{eq:flux_onsager}] with:
\label{eq:mechanial_equations_Landf}
\begin{equation}
\label{eq:L_mech}
    \mathbf{L} = \boldsymbol{\mathcal{R}}^{-1},
\end{equation}
\begin{equation}
    \mathbf{f} = \mathbf{f}^{\rm static}.
\end{equation}
\end{subequations}
We have thus identified the \textit{mechanical origins} of both the driving forces and Onsager transport coefficients without appealing to equilibrium notions. 

Before proceeding further, it is important to first clarify the special case of Galilean-invariant systems. 
Such systems yield a rank deficient tensor $\mathbf{A}$ [see Eq.~\eqref{eq:A_time_matrix}], indicating the existence of modes without an overdamped limit. 
In practical scenarios, strict Galilean invariance may be weakly broken due to coupling with an external heat bath or environment which introduces a corresponding reference velocity. 
However, when this coupling is weak, collective motion still dominates the flux behavior.
In these cases, it can be convenient to define flux relative to the system’s center-of-mass velocity or another suitably transformed reference frame, as described in Appendices~\ref{sec:app_markovian_transport_gi_systems} and~\ref{sec:app_alternative_relative_flux_definitions}.
We can recover a linear relationship for the diffusive flux, relative to the center-of-mass frame, and the static forces:
\begin{align}
    \mathbf{J}^{\rm com} = \mathbf{L}^{\rm com} \cdot \mathbf{f}^{\rm static},
\end{align}
where $\mathbf{L}^{\rm com}$ is the pseudoinverse of $\boldsymbol{\mathcal{R}}$ (or the inverse, in the presence of a weak non-zero coupling), up to projection operators on the left and right. 
For simplicity, in the subsequent analysis, we assume that $\boldsymbol{\mathcal{R}}$ is invertible, but note that this analysis remains applicable to Galilean invariant systems with the appropriate transformations.

\subsection{Transport Properties of Multicomponent Passive Systems}
\label{sec:passive_transport}

The mechanical formulation of linear multicomponent transport provides a rigorous basis for determining the driving forces for species flux without appealing to irreversible thermodynamics and the local equilibrium hypothesis. 
This section investigates the transport properties of passive systems from a mechanical perspective. 
Specifically, we present the form of the resistance tensor (or equivalently, the Onsager transport tensor) and the driving forces for motion for a passive pairwise interacting system.
Additionally, we analyze the general structure of the resistance tensor to ensure the invertibility of $\boldsymbol{\mathcal{R}} $ and recover the expected Onsager reciprocal relations.

We first consider a passive pairwise interacting system at equilibrium within a bath of fixed temperature $T$.
A standard Irving-Kirkwood procedure results in the following effective force acting on particles of species $i$~\cite{Irving1950,Chiu2024}:
\begin{subequations}
\label{eq:f_eff_passive_molecular}
\begin{align}
    &\mathbf{f}_i^{\rm eff} = \mathbf{f}^{\rm ideal}_i  +\mathbf{f}^{\rm int}_i + \mathbf{f}^{\rm dis}_i,\\
    &\mathbf{f}^{\rm ideal}_i(\mathbf{x}) = - \frac{1}{\rho_i(\mathbf{x})} \boldsymbol{\nabla} \left( k_BT \rho_i(\mathbf{x}) \right),\\
    &\mathbf{f}^{\rm int}_i(\mathbf{x}) = \sum_{k}^{n_c} \int_V d \mathbf{x}' \rho_k(\mathbf{x}') g_{ik}(\mathbf{x},\mathbf{x}'; \{\mathbf{v}_i\}) \mathbf{F}_{ik}(\mathbf{x}-\mathbf{x}'),\label{eq:f_int_microscopic_pairwise}
\end{align}
\end{subequations}
where we have divided the effective force into an ideal contribution $\mathbf{f}^{\rm ideal}$, a contribution from pairwise interactions $\mathbf{f}^{\rm int}$, and a dissipative contribution arising from the coupling of the system to a heat bath that we keep general and label $\mathbf{f}^{\rm dis}$. 
For Langevin systems, $\mathbf{f}^{\rm dis}$ manifests as a constant body force proportional to species flux, but for molecular systems, $\mathbf{f}^{\rm dis}$ may exclusively occur at system boundaries. 
In the latter case, it may be useful to either incorporate an effective volumetric drag or incorporate this drag through boundary conditions (e.g., through a ``no-slip'' boundary condition). 
In general, $\mathbf{f}^{\rm dis}$ breaks the Galilean invariance of the system as the relative velocity between the system and its environment now impacts the dynamics (although taken together the system and environment may still retain Galilean invariance). 
We also introduce $g_{ij}(\mathbf{x},\mathbf{x}'; \{\mathbf{v}_i\})$, the pair distribution function, which provides a measure of the likelihood of finding a particle of species $j$ located at $\mathbf{x}'$ given that a particle of species $i$ is located at $\mathbf{x}$. 
This pair distribution function depends on the ensemble, and we explicitly express the dependence of $g_{ij}$ on the species velocities  $\{\mathbf{v}_i\}$ to emphasize their role in setting the ensemble.
The pairwise interaction force between species $i$ and $j$ is denoted as $\mathbf{F}_{ij}(\mathbf{x}- \mathbf{x}')$, which is conservative for passive systems $\mathbf{F}_{ij}(\mathbf{x}- \mathbf{x}') \equiv - \partial U_{ij} / \partial (\mathbf{x}- \mathbf{x}')$.

We can determine the complete resistance tensor and the static force that drives motion [see Eq.~\eqref{eq:effective_force_expansion}] using the effective force defined in Eq.~\eqref{eq:f_eff_passive_molecular}:
\begin{subequations}
\label{eq:resistence_passive_molecular}
    \begin{align}
        &\boldsymbol{\mathcal{R}} = \boldsymbol{\mathcal{R}}^{\rm int} + \boldsymbol{\mathcal{R}}^{\rm dis}, \label{eq:R_passive}\\
        &{\boldsymbol{\mathcal{R}}^{\rm int} = - \left.\frac{\partial \mathbf{f}^{\rm int}}{\partial\mathbf{J}}\right|_{\mathbf{J} = \mathbf{0}}},\\ &\boldsymbol{\mathcal{R}}^{\rm dis} = - \left.\frac{\partial \mathbf{f}^{\rm dis}}{\partial \mathbf{J}}\right|_{\mathbf{J} = \mathbf{0}},\\
        &\mathbf{f}^{\rm static} = \left[\mathbf{f}^{\rm ideal} + \mathbf{f}^{\rm int} + \mathbf{f}^{\rm dis}\right]\Big|_{\mathbf{J} = \mathbf{0}}.
    \end{align}
\end{subequations} 
One consequence of our locality assumption in Eq.~\eqref{eq:effective_force_expansion} is that the strength of the interacting force depends only locally on the value of $\mathbf{J}$.
For systems which satisfy the local equilibrium hypothesis as outlined in Appendix~\ref{app:local_equilibrium_hypothesis}, this static force was identified~\cite{Chiu2024} in the absence of temperature gradients as the anticipated thermodynamic driving force~\cite{DeGroot2013} with $\mathbf{f}^{\rm static} = -\boldsymbol{\nabla}\boldsymbol{\mu}$ where $\boldsymbol{\mu}$ is a vector of species chemical potentials. 
In the case of a pairwise interacting system, we can arrive at an explicit expression for $\boldsymbol{\mathcal{R}}^{\rm int}$ in terms of derivatives of the species pair distribution functions:
\begin{align}
    \label{eq:r_int_pair_distribution}\boldsymbol{\mathcal{R}}^{\rm int}_{ij}(\mathbf{x}) = -\sum_k^{n_c} \int d\mathbf{x}' \rho_k(\mathbf{x}')\left.\frac{\partial g_{ik}(\mathbf{x}, \mathbf{x}')}{\partial \mathbf{J}_j}\right|_{\mathbf{J}=\mathbf{0}}\mathbf{F}_{ik}(\mathbf{x} - \mathbf{x}').
\end{align}
We note that this interaction contribution to the resistance is entirely determined by the response of $g_{ij}(\mathbf{x}, \mathbf{x}')$ to the flux. 
Similar quantities have emerged in theoretical treatments of the conductivity of electrolytes~\cite{Debye1923, Fuoss1955, onsager1969, Blum1975, Bernard1992, Lesnicki2020, Lesnicki2021, Bernard2023} as well as in the context of microrheology~\cite{Squires2005, Squires2010}.
If we now consider a \textit{Langevin} system, where dissipation occurs through a drag that an implicit medium exerts on particles of species $i$, $ \mathbf{f}^{\rm dis}_i$ takes the form:
\begin{align}
\label{eq:f_dis}
    \mathbf{f}^{\rm dis}_i = -\boldsymbol{\zeta}_i\cdot\left(\mathbf{v}_i - \mathbf{v}^{\rm med}\right),
\end{align}
where $\boldsymbol{\zeta}_i$ is the single-particle resistance for species $i$ (assumed to act on each particle independently for simplicity), and $\mathbf{v}^{\rm med}$ is the velocity of the background medium which breaks the system Galilean invariance and is taken to be $\mathbf{0}$ for convenience.
The dissipative force can now equivalently be expressed as:
\begin{subequations}
\label{eq:f_eff_passive_langevin}
\begin{align}
    &\mathbf{f}^{\rm dis} = -\boldsymbol{\mathcal{R}}^{\rm dis} \cdot \mathbf{J}, \\
    &\boldsymbol{\mathcal{R}}^{\rm dis}_{ij} = \boldsymbol{\zeta}_i /\rho_i \delta_{ij} \label{eq:R_dis_langevin}. 
\end{align}
\end{subequations}
With the explicit forms of the interaction resistance tensor $\boldsymbol{\mathcal{R}}^{\rm int}$ [Eq.~\eqref{eq:r_int_pair_distribution}], the dissipative resistance tensor $\boldsymbol{\mathcal{R}}^{\rm dis}$ [Eq.~\eqref{eq:R_dis_langevin}], and the total resistance tensor in Eq.~\eqref{eq:R_passive}, we now obtain the complete expression for the resistance tensor in a Langevin passive system.
We can appreciate that the drag from the medium simply adds the $\boldsymbol{\mathcal{R}}^{\rm dis}$ contribution to the resistance tensor.
By examining the general tensor form of $\boldsymbol{\mathcal{R}}$ in Eq.~\eqref{eq:R_passive}, we conclude that if $\boldsymbol{\mathcal{R}}^{\rm int}$ is positive semi-definite, and every component of $\boldsymbol{\zeta}_i$ is positive, the resistance tensor $\boldsymbol{\mathcal{R}}$ remains positive definite, ensuring its invertibility, the existence of an overdamped limit, and a well-defined Onsager mobility $\mathbf{L}$.

The Onsager reciprocal relations state that we expect $\mathbf{L}$ to be symmetric in the passive limit, so the resistance tensor $\boldsymbol{\mathcal{R}}$ must also be symmetric.
For the systems and resistance tensor under consideration, it is clear that $\boldsymbol{\mathcal{R}}^{\rm dis}$ [Eq.~\eqref{eq:R_dis_langevin}] is symmetric. 
The symmetry of $\boldsymbol{\mathcal{R}}$ thus rests on the symmetry of $\boldsymbol{\mathcal{R}}^{\rm int}$. 
The necessary and sufficient condition for $\mathbf{L}$ to be symmetric is therefore:
\begin{equation}
\label{eq:symmetry_relation}
    \left.\frac{\partial \mathbf{f}^{\rm int}_i}{\partial \mathbf{J}_j}\right|_{\mathbf{J}=\mathbf{0}} = \left. \frac{\partial \mathbf{f}^{\rm int}_j}{\partial \mathbf{J}_i}\right|_{\mathbf{J} = \mathbf{0}}.
\end{equation}
These derivatives can be explicitly computed through the microscopic expression for $\mathbf{f}^{\rm int}$ provided in Eq.~\eqref{eq:f_int_microscopic_pairwise}.
However, it is often easier to make statements about the symmetry of transport coefficients by deriving Green-Kubo formulae for the coefficients in question. 
This connection between system symmetries and the symmetry of transport coefficients, even out of equilibrium,  has been emphasized and demonstrated in recent years by Mandadapu and co-workers in the context of odd transport phenomena~\cite{Epstein2020,Hargus2021, Hargus2025}. 
With our focus now on \textit{passive} systems near equilibrium, we leverage standard linear response theory to derive a Green-Kubo formula that allows us to straightforwardly understand the symmetry of $\boldsymbol{\mathcal{R}}$.

As we consider systems in contact with their environment (through an implicit \textit{equilibrium} medium), the equilibrium probability distribution for particle configurations is the Boltzmann distribution:
\begin{equation}
    f_0 \propto \exp\left[-\beta \mathcal{U}(\{\mathbf{r}_\alpha\})\right],
\end{equation}
where $\mathcal{U}(\{\mathbf{r}_\alpha\})$ is the potential energy of a configuration defined by the positions of all particles $\{\mathbf{r}_\alpha\}$, and $\beta \equiv 1/k_BT$ is the inverse thermal energy.
We are ultimately interested in computing the terms appearing in Eq.~\eqref{eq:symmetry_relation} to verify the symmetry of $\boldsymbol{\mathcal{R}}$ and thus require a description of $\mathbf{f}^{\rm int}$ out of equilibrium. 
For a system with translational invariance, we can relate $\mathbf{f}^{\rm int}$ to the total interaction force ${\mathbf{F}_i = \sum_{\alpha }^{N_i} \mathbf{F}_i^\alpha}$ exerted on all $N_i$ particles of species $i$ where ${\mathbf{F}_i^\alpha = -\frac{\partial \mathcal{U}}{\partial \mathbf{r}_i^\alpha}}$ are the conservative interactions force. 
The connection between $\mathbf{F}_i$ and the interaction contribution to the effective force $\mathbf{f}_i^{\rm int}$ for a translationally invariant system is then simply:
\begin{align}
    \mathbf{f}^{\rm int}_i = \frac{1}{V\rho_i} \langle \mathbf{F}_i\rangle,
\end{align}
where the expectation is over the appropriate distribution for the ensemble under consideration. 
As detailed in Appendix~\ref{sec:app_linear_response}, the derivative of $\mathbf{f}^{\rm int}_i$ with respect to $\mathbf{J}_j$ can be interpreted as the difference between the expectation of $\mathbf{F}_i$ in a perturbed ensemble (where the velocities of species $j$ are shifted by a small amount) and its equilibrium expectation.
We can therefore arrive at a familiar Green-Kubo form for $\boldsymbol{\mathcal{R}}^{\rm int}$:
\begin{align}
\label{eq:interaction_symmetry}
    -\boldsymbol{\mathcal{R}}^{\rm int}_{ij} = \frac{\partial \mathbf{f}^{\rm int}_i}{\partial \mathbf{J}_j} = -\frac{\beta}{V\rho_i\rho_j}\int_0^\infty \langle \mathbf{F}_i(t)\mathbf{F}_j(0)\rangle_0 dt,
\end{align}
where the expectation is taken over the equilibrium distribution (see Appendix~\ref{sec:app_linear_response}). 
From Eq.~\eqref{eq:interaction_symmetry}, we can conclude that under time-reversal symmetry, the relation in Eq.~\eqref{eq:symmetry_relation} is indeed satisfied and therefore consistent with the expected symmetry of the Onsager tensor ($\mathbf{L}_{ij} = \mathbf{L}_{ji}$) for passive systems.
If the system additionally satisfies spatial parity symmetry, this implies:
\begin{align}
    \mathbf{L}_{ij} = \mathbf{L}_{ij}^\intercal,
\end{align}
with $\mathbf{B}^\intercal$ denoting a transpose of $\mathbf{B}$ over spatial dimensions.
Combining the results for parity and time-reversal symmetry, we find:
\begin{align}
    \mathbf{L}_{ij} = \mathbf{L}_{ji}^\intercal.
\end{align}
\emph{Spatially} odd transport is thus absent in passive systems satisfying spatial parity, as anticipated~\cite{Hargus2025}. 
With this derivation, we find that the proposed mechanical framework results in an Onsager transport tensor that recovers the Onsager reciprocal relations for parity-satisfying passive systems.

To conclude the discussion on passive systems, we note that when there is no net force on the system, the interaction contribution to the resistance tensor $\boldsymbol{\mathcal{R}}^{\rm int}$ satisfies:
\begin{align}
    \label{eq:null_space_of_interactions}
    \sum_j^{n_c} \boldsymbol{\mathcal{R}}_{ij}^{\rm int} \rho_j = \frac{\beta}{V\rho_i}\int_0^\infty \left\langle \mathbf{F}_i(t)\sum_j^{n_c} \mathbf{F}_j(0)\right\rangle dt= \mathbf{0},
\end{align}
which results from $\sum_j^{n_c}\mathbf{F}_j = \mathbf{0}$ in the absence of external forces. 
Indeed, we can observe from Eq.~\eqref{eq:null_space_of_interactions} that $\boldsymbol{\mathcal{R}}^{\rm int}$ possesses a null space corresponding to coordinated motion of the form ${\mathbf{J}_i = \rho_i \mathbf{u}}$ for any constant velocity vector $\mathbf{u}$, characteristic of Galilean invariance, as discussed in Appendix~\ref{sec:app_markovian_transport_gi_systems}.
Therefore, for equilibrium systems, any breaking of Galilean invariance is expected to arise from dissipation rather than through $\boldsymbol{\mathcal{R}}^{\rm int}$.
Consequently, the invertibility of the resistance tensor $\boldsymbol{\mathcal{R}}$ is determined by the structure of $\boldsymbol{\mathcal{R}}^{\rm dis}$.

The derived force correlation relationship implies symmetry of $\boldsymbol{\mathcal{R}}$ in the case of passive systems, but we can use mechanical expressions to understand the general structure of $\boldsymbol{\mathcal{R}}$ in nonequilibrium contexts where we still expect linear transport relations to hold, such as active matter systems.
In many of these nonequilibrium systems, time-reversal symmetry of the microscopic dynamics is not satisfied, and the lack of this symmetry plays a key role in the departure from the Onsager reciprocal relations.
In Sec.~\ref{sec:application}, we demonstrate this departure through two examples: mixtures with nonreciprocal interactions, and a collection of chiral active Brownian particles.

\subsection{Mutual Diffusion}
\label{sec:mutual_diffusion}
Equation~\eqref{eq:mechanial_equations_Landf} represents a mechanical description of linear species transport without asserting a local equilibrium hypothesis. 
A subset of forces that can drive species motion includes those proportional to density gradients. 
These \textit{diffusive} contributions to the flux take the form:
\begin{equation}
\label{eq:diffusion_equation}
    \mathbf{J}^{\rm diff}_i = -\sum_j^{n_c} \mathbf{D}_{ij} \cdot \boldsymbol{\nabla}\rho_j,
\end{equation}
where we have introduced the mutual diffusion tensor, $\mathbf{D}_{ij}$, that couples the diffusive flux of species $i$ with the density gradients of species $j$, and $\mathbf{J}^{\rm diff}$ is the flux response.
The mutual diffusion tensor is itself a linear transport coefficient that is central to determining how density fluctuations evolve over space and time in the absence of any other contributions to the flux.
In fact, straightforward linear stability analysis~\cite{Cross1993, kato1995, Hoyle2006, Cross2009} about uniform concentrations reveals that the mutual diffusion tensor entirely governs the linear stability of density fluctuations.
Recent studies further demonstrate that such analyses can predict the emergence of traveling states for active systems, including those with mixtures of active passive particles and systems with nonreciprocal interactions~\cite{Wittkowski2017, Saha2020, You2020, Fruchart2021, Greve2025}. 
Despite its centrality in understanding the stability of far-from-equilibrium mixtures, its structure has often been an input into phenomenological models~\cite{Saha2020, You2020}.

The mechanical perspective now allows us to straightforwardly relate this diffusion tensor to a product of the Onsager tensor $\mathbf{L}$ and contributions to the static forces proportional to density gradients.
To see this, we first expand $\mathbf{f}^{\rm static}$ with respect to density gradients about a species flux-free uniform density state:
\begin{subequations}
\label{eq:force_jacobian_definition}
\begin{align}
    &\mathbf{f}^{\rm static} = -\boldsymbol{\mathcal{F}}\cdot \boldsymbol{\nabla}\boldsymbol{\rho} + \mathcal{O}\big((\boldsymbol{\nabla} \boldsymbol{\rho})^2\big), \\
    &\boldsymbol{\mathcal{F}}_{kj} = -\left.\frac{\partial\mathbf{f}^{\rm static}_k}{\partial\boldsymbol{\nabla}\rho_j}\right|_{\boldsymbol{\nabla}\boldsymbol{\rho}=\mathbf{0}},
\end{align}
\end{subequations}
where we have defined a ``force Jacobian'' tensor $\boldsymbol{\mathcal{F}}$ which relates the linear static force response to density gradients, and $\boldsymbol{\rho} \equiv \{\rho_1, \rho_2, \cdots \rho_{n_c}\}$ is a vector of species density. 
Substitution of this expanded $\mathbf{f}^{\rm static}$ into Eq.~\eqref{eq:flux_beff} and comparing the resulting flux with Eq.~\eqref{eq:diffusion_equation} allows us to identify:
\begin{align}
\label{eq:mutual_diffusion}
    \mathbf{D}_{ij} = \sum_k^{n_c} \mathbf{L}_{ik} \cdot \boldsymbol{\mathcal{F}}_{kj}.
\end{align}
We can now appreciate that the diffusion tensor is the product of an Onsager tensor and $\boldsymbol{\mathcal{F}}_{kj}$. 

For passive systems, both the Onsager tensor and force Jacobian matrices are symmetric.
The symmetry of the force Jacobian for passive systems can be appreciated through its connection to thermodynamics. 
For uniform temperature systems, the static forces driving motion are generated by chemical potential gradients with ${\mathbf{f}_k^{\rm static} = -\boldsymbol{\nabla}\mu_k}$. 
For transport driven by small density field gradients, we can equivalently express this driving force in terms of derivatives of the bulk chemical potential ${\mathbf{f}_k^{\rm static} = -\sum_j^{n_c} \partial \mu_k / \partial\rho_j \cdot\boldsymbol{\nabla}\rho_j}$.
This allows us to identify the force Jacobian of passive systems to simply be the Hessian matrix of the bulk free energy density $\mathpzc{a}^{\rm o}$ with:
\begin{equation}
\label{eq:static_force_hessian}
    \boldsymbol{\mathcal{F}}_{kj} \stackrel{\rm eq}{=\joinrel=} \frac{\partial^2 \mathpzc{a}^{\rm o}}{\partial \rho_k \partial \rho_j}\mathbf{I}_d,
\end{equation}
where $\mathbf{I}_d$ is the $d$-dimensional identity tensor.
We can observe that $\boldsymbol{\mathcal{F}}$ is necessarily symmetric in equilibrium.
For active systems, such as those with nonreciprocal interactions or other nonconservative forces, we demonstrate in Sec.~\ref{sec:application} that neither the Onsager transport tensor nor the force Jacobian tensor is generally expected to be symmetric.

It is instructive to compare the expression for the mutual diffusion tensor obtained from our mechanical perspective with the Green-Kubo (GK) expression for $\mathbf{D}$~\cite{Green1954, Hargus2025}. 
Green-Kubo relations for transport coefficients for active systems can be found by invoking Onsager's regression hypothesis at the level of the flux, as recently proposed in Ref.~\cite{Hargus2025}.
The generalized GK relation derived with the flux hypothesis for mutual diffusion coefficients takes the form~\cite{Hargus2025}:
\begin{multline}
\label{eq:GK_D_1}
     \lim_{\mathbf{k}\rightarrow\mathbf{0}} \Bigg[ \mathbf{E}_{im} \cdot \mathrm{i} \mathbf{k}
     + \int_0^t dt' \left\langle\hat{\mathbf{J}}_i(\mathbf{k},t')\hat{\mathbf{J}}_m(-\mathbf{k},0)\right\rangle \cdot {\rm{i}} \mathbf{k} \Bigg] \\
     = \sum_j^{n_c} \mathbf{D}_{ij} \lim_{\mathbf{k}\rightarrow\mathbf{0}}\left\langle\hat{\rho}_j(\mathbf{k},0)\hat{\rho}_m(-\mathbf{k},0)\right\rangle\cdot \mathrm{i} \mathbf{k}. 
\end{multline}
where $\mathbf{k}$ represents the wave vector associated with the spatial Fourier transform, $\hat{\mathbf{J}}_i$ and $\hat{\rho}_i$ are the microscopic definitions of the flux and density of species $i$, respectively, and ${\mathbf{E}_{im} \cdot \mathrm{i} \mathbf{k} \equiv \left\langle \hat{\mathbf{J}}_i(\mathbf{k},0)\hat{\rho}_m(-\mathbf{k},0)\right\rangle}$ is a static flux-density correlator.\footnote{
Equation~\eqref{eq:GK_D_1} corresponds to the more general form of the Green-Kubo relation (which includes the ${\mathbf{E}_{im}\cdot\mathrm{i}\mathbf{k}}$ term) derived in the Supplemental Material of Ref.~\cite{Hargus2025}. 
Here, the mutual diffusion tensor $D_{ij}$ plays the role of $M_{ij}$ in Ref.~\cite{Hargus2025}.}
All expectations in this GK relation are computed over a steady-state distribution, and here we consider a uniform, flux-free ensemble with both spatial and temporal translational invariance.
In equilibrium systems, velocities are uncorrelated with positions for finite particle inertia such that ${\mathbf{E}_{im}\cdot\mathrm{i}\mathbf{k}}$ vanishes~\cite{Kubo1957I}.
However, in systems with overdamped and/or nonequilibrium dynamics, $\mathbf{E}\cdot\mathrm{i}\mathbf{k}$ does not necessarily vanish~\cite{Chun2021}, and therefore, we retain this term to preserve the generality of the Green–Kubo relation in these contexts.

The connection between this GK relation and our mechanical perspective can be made clear for \textit{passive systems} by recognizing that the static density correlation function is also related to the free energy Hessian with:
\begin{subequations}
\label{eq:GK_hessian}
\begin{align}
    &\left(S\right)^{-1}_{mj} \stackrel{\rm eq}{=\joinrel=} \frac{\sqrt{\rho_m\rho_j}}{k_BT} \frac{\partial^2 \mathpzc{a}^{\rm o}}{\partial \rho_m \partial \rho_j},\\
    &S_{mj}  = \frac{1}{\sqrt{N_m N_j}} \lim_{\mathbf{k}\rightarrow \mathbf{0}}\left\langle \hat{\rho}_m(\mathbf{k}, 0) \hat{\rho}_j(-\mathbf{k}, 0)\right\rangle. \label{eq:structure_factor}
\end{align}
\end{subequations}
In equilibrium, the force Jacobian $\boldsymbol{\mathcal{F}}$ is thus directly related to the large wavelength limit of the structure factor ($S$), resulting in an Einstein relation connecting mutual diffusion and the Onsager transport. 
By assuming that all terms in Eq.~\eqref{eq:GK_D_1} become linear in $\mathbf{k}$ in the low-$\mathbf{k}$ limit, we can identify that the Onsager tensor $\mathbf{L}^{\rm GK}$ and force Jacobian $\boldsymbol{\mathcal{F}}^{\rm GK}$ in the GK relation for the diffusion tensor as:
\begin{subequations}
\label{eq:mutual_diffusion_gk}
\begin{align}
    &\mathbf{D}_{ij} = \sum_m^{n_c} \mathbf{L}^{\rm GK}_{im} \cdot \boldsymbol{\mathcal{F}}^{\rm GK}_{mj}, \label{eq:D_GK}\\
    &\mathbf{L}^{\rm GK}_{im} = \frac{1}{V} \lim_{\mathbf{k}\rightarrow\mathbf{0}} \Bigg[
    \mathbf{E}_{im}
    + \int_0^t dt' \left\langle\hat{\mathbf{J}}_i(\mathbf{k},t')\hat{\mathbf{J}}_m(-\mathbf{k},0)\right\rangle \Bigg], \label{eq:L_GK}\\
    &(\boldsymbol{\mathcal{F}}^{\rm GK})^{-1}_{mj} = \sqrt{\rho_m\rho_j}S_{mj}\mathbf{I}_d, \label{eq:F_GK}
\end{align}
\end{subequations}
where $V$ is the system volume.
We have defined $\mathbf{L}^{\rm GK}$ and $\boldsymbol{\mathcal{F}}^{\rm GK}$ such that we recover in equilibrium $k_BT{\boldsymbol{\mathcal{F}}^{\rm GK} \stackrel{\rm eq}{=\joinrel=} \boldsymbol{\mathcal{F}}}$.
We note that the familiar Green-Kubo relation for the Onsager tensor has the form $\mathbf{L}^{\rm GK}/k_BT$.
Here, we have chosen our definition of $\mathbf{L}^{\rm GK}$ so that it may be applied to athermal systems. 

We now have two expressions for the mutual diffusion tensor $\mathbf{D}_{ij}$ [Eqs.~\eqref{eq:mutual_diffusion} and \eqref{eq:mutual_diffusion_gk}] which we can equate to identify the general relation between $\mathbf{L}^{\rm GK}$ and $\mathbf{L}$:
\begin{align}
    \mathbf{L}^{\rm GK}_{im} = \sum_{k}^{n_c}\sum_{l}^{n_c} \mathbf{L}_{ik}\cdot\boldsymbol{\mathcal{F}}_{kl}\cdot(\boldsymbol{\mathcal{F}}^{\rm GK})^{-1}_{lm}.
\end{align}
In equilibrium, this relation simplifies using Eqs.~\eqref{eq:static_force_hessian} and \eqref{eq:GK_hessian}, yielding:
\begin{align}
\label{eq:L_mech_L_GK_eq}
    \mathbf{L}^{\rm GK} \stackrel{\rm eq}{=\joinrel=} k_BT\mathbf{L}.
\end{align}
However, out of equilibrium, $\mathbf{L}^{\rm GK}$ differs from $\mathbf{L}$ by a factor that captures the extent to which steady-state density fluctuations no longer encode the system’s mechanical response to density gradients.
For some nonequilibrium systems, there may still be an energy scale such that \emph{each component} of $\mathbf{L}$ is proportional to $\mathbf{L}^{\rm GK}$ multiplied by this energy.
This is an indication that the constant of proportionality plays the role of an effective temperature, and fluctuations still encode system response.
For general nonequilibrium systems, we do not expect this simple proportional relationship to hold, and in the absence of this relationship the Einstein relations which couple fluctuations (encoded by $\boldsymbol{\mathcal{F}}^{\rm GK}$) to forces (encoded by $\boldsymbol{\mathcal{F}}$) break down.
While we still expect the regression hypothesis to hold such that Green-Kubo relations correctly capture the effects of mutual diffusion, $\mathbf{L}^{\rm GK}$ can no longer be used in place of $\mathbf{L}$ in Eq.~\eqref{eq:flux_onsager} to capture system response to general forces applied to a given species.
Nevertheless, these two types of transport remain connected through the force Jacobian $\boldsymbol{\mathcal{F}}$.

\begin{figure}
    \centering
    \includegraphics[width=1.0\linewidth]{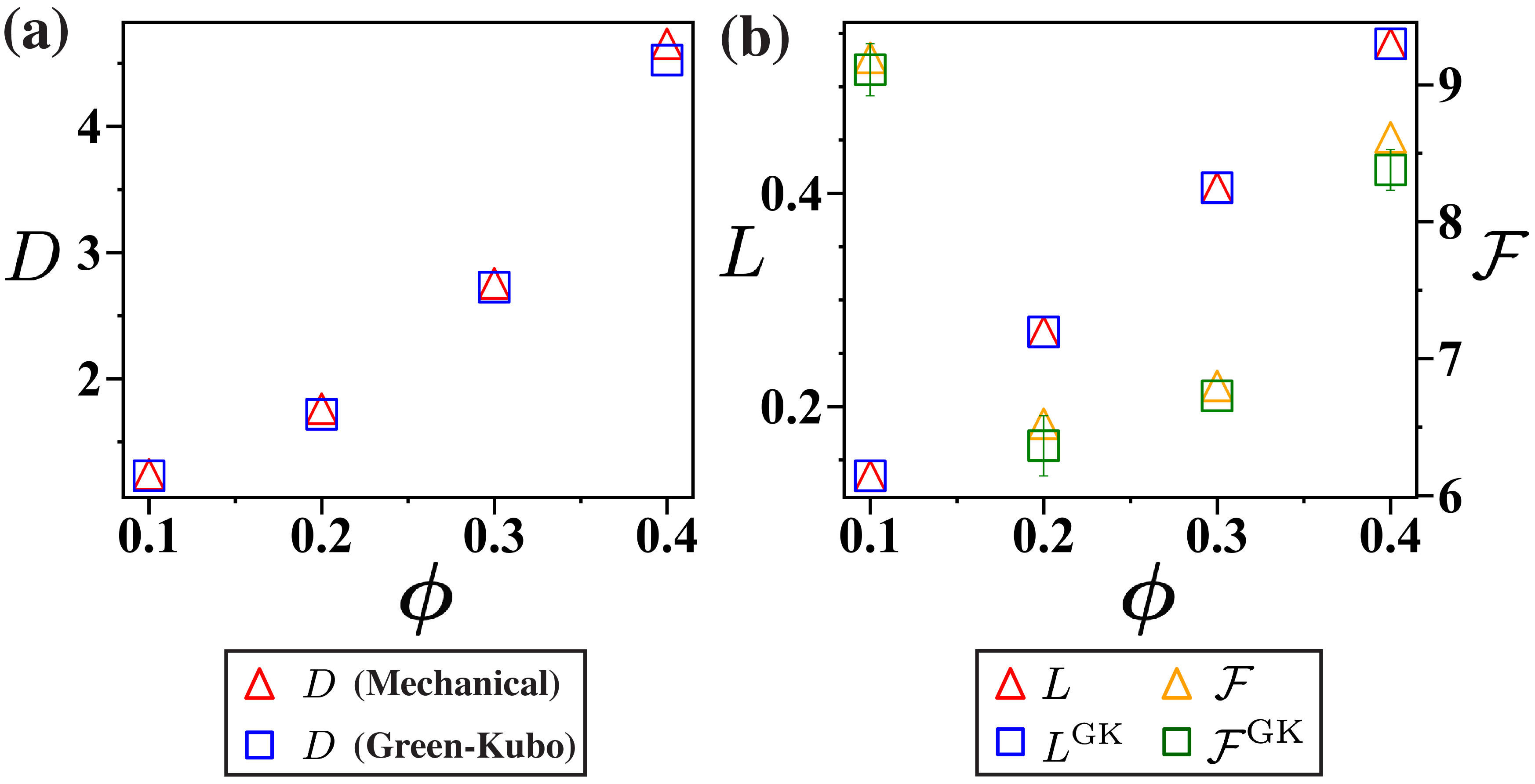}
    \caption{(a) Collective diffusion coefficient, (b) Onsager transport coefficient (left axis), and force Jacobian (right axis) for a passive one-component Langevin system, as computed using mechanical expressions [Eqs.~\eqref{eq:D_one_component_passive},~\eqref{eq:l_one_component_passive}, and~\eqref{eq:force_jacobian_one_component_passive}] and Green-Kubo expressions [Eqs.~\eqref{eq:D_GK},~\eqref{eq:L_GK}, and~\eqref{eq:F_GK}].
    We plot analytical results for the Onsager coefficients with $L=\rho/\zeta$ and $L^{\rm GK}=k_BT\rho/\zeta$, while $\mathcal{F}$ and $\mathcal{F}^{\rm GK}$ are determined numerically, as described in Appendix~\ref{sec:simulation_detail}.
    Here $D$, $L$, $L^{\rm GK}$, $\mathcal{F}$, and $\mathcal{F}^{\rm GK}$ are measured in units of $k_BT/\zeta$, $(\sigma^3\zeta)^{-1}$, $k_BT(\sigma^3\zeta)^{-1}$, $k_BT\sigma^3$, and $\sigma^3$ respectively.
    Error bars indicate standard deviations, with those smaller than the symbols omitted for clarity.
    }
    \label{fig:one_component}
\end{figure}

To demonstrate how our mechanical perspective can be used to understand gradient-driven transport more generally, we build a theory for diffusion in a passive one-component Langevin system with pairwise interactions.
The transport coefficient $\mathbf{D}$ in this system is also referred to as the \textit{collective diffusivity} and is frequently the subject of interest in colloidal suspensions~\cite{Dhont1996} and polymer solutions~\cite{deGennes1979}.
Using Eq.~\eqref{eq:mutual_diffusion}, we can decompose $\mathbf{D}$ into the product of the mechanical transport coefficient, $\mathbf{L}$, and a force Jacobian, $\boldsymbol{\mathcal{F}}$.
To formally derive both of these quantities, we must begin by describing $\mathbf{f}^{\rm eff}$ for this system as provided in Eq.~\eqref{eq:f_eff_passive_molecular} and Eq.~\eqref{eq:f_eff_passive_langevin}.
With the effective force, we proceed to determine the Onsager transport coefficient:
\begin{subequations}
\begin{align}
    &\mathbf{L} = \boldsymbol{\mathcal{R}}^{-1}, \\
    &\boldsymbol{\mathcal{R}} = \boldsymbol{\mathcal{R}}^{\rm dis} + \boldsymbol{\mathcal{R}}^{\rm int}. 
\end{align}
\end{subequations}
The even spatial symmetry of the pair distribution function in a homogeneous one-component system allows us to immediately identify [using Eq.~\eqref{eq:r_int_pair_distribution}] that ${\boldsymbol{\mathcal{R}}^{\rm int}=\mathbf{0}}$.
This can also be simply understood by noting that a translationally-invariant one-component system can never have a net conservative interaction force.
The Onsager transport coefficient for this system is then:
\begin{align}
\label{eq:l_one_component_passive}
    \mathbf{L} = \rho\boldsymbol{\zeta}^{-1},
\end{align}
where $\boldsymbol{\zeta}$ is the (potentially anisotropic) drag of the Langevin bath.

We now use Eq.~\eqref{eq:force_jacobian_definition} to determine $\boldsymbol{\mathcal{F}}$.
We use the local gradient expansion of $\mathbf{f}^{\rm int}$ as derived in Ref.~\cite{Chiu2024}, valid for systems near a spatially uniform state, to find $\frac{\partial \mathbf{f}^{\rm int}}{\partial \boldsymbol{\nabla} \rho}$:
\begin{subequations}
\label{eq:force_jacobian_one_component_passive}
\begin{align}
    &\boldsymbol{\mathcal{F}} = \frac{k_BT}{\rho}\mathbf{I}_d + \boldsymbol{\mathcal{F}}^{\rm int}, \\
    &\boldsymbol{\mathcal{F}}^{\rm int} = \int_V d\Delta\mathbf{x} \mathbf{F}(\Delta\mathbf{x})\Delta \mathbf{x}\left(g^0(\Delta\mathbf{x}) + \frac{\rho}{2}\frac{\partial g^0}{\partial \rho}(\Delta\mathbf{x})\right),
\end{align}
\end{subequations}
where $g^0$ is the single-component pair distribution function for a system with uniform density $\rho$.
The $\frac{\partial g}{\partial \rho}$ term arises from the density dependence of the pair distribution function, a contribution previously recognized as significant for capturing thermodynamic and critical behavior~\cite{Fixman1960}.
Indeed, our expectation that $\boldsymbol{\mathcal{F}}$ is related to thermodynamic forces for passive systems is borne out as it entirely expressed in terms of conservative forces and properties of an equilibrium pair distribution function.

Now we can use Eqs.~\eqref{eq:l_one_component_passive} and~\eqref{eq:force_jacobian_one_component_passive} to determine a microscopic expression for the collective diffusion with $\mathbf{D}=\mathbf{L}\cdot\boldsymbol{\mathcal{F}}$:
\begin{multline}
\label{eq:D_one_component_passive}
    \mathbf{D} = k_BT\boldsymbol{\zeta}^{-1} \\
    + \rho\boldsymbol{\zeta}^{-1}\cdot\int_V d\Delta\mathbf{x} \mathbf{F}(\Delta\mathbf{x})\Delta \mathbf{x}\left(g^0(\Delta\mathbf{x}) + \frac{\rho}{2}\frac{\partial g^0}{\partial \rho}(\Delta\mathbf{x})\right),
\end{multline}
where the first term is the ideal Stokes-Einstein self diffusivity and the second term arises from interactions.

We compute the collective diffusion coefficient via the Onsager tensor and force Jacobian for a one-component Langevin system using both the mechanical framework and the Green-Kubo relations. 
Here, we consider a three-dimensional system with $\boldsymbol{\zeta} = \zeta\mathbf{I}_3$ interacting with a Lennard-Jones (LJ) 6-12 potential with LJ diameter $\sigma$, well depth $\varepsilon = 0.25 k_BT$, and cutoff distance $2.5\sigma$ (see Appendix~\ref{sec:simulation_detail} for further details).
The isotropic drag and interactions result in $\mathbf{L} = L\mathbf{I}_3$ and $\boldsymbol{\mathcal{F}} = \mathcal{F}\mathbf{I}_3$.
As shown in Fig.~\ref{fig:one_component}, the two approaches yield consistent results,
with panel (a) showing good agreement for the collective diffusion coefficient.
The nonmonotonic behavior of $\mathcal{F}$ (or $\mathcal{F}^{\rm GK}$) depicted in panel (b) arises from the competition between ideal and interaction contributions.
At low volume densities $\rho$ (or volume fraction $\phi\equiv (2^{1/6}\sigma)^3\pi\rho/6$), the ideal term $k_BT/\rho$ dominates, resulting in a large initial value of $\mathcal{F}$.
As $\rho$ increases, this ideal term decreases monotonically, while the structural correlations captured by the pair distribution function $g^0(\Delta \mathbf{x})$ and its density derivative $\partial g^0 / \partial \rho$ begin to grow.
This shift leads to an initial decrease in $\mathcal{F}$ followed by an increase at higher $\rho$ as excluded volume interactions suppress density fluctuations.
While the static flux-density correlator term, $E$, is not expected to be identically zero, we anticipate that it is negligible and have omitted it from our calculation.
The agreement between $L$ and $L^{\rm GK}$ (and $\mathcal{F}$ and $\mathcal{F}^{\rm GK}$) confirms the validity of Einstein relations for this system.
A more detailed numerical investigation of the breakdown of Einstein relations in nonequilibrium systems is left for future work.

As previously alluded to, one motivation for examining the mutual diffusion tensor is that it plays a crucial role in determining the stability and dynamics of density fluctuations.
For simplicity, we consider a spatially isotropic system, such that $\mathbf{D}_{ij} = D_{ij} \mathbf{I}_d$ (spatially antisymmetric contributions to $\mathbf{D}_{ij}$ do not impact our analysis below). 
By combining Eqs.~\eqref{eq:density_dynamic} and~\eqref{eq:diffusion_equation}, we obtain the evolution equation of ${\delta\tilde{\rho}_i}(\mathbf{k};t)$, the Fourier-transformed density perturbation of species $i$:
\begin{equation}
\label{eq:density_evol_eq}
    \frac{\partial {\delta\tilde{\rho}_i}(\mathbf{k},t) }{\partial t} = -k^2 \sum_j^{n_c} D_{ij} {\delta\tilde{\rho}_j}(\mathbf{k},t),
\end{equation}
where we decompose the density field ${\rho_i (\mathbf{x};t) = \rho^0_i + \delta\rho_i(\mathbf{x};t)}$, with $\rho^0_i$ representing the uniform average density that $D_{ij}$ is evaluated at. 
This result is obtained by substituting the decomposed density into the linearized evolution equations and applying a spatial Fourier transform.
We can then solve Eq.~\eqref{eq:density_evol_eq} to obtain the general solution for the time evolution of density fluctuations in Fourier space:
\begin{equation}
\label{eq:density_mutual_diffusion}
    {\delta\tilde{\rho}_i}(\mathbf{k},t) = \sum_j^{n_c} \sum_m^{n_c} P_{ij} e^{-k^2\lambda_j t} (P^{-1})_{jm} \delta\tilde{\rho}_m (\mathbf{k},0),
\end{equation}
where $P_{ij}$ is the $i$-th component of the $j$-th right eigenvector of the mutual diffusion tensor $\mathbf{D}$, and $\lambda_j$ is the corresponding eigenvalue.
Examining the structure of Eq.~\eqref{eq:density_mutual_diffusion}, we see that the sign of the eigenvalues of $\mathbf{D}$ determines whether perturbations of species density decay or amplify, directly determining the (linear) stability of the system.
If all eigenvalues have positive real parts, all fluctuations decay over time, ensuring the system is linearly stable.
Conversely, if at least one eigenvalue has a negative real part, the corresponding fluctuation mode grows over time, indicating an instability.
When the eigenvalue of an instability is purely real, the instability is typically referred to as stationary (or Cahn–Hilliard) instability, where localized perturbations grow at their point of origin.
If any of the unstable eigenvalues have nonzero imaginary components, the system undergoes a traveling instability (or conserved-Hopf bifurcation~\cite{Greve2024}), which is a generic large-scale oscillatory instability possible for systems with at least two conserved quantities. 
This bifurcation gives rise to perturbations that not only grow in amplitude but also propagate (or travel) through space.
While this linear stability analysis can predict traveling states in response to long wavelength density perturbations, nonlinear effects may ultimately result in the system settling into a stationary state, where the initially traveling states reach a stationary configuration~\cite{Greve2024}.

In equilibrium, the Onsager transport tensor $\mathbf{L}$ must be symmetric and positive semidefinite.
Furthermore, in equilibrium, the force Jacobian exactly corresponds to the necessarily symmetric Hessian of the free energy.
Since mutual diffusion is given by the product of the Onsager transport tensor and the force Jacobian [Eq.~\eqref{eq:mutual_diffusion}], the symmetry of $\boldsymbol{\mathcal{F}}$ and positive (semi)definite character of $\mathbf{L}$ in equilibrium allows us to characterize the eigenspectrum of $\mathbf{D}$. 
To see this, we observe first that $\mathbf{L}\cdot\boldsymbol{\mathcal{F}} \simeq \mathbf{L}^{1/2}\cdot\boldsymbol{\mathcal{F}}\cdot\mathbf{L}^{1/2}$, where $\simeq$ denotes matrix similarity, on the subspace where $\mathbf{L}$ is invertible.
Consequently, for passive systems, $\mathbf{D}$ has no imaginary eigenvalues because it is similar to a symmetric matrix.
By further noting that $\mathbf{L}^{1/2}\cdot\boldsymbol{\mathcal{F}}\cdot\mathbf{L}^{1/2}$ is congruent as a matrix to $\boldsymbol{\mathcal{F}}$, by Sylvester's Law of Inertia~\cite{Carrell2017}, $\mathbf{L}\cdot\boldsymbol{\mathcal{F}}$ has at most as many negative eigenvalues as $\boldsymbol{\mathcal{F}}$.
This implies that the stability of the system as determined by the spectrum of $\mathbf{D}$ is entirely determined by the eigenvalues of $\boldsymbol{\mathcal{F}}$.
Consequently, the stability of density fluctuations in the system is dictated by $\boldsymbol{\mathcal{F}}$, which aligns with thermodynamic expectations: a system is stable if the free energy Hessian $\boldsymbol{\mathcal{F}}$ is positive definite (i.e.,~all eigenvalues are positive), ensuring that perturbations decay over time.
This analysis implies that the emergence of traveling or oscillatory states is fundamentally precluded in passive systems, as the equilibrium free energy Hessian cannot have imaginary eigenvalues.
This contrasts with active systems, where departing from equilibrium relaxes the constraints on the Onsager transport tensor $\mathbf{L}$ and the force Jacobian tensor $\boldsymbol{\mathcal{F}}$ such that neither is generally symmetric. 
As a result, $\mathbf{D}$ can acquire complex eigenvalues, allowing for distinct signatures of nonequilibrium dynamics such as traveling states~\cite{Wittkowski2017, You2020, Saha2020}.
In Sec.~\ref{sec:nonreciprocal_interacting_systems}, we study the specific case of systems with nonreciprocal interactions, where traveling states have been explicitly observed~\cite{You2020, Saha2020, Fruchart2021,Chiu2023}. 
We demonstrate numerically that the symmetry of $\mathbf{L}$ is broken such that traveling states are allowed and explore the mechanical origins of this broken symmetry.

\subsection{Nonequilibrium Color Field Theory}
\label{sec:color_field}
The mechanical description of multicomponent transport suggests a route to computing the Onsager transport coefficients that should be applicable to both passive and active systems. 
The approach is to add a species-dependent \textit{constant} external force $\mathbf{f}^{\rm ext}$ to the particle equations-of-motion that will then appear as an additional term in the effective force balance. 
The independence of this applied external force on the flux ensures that it will not modify the resistance tensor, but this force \emph{will} contribute to the static forces. 
If we further ensure conditions such that this ``color'' force is the only static force driving particle motion, then we can arrive at:
\begin{equation}
\label{eq:color_field}
    \mathbf{J} = \mathbf{L} \cdot \mathbf{f}^{\rm ext},
\end{equation}
ensuring that the sole driving force is indeed $\mathbf{f}^{\rm ext}$ may appear to require an understanding of the other static forces. 
However, since other static forces must be independent of flux, and the applied external field is assumed to be independent of all other field variables, we do not expect constant external forces to indirectly generate additional static forces.
Computationally, appropriate boundary conditions must be imposed along the direction of the flux to maintain this constant flux in the steady state. 
The flux needs to be measured after some initial transience (set by the maximum of the timescales $\tau^{\rm NL}$ and $\tau$) so that the system is truly in a steady state with respect to $\mathbf{J}$.
Moreover, several computational experiments will be needed to obtain $\mathbf{L}_{ij}$ and all of its spatial components.
The simplest procedure is to apply the external force on a single species in $d$ orthogonal directions and repeat this procedure for all $n_c$ species.
This ensures the applied forces in each experiment are linearly independent, and thus we only perform the minimum number of experiments, $dn_c$.
Crucially, the magnitude of the applied forces will need to be sufficiently small to ensure that we are in the linear regime. 
For each applied force, the force should be varied to ensure that the measured fluxes exhibit linear dependence on the force, with an intercept coinciding with vanishing flux in the absence of force. 

The color field approach described above is similar to methods described in the nonequilibrium molecular dynamics literature (NEMD)~\cite{Ciccotti1975, Evans1982, MacGowan1986, Sarman1992, Ciccotti2005, Evans2008, Wheeler2004_2,Sasaki2025}.
NEMD methods are often based upon linear response theory and, while frequently derived using the Hamiltonian formalism for the particle dynamics~\cite{Evans2008}, have been derived for non-Hamiltonian systems as well~\cite{Ciccotti2005}.
NEMD methods have long been used to find both mechanical and gradient transport coefficients, including thermal conductivity, Onsager transport coefficients, diffusion coefficients, and electrical conductivity~\cite{Evans1982,Sarman1992,Wheeler2004_2,Mandadapu2009, Mandadapu2010, Lesnicki2020,Tripathi2024,Sasaki2025}.
In the color field perspective, we interrogate transport in the Thévenin ensemble, defined by the application of a constant external field, due to its simplicity and ease of implementation~\cite{Evans1993}. 
A more direct approach to measuring $\boldsymbol{\mathcal{R}}$ is to take measurements in the Norton ensemble, defined by imposing a constant flux.
Prior work has shown the Norton ensemble to be equivalent to the Thévenin ensemble under certain relatively general conditions, and explores how to perform simulations in the Norton ensemble~\cite{Evans1993,Blassel2024}.
Thermostatting issues in true molecular systems with an applied external ``color'' field have been previously discussed in the context of equilibrium systems~\cite{Nose1984,Hoover1985,Evans2008}, and similar considerations are expected to apply when extending our nonequilibrium theory to such systems.
For example, depending on the choice of thermostat, it may be necessary to apply only combinations of color fields that yield zero net force on the system, thereby conserving momentum and ensuring correct thermostat behavior~\cite{MacGowan1986}.
However, in Langevin systems such as those considered in this work, the presence of dissipative forces prevents sustained acceleration and ensures that no net force accumulates on the system, and as a result, no explicit constraint on the total applied force is required.

To demonstrate the determination of the Onsager tensor using our mechanical approach, we consider a two-component system consisting of species $A$ and $B$ with passive Langevin dynamics. 
The particles interact with a Lennard-Jones 6-12 potential that is distinct for different pairs of species (see Appendix~\ref{sec:simulation_detail} for simulation details).
The parameters are chosen such that the system is, on average, spatially uniform.
The number ratio of $A$ to $B$ particles is fixed to $1:3$ while the overall volume fraction, $\phi$, is varied.
We obtain the Onsager tensor through direct measurement of the steady state flux of species $A$ and $B$ in response to our external forces using Eq.~\eqref{eq:color_field}.
As our system satisfies spatial parity and is isotropic, we anticipate that the Onsager transport tensor takes the form $\mathbf{L}_{ij} = L_{ij} \mathbf{I}_d$.
We verify that the fluxes are indeed linear with the forces by varying the force amplitude [shown in Fig.~\ref{fig:flux_force_response}] and confirming linearity of the flux.
Additionally, we compute the Onsager tensor with Green-Kubo relations to validate the accuracy of the color field approach in equilibrium [see Sec.~\ref{sec:mutual_diffusion}].
We used the displacement form of the Green-Kubo relations\footnote{While this form of the Green-Kubo relation is convenient to calculate, it precludes us from independently verifying the symmetry of the off-diagonal of $\mathbf{L}$.
To explicitly see this symmetry numerically from two independent calculations in equilibrium, one needs to use the more general form of the Green-Kubo relation~\eqref{eq:L_GK}.}, which requires microscopic time-reversal symmetry, as detailed in Appendix~\ref{sec:simulation_detail}.

\begin{figure}
    \centering
    \includegraphics[width=1.0\linewidth]{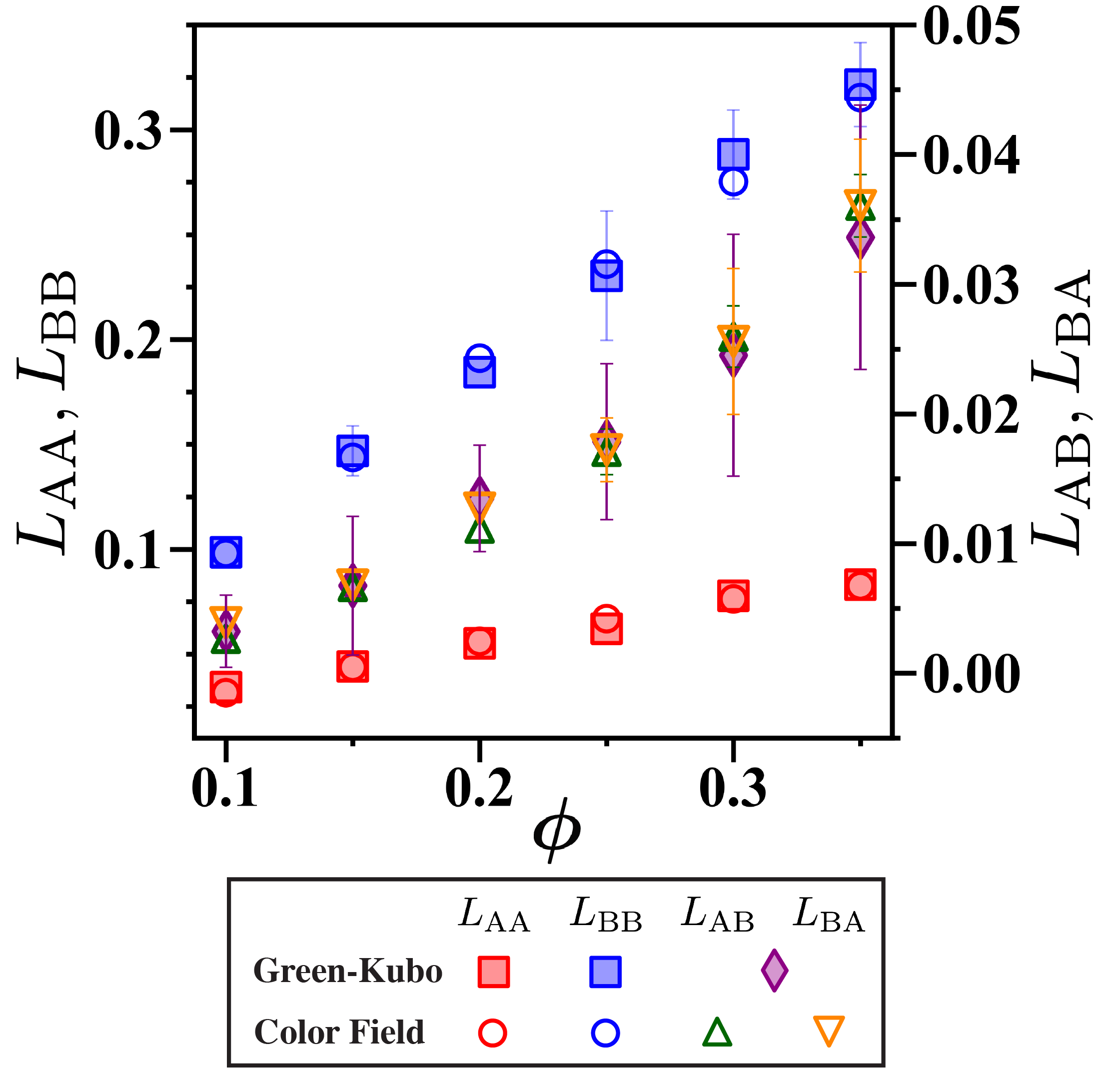}
    \caption{Comparison of the Onsager transport tensor components $L_{\rm AA}$ and $L_{\rm BB}$ (left axis) and $L_{\rm AB}$ and $L_{\rm BA}$ (right axis) for a passive pairwise interacting Langevin system, obtained using the Green-Kubo relation [Eq.~\eqref{eq:GK_msd}] and the color field expression [Eq.~\eqref{eq:color_field}].
    Here $L$ and $L^{\rm GK}$ are measured in units of $(\sigma^3\zeta)^{-1}$ and $k_BT(\sigma^3\zeta)^{-1}$.
    Error bars indicate standard deviations, with those smaller than the symbols omitted for clarity.}
    \label{fig:Onsager_comparison}
\end{figure}

Figure~\ref{fig:Onsager_comparison} presents the Onsager transport tensor obtained from the color field method [Eq.~\eqref{eq:color_field}] and the Green-Kubo relations [Eq.~\eqref{eq:GK_msd}]. 
The two methods produce identical results within statistical error across all examined densities.
Moreover, the off-diagonal mobilities are statistically identical, providing an additional numerical demonstration that Onsager's reciprocal relations hold for this passive system.
For \textit{noninteracting} particles in a Langevin bath (with a dissipative force of the form of Eq.~\eqref{eq:f_dis} with spatially isotropic species resistance), we expect the Onsager tensor to have components: $L_{\rm AA} = \rho_A/\zeta_A$, $L_{\rm BB} = \rho_B /\zeta_B$ and $L_{\rm AB} = L_{\rm BA} = 0$.
The increase in these off-diagonal mobilities with the overall concentration is consistent with our expectation of increased inter-species correlation as we depart from the ideal limit where $L_{\rm BA} = L_{\rm AB} \approx 0$. 
It should be noted that the increase in the diagonal contributions with increasing density is also to be expected, with the observed sublinear increase indicative of an \textit{increase} in the effective single-particle resistance, defined as $\zeta_i^{\rm eff} = \rho_i/L_{ii}$.

\section{Applications}
\label{sec:application}
We have developed the mechanical transport perspective and can now apply it to recover expressions for linear transport coefficients in systems independent of the validity of the local equilibrium hypothesis.
In Sec.~\ref{sec:nonreciprocal_interacting_systems}, we investigate how nonreciprocal interactions lead to a breaking of the Onsager reciprocal relations, and use the color field method to numerically confirm the resulting asymmetry of the Onsager matrix $\mathbf{L}$.
Then, in Sec.~\ref{sec:chiral_active_brownian_particles} we explore chiral active Brownian particles and show how their odd spatial transport arises from an underlying activity mechanism that breaks both parity and time-reversal symmetry.
Finally, in Sec.~\ref{sec:nonlinear_transport} we demonstrate the applicability of the mechanical perspective in understanding the nonlinear ``differential'' conductivity of electrolytes and connect our mechanical perspective with prior work in calculating mechanical transport coefficients.
Together, these applications demonstrate the broad applicability of the proposed framework and establish a foundation for future investigations of linear transport in far-from-equilibrium systems.

\subsection{Nonreciprocal Interacting Systems}
\label{sec:nonreciprocal_interacting_systems}
Traveling states have previously been observed in systems with nonreciprocal interactions (systems in which the coarse-grained interactions between constituents appear to violate Newton’s third law)~\cite{You2020, Saha2020, Fruchart2021, Dinelli2023, Chiu2023}, and their emergence can be understood from a stability analysis involving the mutual diffusion tensor, $\mathbf{D}$, as discussed in detail in Sec.~\ref{sec:mutual_diffusion}.
Specifically, the breakdown of Onsager reciprocal relations of $\mathbf{L}$ and the potential asymmetry of the force Jacobian $\boldsymbol{\mathcal{F}}$ in active systems can produce imaginary eigenvalues in $\mathbf{D}$, enabling traveling states—an outcome \textit{not possible} in passive systems.
We now apply our theory to a system with nonreciprocal interactions, elucidating the mechanical origins of the breaking of the passive structures of $\mathbf{L}$ and $\boldsymbol{\mathcal{F}}$.

We consider \textit{pairwise} interaction forces between distinct species that violate Newton's Third Law with ${\mathbf{F}_{ij} (\mathbf{r}) \neq  \mathbf{F}_{ji}(-\mathbf{r}) \forall i \neq j}$, where $\mathbf{r}$ denotes the interparticle separation.
The force can be decomposed into reciprocal and nonreciprocal contributions: ${\mathbf{F}_{ij}(\mathbf{r}) = \mathbf{F}_{ij}^{\rm R}(\mathbf{r}) + \mathbf{F}_{ij}^{\rm NR}(\mathbf{r})}$ with the reciprocal contribution defined as ${2\mathbf{F}_{ij}^{\rm R}(\mathbf{r}) = \mathbf{F}_{ij}(\mathbf{r})+\mathbf{F}_{ji}(-\mathbf{r})}$ and a nonreciprocal force of ${2\mathbf{F}_{ij}^{\rm NR}(\mathbf{r}) = 2\mathbf{F}_{ji}^{\rm NR}(-\mathbf{r}) = \mathbf{F}_{ij}(\mathbf{r}) - \mathbf{F}_{ji}(-\mathbf{r})}$.
For the following analysis, we are interested in the system response to weak density gradients, so we take ${\rho_i (\mathbf{x}) = \rho^0_i + \delta\rho_i(\mathbf{x})}$.
The expression for $\mathbf{f}^{\rm int}$ provided in Eq.~\eqref{eq:f_int_microscopic_pairwise} is the general form of the interaction contribution to the effective force for pairwise interacting systems, in or out of equilibrium. 
Therefore, we can express $\mathbf{f}^{\rm int}$ as:
\begin{multline}
\label{eq:f_int_nonreciprocal}
    \mathbf{f}^{\rm int}(\mathbf{x}) = \sum_k^{n_c}\int_V d\mathbf{x}' \rho_k^0 g_{ik}(\mathbf{x}, \mathbf{x}')\left(\mathbf{F}_{ik}^{\rm R}(\mathbf{x} - \mathbf{x}')\right. \\
    \left.+\mathbf{F}_{ik}^{\rm NR}(\mathbf{x} - \mathbf{x}')\right) + \mathcal{O}(\delta \rho).
\end{multline}
It should be noted that the ideal contribution of the effective force does not have any explicit dependence on the species flux and, therefore, does not contribute to the resistance tensor.
Additionally, we will only focus on the interaction contribution to the resistance tensor $\boldsymbol{\mathcal{R}}^{\rm int}$, as $\boldsymbol{\mathcal{R}}^{\rm dis}$ will not change with the introduction of nonreciprocal forces and so the asymmetry of $\mathbf{L}$ outside of equilibrium originates in $\boldsymbol{\mathcal{R}}^{\rm int}$ which is determined through differentiation of Eq.~\eqref{eq:f_int_nonreciprocal} with species flux.

We consider the system in the \textit{dilute} limit, such that we can make an approximation of the dependence of the derivative of the pair distribution function with respect to species velocity.
This assumption will be important in simplifying the expression for the resistance tensor. 
Specifically, we expect that any variation in the pair correlation function $g_{ik}$ due to changes in species velocity $\mathbf{v}_j$ originates from particle interactions. 
Moreover, we make the ansatz that in the dilute limit we can make the following approximation:
\begin{align}
\label{eq:g_v_sum}
    \sum_k^{n_c} \frac{\partial g_{ik}}{\partial \mathbf{v}_j} \approx \sum_{k\neq i}^{n_c}\delta_{ij}\frac{\partial g_{ik}}{\partial \mathbf{v}_i} + \frac{\partial g_{ij}}{\partial \mathbf{v}_j}.
\end{align}
Physically, this ansatz asserts that in the dilute limit the pair distribution function only depends on the relative velocities of the species directly involved in $g_{ik}$, and that if ${j\neq i}$ and ${j \neq k}$ the derivative is of higher order in density.
Thus, by utilizing the form of $\mathbf{f}^{\rm int}$ in Eq.~\eqref{eq:f_int_nonreciprocal} and the dilute assumption which culminates in Eq.~\eqref{eq:g_v_sum}, the interaction contribution takes the form:
\begin{subequations}
\label{eq:R_nonreciprocal}
\begin{align}
    \boldsymbol{\mathcal{R}}^{\rm int}_{ij} &= \boldsymbol{\mathcal{R}}_{ij}^{\rm S} + \boldsymbol{\mathcal{R}}_{ij}^{\rm A} + \mathcal{O}(\delta\rho ),\\
    \boldsymbol{\mathcal{R}}_{ij}^{\rm S} &= 
    \begin{dcases}
        -\sum_{k \neq i}^{n_c} \frac{\rho_k^0}{\rho_i^0} \left(\int_V d\mathbf{x}'\frac{\partial g_{ik}^0(\Delta\mathbf{x})}{\partial \mathbf{v}_i}\mathbf{F}_{ik}(-\Delta\mathbf{x}) \right): i = j\\
      -\int_V d\mathbf{x}' \mathbf{G}_{ij}^{\rm R}\mathbf{F}_{ij}^{\rm R}(-\Delta\mathbf{x}) + \mathbf{G}_{ij}^{\rm NR} \mathbf{F}_{ij}^{\rm NR}(-\Delta\mathbf{x}): i \neq j
    \end{dcases},\\
    \boldsymbol{\mathcal{R}}_{ij}^{\rm A} &= 
    \begin{dcases}
        \boldsymbol{0}: i = j\\
      -\int_V d\mathbf{x}' \mathbf{G}_{ij}^{\rm NR} \mathbf{F}_{ij}^{\rm R}(-\Delta\mathbf{x}) + \mathbf{G}_{ij}^{R}\mathbf{F}_{ij}^{\rm NR}(-\Delta\mathbf{x}): i \neq j
    \end{dcases},
\end{align}
\end{subequations}
where we decompose the resistance tensor into its symmetric ${\boldsymbol{\mathcal{R}}^{\rm S}_{ij} = \frac{1}{2}(\boldsymbol{\mathcal{R}}_{ij}^{\rm int} + \boldsymbol{\mathcal{R}}_{ji}^{\rm int})}$ and antisymmetric ${\boldsymbol{\mathcal{R}}^{\rm A}_{ij} = \frac{1}{2}(\boldsymbol{\mathcal{R}}^{\rm int}_{ij} - \boldsymbol{\mathcal{R}}^{\rm int}_{ji})}$ contributions, and $\Delta \mathbf{x} \equiv \mathbf{x}' - \mathbf{x}$.
We have now eliminated the spatial dependency of $\boldsymbol{\mathcal{R}}^{\rm int}$.
Here, $g_{ij}^0(\Delta \mathbf{x}) \equiv g_{ij}^0(\mathbf{x}, \mathbf{x}';\boldsymbol{\rho}^0)$ is the homogeneous radial distribution with species density $\boldsymbol{\rho}^0$ in the absence of \textit{both} net species fluxes and density gradients.
From the form of Eq.~\eqref{eq:R_nonreciprocal}, we see that $\boldsymbol{\mathcal{R}}$ is independent of $\mathbf{x}$ for this system because it depends only on functions which are translationally invariant in their arguments.
The two indexed tensor components $\mathbf{G}^{\rm R}_{ij}$ and $\mathbf{G}^{\rm NR}_{ij}$ are defined as:
\begin{align}
    \mathbf{G}^{\rm R}_{ij}(\Delta\mathbf{x}) = \frac{1}{2} \left(\frac{\partial g_{ij}^0(\Delta\mathbf{x})}{\partial \mathbf{v}_j} + \frac{\partial g_{ji}^0(\Delta\mathbf{x})}{\partial \mathbf{v}_i}\right), \\
    \mathbf{G}^{\rm NR}_{ij}(\Delta\mathbf{x}) = \frac{1}{2} \left(\frac{\partial g_{ij}^0(\Delta\mathbf{x})}{\partial \mathbf{v}_j} - \frac{\partial g_{ji}^0(\Delta\mathbf{x})}{\partial \mathbf{v}_i}\right),
\end{align}
where we suppress the $\Delta \mathbf{x}$ dependence of $\mathbf{G}_{ij}^{\rm R}$ and $\mathbf{G}_{ij}^{\rm NR}$ in Eq.~\eqref{eq:R_nonreciprocal} for brevity.
If nonreciprocity is present in the system, $\boldsymbol{\mathcal{R}}^{\rm A}$ is generally non-zero unless the asymmetric modification of the pair distribution function precisely counterbalances the nonreciprocal forces.
We can also express the force Jacobian of the dilute systems with nonreciprocal pairwise force using the dilute limit of the nonlocal density gradient expansion adopted from Ref.~\cite{Chiu2024}:
\begin{subequations}
\begin{align}
    \boldsymbol{\mathcal{F}}_{ij} &= \boldsymbol{\mathcal{F}}_{ij}^{\rm S} + \boldsymbol{\mathcal{F}}_{ij}^{\rm A},\\
    \boldsymbol{\mathcal{F}}_{ij}^{\rm S} &= k_BT/\rho_i^0  \delta_{ij}\mathbf{I}_d \nonumber\\
    &+ \int_V d\mathbf{x}' \mathbf{F}^{\rm R}_{ij}(\Delta\mathbf{x}) \Delta\mathbf{x} g_{ij}^0(|\Delta\mathbf{x}|),\\
    \boldsymbol{\mathcal{F}}_{ij}^{\rm A} &= \int_V d\mathbf{x}' \mathbf{F}^{\rm NR}_{ij}(\Delta\mathbf{x}) \Delta\mathbf{x} g_{ij}^0(|\Delta\mathbf{x}|).
\end{align}
\end{subequations}
Just as in the case of $\boldsymbol{\mathcal{R}}^{\rm int}$, we can separate out the antisymmetric contribution to the force Jacobian, which allows us to observe that the symmetry of $\boldsymbol{\mathcal{F}}_{ij}$ is broken if there are any nonreciprocal interactions.

To confirm that nonreciprocal interactions lead to an asymmetric Onsager transport tensor $\mathbf{L}$, we simulated a two-component system consisting of species $A$ and $B$, incorporating \textit{nonreciprocal interactions} between dissimilar species.
We define the pairwise force exerted on a particle of species $i$ by a particle of species $j$ based on the nonreciprocal model from Ref.~\cite{Chiu2023}:
\begin{equation}
\label{eq:modelforces}
\mathbf{F}_{ij} =
\mathbf{F}_{ij}^{\rm C}\times
\begin{cases}
1 - \Delta(r), & \text{if } (ij) = (AB) \\
1 + \Delta(r), & \text{if } (ij) = (BA) \\
1, & \text{if } (ij) = (AA) \text{ or } (BB)
\end{cases},
\end{equation}
where $\mathbf{F}_{ij}^{\rm C}$ is the conservative (reciprocal) attractive interaction force, chosen to be derived from the Lennard-Jones potential, and $\Delta(r)$ is the nonreciprocity parameter, with $r$ representing the interparticle distance.
For simplicity, we adopted the step-function form for the nonreciprocity parameter~\cite{Chiu2023}, $\Delta(r) = \Delta\Theta(r - d_{\rm rec})$, where $\Theta$ is the Heaviside function such that interactions remain fully reciprocal for separations less than the reciprocity diameter $d_{\rm rec}$, and have a constant nonreciprocity strength of $\Delta$ for separation distance greater than $ d_{\rm rec}$.
When $\Delta = 0$, the interactions are fully reciprocal, and the system behaves as a passive system.
As $\Delta$ increases from zero, nonreciprocal interactions are introduced between dissimilar species, driving the system out of equilibrium.
We set the number ratio of $A$ to $B$ particles to $1:3$, and the overall volume fraction of the system at $\phi = 0.2$.
We further ensure that all system parameters are identical to those in the passive case presented in Sec.~\ref{sec:color_field} to isolate the impact of the introduction of nonreciprocal interactions through the nonreciprocity parameter $\Delta$.
We use the nonequilibrium color field theory [Eq.~\eqref{eq:color_field}] to obtain the Onsager tensor through direct measurement of the flux of species $A$ and $B$ in response to external forces, detailed in Appendix~\ref{sec:simulation_detail}. 
For all simulations, we verify that the system remains isotropic as $\Delta$ increases and the Onsager transport tensor is again anticipated to take the form $\mathbf{L}_{ij} = L_{ij} \mathbf{I}_d$.

\begin{figure}
    \centering
    \includegraphics[width=1.0\linewidth]{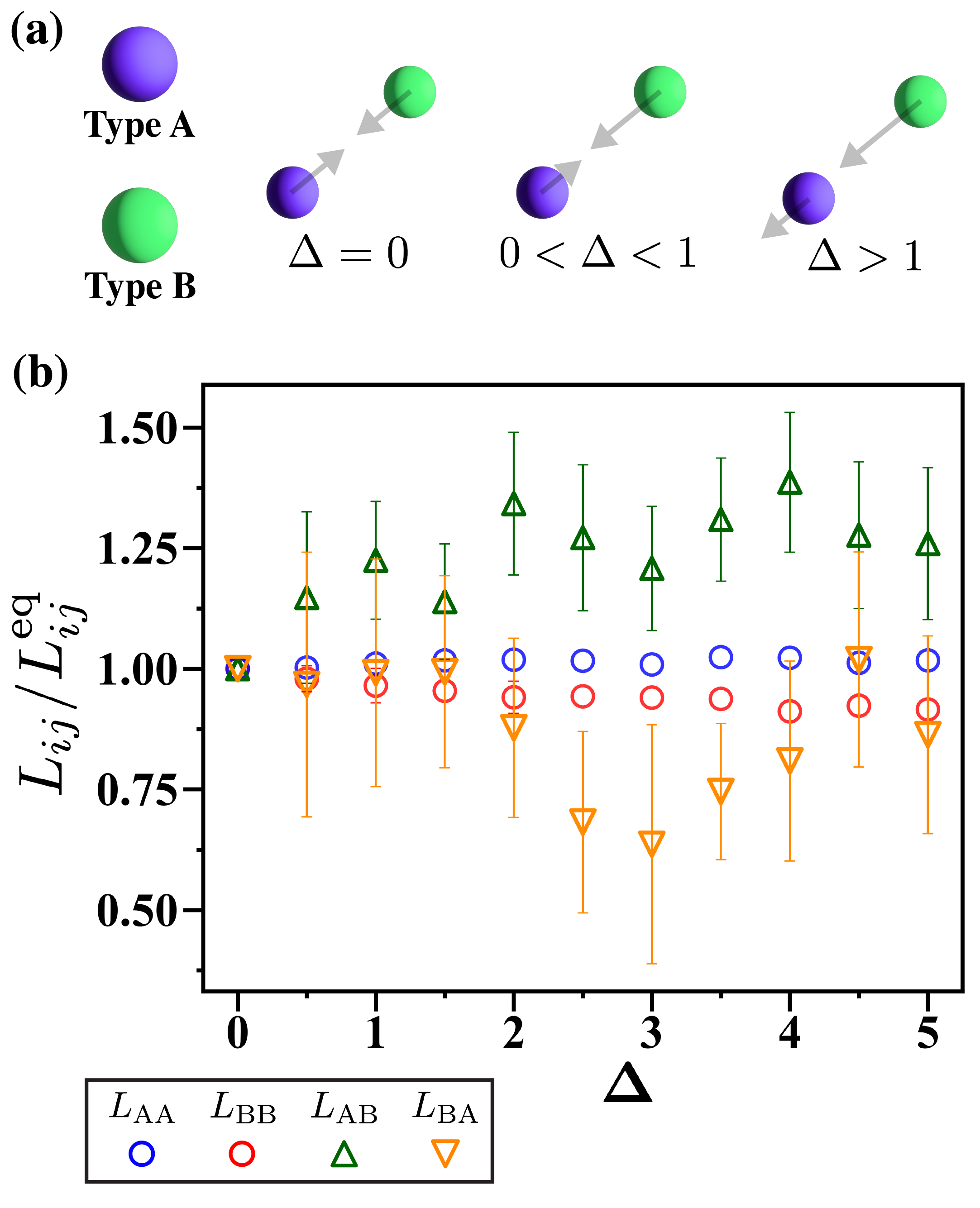}
    \caption{(a) Schematic of pairwise nonreciprocal interactions between species $A$ and $B$ and (b) nonreciprocity dependence of the Onsager transport tensor components for a nonreciprocal pairwise interacting Langevin system with $\phi = 0.2$, obtained using the color field expression [Eq.~\eqref{eq:color_field}].
    Each component of $\mathbf{L}$ is normalized by its respective equilibrium value, $L^{\rm eq}_{ij} \equiv L_{ij}(\Delta = 0)$.
    Error bars indicate standard deviations, with those smaller than the symbol size omitted for clarity.}
    \label{fig:nonreciprocal_L}
\end{figure}

Figure~\ref{fig:nonreciprocal_L} shows the dependence of the Onsager transport tensor on the nonreciprocity parameter $\Delta$.
To highlight deviations from equilibrium behavior, each component of $\mathbf{L}$ is normalized by its equilibrium value, defined as the value at $\Delta = 0$, i.e., $L^{\rm eq}_{ij} \equiv L_{ij}(\Delta = 0)$.
While the self-terms ($L_{\rm AA}$, $L_{\rm BB}$) remain relatively unchanged as $\Delta$ increases, we observe a growing asymmetry between $L_{\rm AB}$ and  $L_{\rm BA}$, indicating the breaking of Onsager reciprocity as result of the nonreciprocal interactions.
This result aligns with our earlier analysis of a dilute nonreciprocal mixture and the numerical simulations demonstrate that the asymmetry persists beyond the dilute limit.

The broken reciprocal relations between $L_{\rm AB}$ and $L_{\rm BA}$ can be understood intuitively by considering how the interactions are modified by $\Delta$.
We build this intuition by using the Onsager-Machlup framework~\cite{Gao2019, Cugliandolo2019, Maes2020} to understand how nonreciprocal forces modify transport close to equilibrium in Appendix~\ref{app:onsager_machlup}.
From Eq.~\eqref{eq:modelforces}, when $\Delta > 1$, the system enters the “predator-prey” regime [see Fig.~\ref{fig:nonreciprocal_L}(a)], where the force exerted by species $B$ on species $A$ reverses sign, such that instead of attracting species $A$, species $B$ \textit{repels} it. 
In order to get a sense of how $\mathbf{L}_{AB}$ and $\mathbf{L}_{BA}$ change with the presence of a nonreciprocal force, we can analyze the effect of an applied force on a particle of species $A$ ``chased'' by a particle of species $B$.
If an additional force is applied to species $B$ in the direction of motion of $A$, the interparticle separation between the particles will decrease, the repulsive force (and therefore velocity) on $A$ will be larger, and so $\mathbf{L}_{AB}$ is larger than would be the case in the absence of the nonreciprocal force.
Conversely, if a force is applied to species $A$ in the direction of motion of species $B$, the interparticle distance will increase, leading to a weaker attractive force in the direction of motion of $B$ and therefore a smaller overall velocity, meaning $\mathbf{L}_{BA}$ will be smaller than in equilibrium.
Consequently, we expect the transport coefficient $L_{\rm BA}$ to decrease with increasing nonreciprocal forces, while $L_{\rm AB}$ correspondingly increases.

\subsection{Chiral Active Brownian Particles}
\label{sec:chiral_active_brownian_particles}
Chiral active matter describes systems composed of particles driven by microscopic torques. 
These systems exhibit a range of nonequilibrium behaviors, including what has been termed ``odd'' diffusion, where the diffusion tensor contains spatially antisymmetric components that generate fluxes perpendicular to concentration gradients, as an example of odd transport phenomena~\cite{Epstein2020, Hargus2021,Banarjee2017, Hargus2020, Han2021, Lou2022, Poggioli2023, Fruchart2023}. 
We emphasize here that odd diffusion/transport refers to \emph{spatial} asymmetry of the diffusion tensor. 
(In general, we see that even for isotropic systems in equilibrium, $\mathbf{D}$ is the product of two symmetric matrices and is not itself generally symmetric at the \emph{species} level.)
Odd transport coefficients arising from the presence of an applied magnetic field have previously been described by the Onsager-Casimir relations~\cite{Casimir1945, DeGroot2013}.
In the following analysis, we demonstrate that odd diffusion is recovered with our mechanical transport perspective by examining the transport properties of a single-component system of pairwise interacting chiral active Brownian particles (cABP) in two spatial dimensions.

The equation of motion governing the cABPs are~\cite{Hargus2021, Langford2025}:
\begin{subequations}
\begin{align}
    \dot{\mathbf{r}}_\alpha = U_0 \mathbf{q}_\alpha + \frac{1}{\zeta}\sum_{\beta\neq\alpha}^N\mathbf{F}_{\alpha\beta},\\
    \dot{\mathbf{q}}_\alpha =  \boldsymbol{\omega}_0 \times \mathbf{q}_\alpha +   \mathbf{\Omega}_\alpha \times \mathbf{q}_\alpha,
\end{align}
\end{subequations}
where $\mathbf{q}_\alpha$ ($\dot{\mathbf{q}}_\alpha$) is the orientation (time derivative of orientation) of the $\alpha$th particle, $\mathbf{F}_{\alpha\beta}$ is a conservative pairwise force, $U_0$ and $\boldsymbol{\omega}_0$ are the intrinsic active speed and angular velocity, respectively.
The orientation is further modified by $\mathbf{\Omega}_\alpha$, the stochastic angular velocity with zero mean and variance of ${\langle \mathbf{\Omega}_\alpha(t) \mathbf{\Omega}_\beta(t') \rangle = 2D_r\delta_{\alpha\beta}\delta(t-t')\mathbf{I}_d}$, where $D_r$ is the rotational diffusivity.
Since we are considering a two-dimensional system, we only consider the out-of-plane component of $\boldsymbol{\omega}_0 \equiv \omega_0 \mathbf{e}_z$ as it is the only component relevant to the orientation dynamics, where $\mathbf{e}_z$ is the unit vector in the out-of-plane direction ($\mathbf{e}_x$ and $\mathbf{e}_y$ represent the unit vectors of orthogonal in-plane directions). 
Here, the sign of $\omega_0$ determines the handedness of the chirality of the system, breaking time-reversal and parity symmetries.

Following the Irving-Kirkwood procedure~\cite{Irving1950}, we can find the exact form of the effective force of the cABP system as:
\begin{subequations}
\label{eq:chiral_eff_f}
\begin{align}
    &\mathbf{f}^{\rm eff} = \mathbf{f}^{\rm act} + \mathbf{f}^{\rm int} + \mathbf{f}^{\rm dis}, \\
    &\mathbf{f}^{\rm act}(\mathbf{x}) = \frac{\zeta U_0}{\rho(\mathbf{x})} \mathbf{m}(\mathbf{x}), \\
    &\mathbf{f}^{\rm int}(\mathbf{x}) = \int_V d\mathbf{x}' \rho(\mathbf{x}')g(\mathbf{x}, \mathbf{x}') \mathbf{F}(\mathbf{x} - \mathbf{x}'), \\
    &\mathbf{f}^{\rm dis}(\mathbf{x}) = -\frac{\zeta}{\rho(\mathbf{x})}\mathbf{J}(\mathbf{x}),
\end{align}
\end{subequations}
where $\mathbf{f}^{\rm int}$ and $\mathbf{f}^{\rm dis}$ are familiar from our treatment of Langevin passive systems, $\mathbf{f}^{\rm act}$ is the effective active force and $\mathbf{m}(\mathbf{x})$ is the polar order density of the system.
We can determine the complete resistance tensor and the static force that drives the motion given the effective force [see Eq.~\eqref{eq:chiral_eff_f}] for the cABP system near a translationally and rotationally-invariant state:
\begin{subequations}
\label{eq:resistence_cABP}
\begin{align}
&\boldsymbol{\mathcal{R}} = \boldsymbol{\mathcal{R}}^{\rm dis} + \boldsymbol{\mathcal{R}}^{\rm act}, \\
&\mathbf{f}^{\rm static} = (\mathbf{f^{\rm act}} + \mathbf{f}^{\rm int} + \mathbf{f}^{\rm dis})\rvert_{\mathbf{J} = \mathbf{0}},
\end{align}
\end{subequations}
where we define the dissipative and active contributions to the resistance tensor as $\boldsymbol{\mathcal{R}}^{\rm dis}= \frac{\zeta}{\rho}\mathbf{I}_d$ and $\boldsymbol{\mathcal{R}}^{\rm act} = - \frac{\zeta U_0}{\rho} \frac{\partial \mathbf{m}}{\partial \mathbf{J}} \big|_{\mathbf{J} = \mathbf{0}}$, respectively.
We note that in a system with translational and rotational invariance, the interaction contribution to the resistance tensor ${\boldsymbol{\mathcal{R}}^{\rm int} = \mathbf{0}}$ as the pair distribution function for a single-component is spatially even.

Following the perspective in Sec.~\ref{sec:mutual_diffusion}, we can identify the force Jacobian $\boldsymbol{\mathcal{F}}$ for cABPs as:
\begin{subequations}
\label{eq:force_jacobian_cABP}
\begin{align}
&\boldsymbol{\mathcal{F}} = \boldsymbol{\mathcal{F}}^{\rm act} + \boldsymbol{\mathcal{F}}^{\rm int}, \\
&\boldsymbol{\mathcal{F}}^{\rm act} = -\frac{\zeta U_0}{\rho} \frac{\partial \mathbf{m}}{\partial \boldsymbol{\nabla} \rho}, \\
&\boldsymbol{\mathcal{F}}^{\rm int} = \int_V d\mathbf{x}' \mathbf{F}(\Delta \mathbf{x})  \Delta\mathbf{x}\left(g^{0}(\Delta \mathbf{x}) + \frac{1}{2} \rho\frac{\partial g^{0}(\Delta \mathbf{x})}{\partial \rho}\right),
\end{align}
\end{subequations}
where the we again adopt the density gradient expression $\mathbf{f}^{\rm int}$ from Ref.~\cite{Chiu2024}, and $g^0$ is the single component homogeneous pair distribution function at density $\rho$.
The cABP system is invariant to rotation and governed by a central potential, so the internal force Jacobian $\boldsymbol{\mathcal{F}}^{\rm int}$ simplifies to an isotropic form:
\begin{subequations}
\begin{align}
    \boldsymbol{\mathcal{F}}^{\rm int} &= \mathcal{F}^{\rm int}\mathbf{I}_d, \\
    \mathcal{F}^{\rm int} &= \pi\int_0^\infty r^2 F_r(r)\left(g^{0}(r) + \frac{1}{2}\rho\frac{\partial g^{0}(r)}{\partial \rho}\right)dr,
\end{align}
\end{subequations}
where $F_r$ is the radial component of the force.
Then, from Eq.~\eqref{eq:mutual_diffusion} we find that the diffusion coefficient for a single component system in terms of the resistance tensor is: 
\begin{align}
\label{eq:diffusion_resistence_form}
\mathbf{D} = \boldsymbol{\mathcal{R}}^{-1} \cdot \boldsymbol{\mathcal{F}}.
\end{align}
We can substitute in the results we obtain from Eqs.~\eqref{eq:resistence_cABP} and~\eqref{eq:force_jacobian_cABP} into Eq.~\eqref{eq:diffusion_resistence_form} and arrive at the general form of the diffusion tensor for cABPs:
\begin{equation}
\label{eq:cABP_D}
    \mathbf{D} =  \Bigg(\underbrace{\frac{\zeta}{\rho}\mathbf{I}_d}_{\boldsymbol{\mathcal{R}}^{\rm dis}} \underbrace{- \frac{\zeta}{\rho}U_0\frac{\partial\mathbf{m}}{\partial\mathbf{J}}}_{\boldsymbol{\mathcal{R}}^{\rm act}}\Bigg)^{-1} \cdot \Bigg( \underbrace{\vphantom{\displaystyle\frac{\zeta\partial}{\partial\nabla\rho}}\mathcal{F}^{\rm int}\mathbf{I}_d}_{\boldsymbol{\mathcal{F}}^{\rm int}}\underbrace{- \frac{\zeta}{\rho} U_0 \frac{\partial\mathbf{m}}{\partial\boldsymbol{\nabla}\rho}}_{\boldsymbol{\mathcal{F}}^{\rm act}}  \Bigg).
\end{equation}
By analyzing the contributions to $\mathbf{D}$, we observe that $\boldsymbol{\mathcal{R}}^{\rm dis}$ and $\boldsymbol{\mathcal{F}}^{\rm int}$ are simply scalar multiples of $\mathbf{I}_d$.
To highlight the antisymmetric, or ``odd'', contribution, we now explicitly denote the spatial components of $\mathbf{D}$ and define ${D^{\rm odd} = \frac{1}{2}(D_{xy} - D_{yx})}$.
Using the diffusion tensor expression provided in Eq.~\eqref{eq:cABP_D}, and the rotational invariance of the system (which leads to the constraint that $A_{xx} = A_{yy}$ and $A_{xy} = -A_{yx}$ for physical tensors) we can arrive at:
\begin{align}
\label{eq:odd_diffusion_formula}
    D^{\rm odd} &= \frac{\mathcal{R}_{xx}\mathcal{F}^{\rm act}_{xy} - \mathcal{R}^{\rm act}_{xy}\mathcal{F}_{xx}}{\det ( \boldsymbol{\mathcal{R}})}.
\end{align}
For parity-satisfying systems, we have that ${\mathcal{F}_{xy} = \mathcal{F}_{yx} = 0}$ and ${\mathcal{R}_{xy} = \mathcal{R}_{yx} = 0}$, such that $D^{\rm odd}$ = 0.
We can observe that any odd diffusion contribution to $\mathbf{D}$ must originate from the active components of the resistance tensor and force Jacobian, $\boldsymbol{\mathcal{R}}^{\rm act}$ and $\boldsymbol{\mathcal{F}}^{\rm act}$.
With Eqs.~\eqref{eq:cABP_D} and~\eqref{eq:odd_diffusion_formula}, we can see that the emergence of odd diffusion requires that either density gradients or fluxes induce polarization in the orthogonal spatial direction.

We may further understand the changes in polar order with changes in density gradients or fluxes by analyzing the equation of motion of the polar order field~\cite{Langford2025}:
\begin{align}
    \frac{\partial \mathbf{m}}{\partial t} = - \boldsymbol{\nabla} \cdot \mathbf{J}^{\rm m} - D_r \mathbf{m} + \boldsymbol{\omega}_0 \times \mathbf{m}.
\end{align}
If we are interested in timescales for which the polarization evolves quasistatically such that ${\frac{\partial\mathbf{m}}{\partial t}\approx0}$, we can find a matrix equation for the polar order in terms of the divergence of the polar flux $\mathbf{J}^{\rm m}$:
\begin{align}
    \mathbf{m} = -\frac{1}{D_r^2 + \omega_0^2}\begin{bmatrix}
D_r & \omega_0 \\
-\omega_0 & D_r
\end{bmatrix} \boldsymbol{\nabla} \cdot \mathbf{J}^{\rm m}.
\end{align}
While the expression for polar order does not explicitly contain $\mathbf{J}$, we expect in general that ${\frac{\partial\mathbf{J}^{\rm m}}{\partial \mathbf{J}} \neq \mathbf{0}}$.
As a result, the off-diagonal elements of the active contribution to the resistance tensor, $\boldsymbol{\mathcal{R}}^{\rm act}$, are generally nonzero.
Furthermore, as shown in Ref.~\cite{Langford2025}, $\mathbf{J}^{\rm m}$ exhibits an explicit density dependence and so $\boldsymbol{\nabla} \cdot \mathbf{J}^{\rm m}$ will depend explicitly on $\boldsymbol{\nabla} \rho$.
Therefore, in general, we expect $\frac{\partial \boldsymbol{\nabla} \cdot \mathbf{J}^{\rm m}}{\partial \boldsymbol{\nabla} \rho} \neq \mathbf{0}$, implying that $\boldsymbol{\mathcal{F}}^{\rm act}$ possesses nonzero off-diagonal elements.
In fact, for non-interacting cABPs we find that ${\mathbf{J}^{\rm m}=\frac{\rho}{2}\mathbf{I}_2}$, and we recover the result derived by Hargus et al.~\cite{Hargus2021}:
\begin{align}
    D^{\rm odd} = \frac{U_0^2\omega_0}{2(D_r^2 + \omega^2_0)}.
\end{align}
From this analysis, we conclude that the off-diagonal components of $\boldsymbol{\mathcal{R}}^{\rm act}$ may be non-zero, and those of $\boldsymbol{\mathcal{F}}^{\rm act}$ are generally finite such that cABPs are expected to display odd diffusion.

\subsection{Nonlinear Transport in the Presence of an External Field}
\label{sec:nonlinear_transport}
Several studies have sought to characterize transport beyond the linear response regime~\cite{Onsager1957,Lesnicki2020, Lesnicki2021, Avni2022, Berthoumieux2024}.
While our framework remains linear in form, it is not restricted to expansions around equilibrium states absent irreversible processes.
Instead, it applies more broadly to general steady states, allowing us to analyze how perturbations to forces or fields about these steady states lead to differential changes in flux.
This perspective is particularly relevant for small systems with gradient-driven transport, where both flux and driving forces may become large.
To illustrate this, we develop a mechanical perspective on nonlinear transport under strong external fields by examining a representative case: electric field driven transport in an electrolyte, where nonlinear conductivity arises at high field strength~\cite{Onsager1957}.

In this setting, for an electric field $\mathbf{E}$, we are interested in the \textit{field dependent} differential conductivity defined as:
\begin{subequations}
\label{eq:current_conductivity_definition}
\begin{align}
    \boldsymbol{\sigma}(\mathbf{E}) = \frac{\partial \boldsymbol{\mathcal{I}}}{\partial \mathbf{E}},
\end{align}
with the free charge current density $\boldsymbol{\mathcal{I}}$ given:
\begin{align}
\boldsymbol{\mathcal{I}} = \sum_{i=1}^{n_c}qz_i\mathbf{J}_i, 
\end{align}
\end{subequations}
where $q$ is the unit of fundamental charge and $z_i$ is the valency of species $i$. 
The differential conductivity is equal to the linear conductivity when $\boldsymbol{\sigma}(\mathbf{E})\approx\boldsymbol{\sigma}(\mathbf{0})$.
Higher order corrections to the conductivity as a function of electric field were of interest almost a century ago~\cite{Onsager1957}, and continue to be of interest today~\cite{Lesnicki2020, Lesnicki2021, Avni2022, Berthoumieux2024}.
We first consider a system with uniform densities $\boldsymbol{\rho}$, experiencing a constant electric field $\mathbf{E}$ and species fluxes $\mathbf{J}$.
We introduce a reference state characterized by an electric field $\mathbf{E}^{\rm ref}$ and corresponding species fluxes $\mathbf{J}^{\rm ref}$, chosen such that in this configuration the species momentum balance [see Eq.~\eqref{eq:species_momentum_balance}] takes the form:  
\begin{align}
    \label{eq:nonlinear_mechanical_balance}
    \mathbf{0} = \mathbf{f}^{\rm eff}(\mathbf{E}^{\rm ref}, \mathbf{J}^{\rm ref}, \boldsymbol{\rho}).
\end{align}
For many electrolyte systems, we expect that close to this steady-state the system dynamics meet the requirements for Markovian transport detailed in Appendices~\ref{sec:app_locality} and~\ref{sec:timescale}.
We expand the effective force for fluxes $\mathbf{J}$ and field $\mathbf{E}$ near $\mathbf{J}^{\rm ref}$ and $\mathbf{E}^{\rm ref}$ up to linear order:
\begin{subequations}
\label{eq:f_eff_nonlinear_expansion}
\begin{multline}
    \mathbf{f}^{\rm eff}(\mathbf{E}, \mathbf{J}, \boldsymbol{\rho}) \approx -\boldsymbol{\mathcal{R}}(\mathbf{E}^{\rm ref}, \mathbf{J}^{\rm ref}, \boldsymbol{\rho})\cdot\Delta\mathbf{J} \\
    + \left.\frac{\partial \mathbf{f}^{\rm eff}}{\partial \mathbf{E}}\right|_{(\mathbf{E}^{\rm ref}, \mathbf{J}^{\rm ref}, \boldsymbol{\rho})}\cdot\Delta\mathbf{E},
\end{multline}    
where we define:
\begin{align}
    &\boldsymbol{\mathcal{R}}(\mathbf{E}^{\rm ref}, \mathbf{J}^{\rm ref}, \boldsymbol{\rho}) = -\left.\frac{\partial \mathbf{f}^{\rm eff}}{\partial \mathbf{J}}\right\rvert_{(\mathbf{E}^{\rm ref}, \mathbf{J}^{\rm ref}, \boldsymbol{\rho})}, \\
    &\Delta\mathbf{J} = \mathbf{J} - \mathbf{J}^{\rm ref}, \\
    &\Delta\mathbf{E} = \mathbf{E} - \mathbf{E}^{\rm ref}.
\end{align}
\end{subequations}
Equation~\eqref{eq:f_eff_nonlinear_expansion} is the analog of Eq.~\eqref{eq:effective_force_expansion} for expansions about finite-flux steady states. 
One important difference is that we also expand up to linear order in $\Delta\mathbf{E}$, because we know the explicit form of ${\frac{\partial \mathbf{f}_i^{\rm eff}}{\partial \mathbf{E}}=qz_i\mathbf{I}_d}$.
We can now identify the differential flux response from Eq.~\eqref{eq:f_eff_nonlinear_expansion} as:
\begin{subequations}
\label{eq:nonlinear_flux_response}
    \begin{align}
    &\Delta\mathbf{J}_i = \sum_{j}^{n_c}qz_j\mathbf{L}_{ij}(\mathbf{E}^{\rm ref}, \mathbf{J}^{\rm ref}, \boldsymbol{\rho})\cdot\Delta\mathbf{E}, \label{eq:generalized_flux_response_subeq}\\
    &\mathbf{L}(\mathbf{E}^{\rm ref}, \mathbf{J}^{\rm ref}, \boldsymbol{\rho}) = \boldsymbol{\mathcal{R}}^{-1}(\mathbf{E}^{\rm ref}, \mathbf{J}^{\rm ref}, \boldsymbol{\rho}).
\end{align}
\end{subequations}
For systems that are approximately Galilean invariant and/or exhibit weak dissipation, $\mathbf{L}$ may be constructed with respect to a reference velocity, as discussed in Appendix~\ref{sec:app_markovian_transport_gi_systems}.
Similar to the static case examined in Sec.~\ref{sec:mechanics_of_multicomponent_transport}, we can describe small changes to the now finite steady-state flux.

We are interested in nonlinear transport coefficients with respect to an applied electric field.
In order to find these coefficients, we utilize Eq.~\eqref{eq:nonlinear_flux_response} to build up an understanding of differential transport at each $\mathbf{E}^{\rm ref}$. 
It is then a simple matter to identify in this framework:
\begin{align}
\label{eq:flux_field_derivative}
    \left.\frac{\partial \mathbf{J}_i}{\partial \mathbf{E}}\right\vert_{(\mathbf{E}, \mathbf{J}, \boldsymbol{\rho})} = \sum_{j}^{n_c}qz_j\mathbf{L}_{ij}(\mathbf{E}, \mathbf{J}, \boldsymbol{\rho}),
\end{align}
where this derivative is well-defined for the steady-state flux measured beyond the important timescales $\tau$ and $\tau^{\rm NL}$.
 We can further identify from Eqs.~\eqref{eq:current_conductivity_definition} and~\eqref{eq:flux_field_derivative} a relation between our differential conductivity and finite-field Onsager transport tensor:
\begin{align}
     \boldsymbol{\sigma}(\mathbf{E}, \mathbf{J}, \boldsymbol{\rho}) = \sum_{i}^{n_c}\sum_{j}^{n_c} q^2z_iz_j\mathbf{L}_{ij}(\mathbf{E}, \mathbf{J}, \boldsymbol{\rho}),
\end{align}
which is analogous to the expression for the linear conductivity~\cite{Miller1966,Dufreche2005,Fong2020}.
These results demonstrate the general versatility of this mechanical framework in understanding nonlinear transport through linear, differential transport coefficients.

We can connect these mechanical transport coefficients to prior formulations of differential transport by investigating an overdamped Langevin system. 
For this system, we take the following decomposition of $\mathbf{f}^{\rm eff}$:
\begin{subequations}
    \begin{align}
        &\mathbf{f}^{\rm eff}_i = \mathbf{f}^{\rm int}_i + \mathbf{f}^{\rm dis}_i + qz_i\mathbf{E}, \\
        &\mathbf{f}^{\rm dis}_i = -\frac{\zeta_i}{\rho_i}\mathbf{J}_i,
    \end{align}
\end{subequations}
where the form of $\mathbf{f}^{\rm dis}$ is the same as previously found in Eq.~\eqref{eq:f_eff_passive_langevin}.
The simple form of dissipation in this setting allows us to identify the functional dependence of $\mathbf{J}$ on $\mathbf{E}$ at steady state from the mechanical balance condition [Eq.~\eqref{eq:nonlinear_mechanical_balance}]:
\begin{align}
    \mathbf{J}_i = \frac{\rho_i}{\zeta_i}\mathbf{f}_i^{\rm int} + \frac{\rho_i}{\zeta_i}qz_i\mathbf{E}.
\end{align}
The explicit form of dissipation in overdamped Langevin dynamics allows us to find an expression for flux in terms of an electric field without needing to Taylor expand $\mathbf{f}^{\rm int}$ with respect to the flux.
Specifically, for any external field $\mathbf{E}$, we have:
\begin{align}
    \frac{\partial \mathbf{J}_i}{\partial \mathbf{E}} = \frac{\rho_i}{\zeta_i}\left(\left.\frac{\partial \mathbf{f}_i^{\rm int}}{\partial \mathbf{E}}\right|_{(\mathbf{E}, \mathbf{J}, \boldsymbol{\rho})} + qz_i\mathbf{I}_d\right).
\end{align}
We then observe that for this system:
\begin{align}
    \sum_j^{n_c}qz_j\mathbf{L}_{ij}(\mathbf{E}, \mathbf{J}, \boldsymbol{\rho}) =\frac{\rho_i}{\zeta_i}\left(\left.\frac{\partial \mathbf{f}_i^{\rm int}}{\partial \mathbf{E}}\right|_{(\mathbf{E}, \mathbf{J}, \boldsymbol{\rho})} + qz_i\mathbf{I}_d\right).
\end{align}
For isotropic systems with an applied electric field, we arrive at the expression for differential conductivity as:
\begin{subequations}
\begin{align}
    \sigma(\mathbf{E}) &= \sigma^{\rm id} \nonumber\\
    &+ \sum_i^{n_c}\frac{qz_i\rho_i}{d\zeta_i}\int_V d\Delta\mathbf{x}\sum_j^{n_c}\rho_j\frac{\partial g_{ij}}{\partial \mathbf{E}}(\Delta\mathbf{x};\mathbf{E})\cdot\mathbf{F}_{ij}(-\Delta\mathbf{x}), \\
    \sigma^{\rm id} &= \sum_i^{n_c} \frac{q^2z_i^2\rho_i}{\zeta_i},
\end{align}
\end{subequations}
consistent with prior results such as those in Ref.~\cite{Lesnicki2020}.
It is important to emphasize that the force balance form above relies on the simple linear structure of dissipation inherent in Langevin dynamics. 
However, for many physical systems, this approximation remains practically valid. 
Furthermore, the simple form of dissipation and noise terms in Langevin dynamics enables the use of trajectory ensemble formalism to calculate $\mathbf{L}(\mathbf{E}, \mathbf{J}, \boldsymbol{\rho})$, as performed for differential conductivity in Refs.~\cite{Baesi2013,Lesnicki2021}, and to understand the effect of nonreciprocity on the Onsager reciprocal relations in Appendix~\ref{app:onsager_machlup}.

\section{Discussion and Conclusion}

From the conservation equation for density fields and the species momentum balance, we construct a mechanical framework for macroscopic transport that offers both physical intuition and analytical tools for understanding the dynamical origins of transport relations.
This perspective complements the thermodynamic understanding for systems in which the local equilibrium hypothesis applies, and provides a formalism to extend understanding of transport to active systems and other systems that are intrinsically out of equilibrium. 
From the equations of motion, we identify the resistance tensor $\boldsymbol{\mathcal{R}}$, shown to be the inverse of the mechanical Onsager transport tensor $\mathbf{L}$, and identify the timescales and conditions of applicability for which linear transport laws remain valid. 
To compute the mechanical transport coefficients in nonequilibrium systems, we employed color field theory and demonstrated its validity within our framework, both formally and through numerical simulations.

We applied our mechanical framework to investigate several nonequilibrium systems of interest. 
First, we demonstrated that in nonreciprocal active systems, nonconservative interactions break the Onsager reciprocal relations, resulting in an asymmetric transport tensor $\mathbf{L}$. 
We further showed that the emergence of odd diffusion in chiral active Brownian particles originates entirely from the active driving forces, identifying the mechanical origin of this behavior.
Finally, we extended our framework to analyze nonlinear conductivity, showing that the differential conductivity can be expressed as a function of mechanical transport coefficients evaluated around a nonequilibrium steady state.

This mechanical framework offers potential for extension and application in understanding nonequilibrium transport coefficients beyond those explicitly considered in this work, including the viscosity and elasticity tensors.
For other field variables of importance to dynamics beyond density fields, a similar procedure can be applied by identifying the field evolution equation and relevant flux dynamics, linearizing around a steady state, and extracting the corresponding resistance tensor, which governs the flux dynamics in the overdamped limit.
We note that in our framework, we made no assumptions about the values of other fields (like temperature for equilibrium systems) that may influence the dynamics and introduce new components to the Onsager transport tensor.
This generality allows the mechanical framework to be naturally extended to study transport processes driven by gradients in other fields. 
As an example, we demonstrate how this perspective can be used to understand heat flux and transport due to temperature gradients in Appendix~\ref{sec:heat_flux}.

Importantly, we find that gradient transport coefficients can be decomposed in terms of the mechanical transport coefficient $\mathbf{L}$ and a force Jacobian $\boldsymbol{\mathcal{F}}$.
It is interesting to investigate the explicit breakdown of Einstein relations in active systems by comparing the structure factor matrix with the force Jacobian tensor [see Eq.~\eqref{eq:L_mech_L_GK_eq}].
By identifying discrepancies between these matrices, we can point to where the Einstein relationships fail, as their differences highlight the departure from equilibrium conditions in active systems.
This mechanical framework provides a new path to build up macroscopic transport coefficients from microscopic dynamics, and gives additional analytical and computational approaches to understanding nonequilibrium mechanical and gradient-driven transport with a single unified description.
It is our hope that in future work, this mechanical framework can serve as both analytical and predictive tools for studying the stability of nonequilibrium systems by uncovering transport processes in far-from-equilibrium active multicomponent mixtures.

\appendix
\section{Species Momentum Balance}
\label{sec:appendix_species_momentum_balance}
Our mechanical theory leverages the momentum balance equation for each species. 
To derive its general form, we begin with the microscopic equations of motion.
We consider a system composed of $n_c$ distinct species, where each species $i$ contains $N_i$ particles, such that the total number of particles is given by $N = \sum_{i}^{n_c} N_i$. 
The system’s microstate is given by $\boldsymbol{\Gamma}$, which includes all relevant microscopic degrees of freedom (e.g., particle positions, particle velocities, and orientations), and its statistical behavior is governed by the probability density function  $f(\boldsymbol{\Gamma})$, which describes the likelihood of the system being in a given microstate.
We can describe the general particle dynamics with the following equations of motion:
\begin{subequations}
\begin{align}
    \dot{\mathbf{p}}_i^\alpha = \mathbf{F}_i^\alpha(\boldsymbol{\Gamma}),\\
    \dot{\mathbf{r}}_i^\alpha = \mathbf{p}_i^\alpha/m_i,
\end{align}
\end{subequations}
where $\mathbf{r}_i^\alpha$ ($\mathbf{p}_i^\alpha$) represents the position (momentum) of the $\alpha$th particle of species $i$, $\dot{a}$ denotes the time derivative of $a$, and $\mathbf{F}_i^\alpha(\boldsymbol{\Gamma})$ is the force acting on particle $\alpha$ of species $i$, which can generally depend on the system's entire microstate.

We may define the microscopic expression for the averaged species density as either a spatial average:
\begin{equation}
    \rho_i(\mathbf{x};t) = \sum_\alpha^{N_i} \Delta^{\rm cg}(\mathbf{x} - \mathbf{r}_i^\alpha),
\end{equation}
or an ensemble average:
\begin{subequations}
\begin{align}
     &\rho_i(\mathbf{x};t) = \int_\gamma d\boldsymbol{\Gamma}  \sum_{\alpha}^{N_i}\delta(\mathbf{x} - \mathbf{r}_i^\alpha(t))f(\boldsymbol{\Gamma}),
\end{align}
\end{subequations}
where $\Delta^{\rm cg}(\mathbf{x} - \mathbf{r}_i)$ is a normalized coarse-graining
function, $\delta(\mathbf{x} - \mathbf{r}_i)$ is the Dirac delta function, and $\gamma$ is the phase space volume.
We make the following derivation for the ensemble-averaged density field because it will prove easier to build expressions for, but an analogous argument holds for the spatially coarse-grained density.
The microscopic definition of the species flux is given by:
\begin{subequations}
\label{sieq:species_flux}
\begin{align}
    &\mathbf{J}_i = \sum_\alpha^{N_i}\langle\mathbf{p}_i^\alpha/m_i \delta(\mathbf{x} - \mathbf{r}_i^\alpha)\rangle, \\
    &\langle a\rangle = \int_\gamma d\boldsymbol{\Gamma} a(\boldsymbol{\Gamma})f(\boldsymbol{\Gamma})
\end{align}
\end{subequations}
To derive the time evolution of the species flux, we take the time derivative of Eq.~\eqref{sieq:species_flux}, yielding:
\begin{align}
\label{sieq:species_flux_evol}
&\frac{\partial\mathbf{J}_i}{\partial t} = \frac{1}{m_i}\left \langle\sum_\alpha^{N_i}  \dot{\mathbf{p}}_i^\alpha \delta(\mathbf{x} - \mathbf{r}_i^\alpha) + \sum_\alpha^{N_i} \mathbf{p}_i^\alpha \frac{\partial}{\partial t} \delta(\mathbf{x} - \mathbf{r}_i^\alpha) \right\rangle \nonumber\\
&= \frac{1}{m_i}\left \langle\sum_\alpha^{N_i}  \mathbf{F}_i^\alpha \delta(\mathbf{x} - \mathbf{r}_i^\alpha) + \sum_\alpha^{N_i} \mathbf{p}_i^\alpha \frac{\partial}{\partial\mathbf{r}_i^\alpha} \delta(\mathbf{x} - \mathbf{r}_i^\alpha) \cdot \dot{\mathbf{r}_i^\alpha} \right\rangle\nonumber\\
&= \frac{1}{m_i}\left \langle\sum_\alpha^{N_i}  \mathbf{F}_i^\alpha \delta(\mathbf{x} - \mathbf{r}_i^\alpha) - \sum_\alpha^{N_i} \mathbf{p}_i^\alpha \frac{\partial}{\partial\mathbf{x}} \delta(\mathbf{x} - \mathbf{r}_i^\alpha) \cdot\dot{\mathbf{r}_i^\alpha} \right\rangle\nonumber\\
&= \frac{1}{m_i}\left \langle\sum_\alpha^{N_i}  \mathbf{F}_i^\alpha \delta(\mathbf{x} - \mathbf{r}_i^\alpha) - \frac{\partial}{\partial\mathbf{x}}\cdot \left(\sum_\alpha^{N_i} \frac{\mathbf{p}_i^\alpha \mathbf{p}_i^\alpha}{m_i}  \delta(\mathbf{x} - \mathbf{r}_i^\alpha) \right)\right\rangle,
\end{align}
where we have applied the chain rule and used the fact that the coarse-graining function depends only on the difference between $\mathbf{x}$ and $\mathbf{r}_i^\alpha$ to change the differentiation variable.
By rearrangement of Eq.~\eqref{sieq:species_flux_evol}, we obtain:
\begin{subequations}
\label{eq:derived_species_momentum_balance}
\begin{align}
    &m_i \frac{\partial\mathbf{J}_i}{\partial t} + m_i\boldsymbol{\nabla}\cdot (\mathbf{J}_i\mathbf{J}_i/\rho_i) = \rho_i\mathbf{f}^{\rm eff}_i,\\ 
    &\rho_i\mathbf{f}^{\rm eff}_i \equiv \sum_\alpha^{N_i}  \langle\mathbf{F}_i^\alpha\delta(\mathbf{x}-\mathbf{r}_i^\alpha)\rangle + \rho_i \mathbf{F}^{\rm id}_i(\mathbf{x}),\\
    &\rho_i\mathbf{F}^{\rm id} = -\frac{1}{m_i}\sum_{\alpha}^{N_i} \frac{\partial}{\partial\mathbf{x}}\cdot\left\langle\delta\mathbf{p}_i^\alpha(\mathbf{x})^2\delta(\mathbf{x} - \mathbf{r}_i^\alpha)\right\rangle,\\
    &\delta\mathbf{p}_i^\alpha(\mathbf{x}) = \mathbf{p}_i^\alpha - m_i\mathbf{v}_i,
\end{align}
\end{subequations}
where we define the sum of the force contributions to species $i$ as the effective force, $\mathbf{f}_i^{\rm eff}$, in the main text.
Thus, we have recovered the species momentum balance equation presented in Eq.~\eqref{eq:species_momentum_balance}.

While we construct the species momentum balance using species-specific velocities as the reference frame, it remains possible to recover the overall static momentum balance from our formulation. 
In the static limit ($\mathbf{J} = \mathbf{0})$, the sum of the species-level effective body forces yields the familiar static form of the total momentum balance:
\begin{align}
    \sum_i^{n_c} \rho_i \mathbf{f}_i^{\rm eff} = \boldsymbol{\nabla} \cdot \boldsymbol{\sigma} + \mathbf{b},
\end{align}
where $\boldsymbol{\sigma}$ is the overall system stress tensor and $\mathbf{b}$ is the overall body force.
This relation holds for the total momentum balance regardless of the reference frame, as the convective contributions vanish in the static limit.

\section{Locality Assumption Made Explicit}
\label{sec:app_locality}
We want to understand the characteristic time and length scales that allow us to localize the resistance kernel $\mathbf{R}(\mathbf{x}, \mathbf{x}', t, t')$ in the expansion of effective force $\mathbf{f}^{\rm eff}(\mathbf{x}, t)$.
This constitutes an assumption that $\mathbf{f}^{\rm eff}(\mathbf{x}, t)$ depends only on the values of the fields that define the system at $\mathbf{x}$ and $t$, up to system specific a molecular interaction lengthscale and relaxation timescale. 
To clarify the mathematical implications of this assumption,  we start by examining the general non-local expansion of the effective force given in Eq.~\eqref{eq:f_eff_nonlocal}.
We first note that we can always write:
\begin{align}
    \mathbf{J}(\mathbf{x}',t') = 
    \mathbf{J}(\mathbf{x},t) + \Delta\mathbf{J}(\mathbf{x},\mathbf{x}',t,t').
    \end{align}
Assuming the flux field $\mathbf{J}$ varies smoothly and exhibits small gradients, we expand the flux employing the Lagrange form of the Taylor expansion to write the correction:
\begin{align}
    \Delta\mathbf{J}(\mathbf{x},\mathbf{x}',t,t') = \boldsymbol{\nabla} \mathbf{J}(\mathbf{x}^*, t^*)\cdot(\mathbf{x}' - \mathbf{x}) + \partial_t \mathbf{J}(\mathbf{x}^*, t^*)(t' - t),
\end{align}
for some intermediate point in space and time $(\mathbf{x}^*, t^*)$ located within the domain described by $(\mathbf{x}, t)$ and $(\mathbf{x}', t')$. 
We can now rewrite our expansion in Eq.~\eqref{eq:f_eff_nonlocal} as:
\begin{subequations}
\begin{align}
    \mathbf{f}^{\rm eff}(\mathbf{x}, t) &= \mathbf{f}^{\rm eff}(\mathbf{x}, t)|_{\mathbf{J}=\mathbf{0}} - \boldsymbol{\mathcal{R}}(\mathbf{x}, t)\cdot\mathbf{J}(\mathbf{x}, t) \nonumber \\
    &-\mathbf{R}^x(\mathbf{x}, t)\cdot\boldsymbol{\nabla} \mathbf{J}(\mathbf{x}^*, t^*) - \mathbf{R}^t(\mathbf{x}, t)\cdot\partial_t\mathbf{J}(\mathbf{x}^*, t^*), \\
    \mathbf{R}^x(\mathbf{x}, t) &= -\int_{-\infty}^t dt'\int_V d\mathbf{x}' \mathbf{R}(\mathbf{x}, \mathbf{x}', t, t')(\mathbf{x}'-\mathbf{x}), \\
    \mathbf{R}^t(\mathbf{x}, t) &= -\int_{-\infty}^t dt'\int_V d\mathbf{x}' \mathbf{R}(\mathbf{x}, \mathbf{x}', t, t')(t'-t).
\end{align}
\end{subequations}
This naturally leads to the definition of the characteristic time and length scales over which non-locality becomes important:
\begin{subequations}
\begin{align}
\lambda^{\rm NL} = \sup_{\mathbf{x},t}\frac{\|\mathbf{R}^x(\mathbf{x},t)\|_{\rm op}}{\|\mathbf{R}(\mathbf{x}, t)\|_{\rm op}},\\
\tau^{\rm NL} = \sup_{\mathbf{x},t}\frac{\|\mathbf{R}^t(\mathbf{x},t)\|_{\rm op}}{\|\mathbf{R}(\mathbf{x}, t)\|_{\rm op}},
\end{align}
\end{subequations}
where the norm $|| \cdot ||_{\rm op}$ is the appropriate operator norm for each tensor. 
We expect that as long as the following conditions hold:
\begin{subequations}
\begin{align}
    \lambda^{\rm NL}\ll \sup_{\mathbf{x},\mathbf{x}', t,t'}\frac{\|\mathbf{J}(\mathbf{x}',t')\|}{\|\boldsymbol{\nabla}\mathbf{J}(\mathbf{x},t)\|}, \\
    \tau^{\rm NL}\ll \sup_{\mathbf{x},\mathbf{x}', t,t'}\frac{\|\mathbf{J}(\mathbf{x}',t')\|}{\|\partial_t\mathbf{J}(\mathbf{x},t)\|},
\end{align}
\end{subequations}
the non-local contributions become negligible.
Consequently, the effective force reduces to its local zeroth-moment form:
\begin{align}
    \mathbf{f}^{\rm eff}(\mathbf{x}, t) \approx \mathbf{f}^{\rm eff}(\mathbf{x},t)|_{\mathbf{J}=\mathbf{0}} - \boldsymbol{\mathcal{R}}(\mathbf{x},t)\cdot\mathbf{J}(\mathbf{x},t).
\end{align}
We can conclude that when nonlocality is negligible based on the characteristic scales and the scales of interest, we can consider the effective force to only depend on the local flux, leading to a local-form transport equation.

\section{Inertial Timescales of Transport}
\label{sec:timescale}
In Appendix~\ref{sec:app_locality} we examine system-specific length and timescales which define a local dependence of $\mathbf{f}^{\rm eff}$ on $\mathbf{J}$ in space and time.
We now aim to understand the inertial timescales associated with the response of $\mathbf{J}$ to $\mathbf{f}^{\rm eff}$.
For our analysis, we consider the small flux limit and now seek to understand the timescales over which the system will appear Markovian.
As derived in the main text, under the conditions stated, we can arrive at the species momentum density in the form of Eq.~\eqref{eq:species_momentum_balance_linear}.
By rearrangement of Eq.~\eqref{eq:species_momentum_balance_linear}, we find that up to linear order in $\mathbf{J}_i$, we have the explicit solution of species flux as:
\begin{subequations}
\begin{align}
    \mathbf{J}_i(t) = \sum_j^{n_c} e^{-\mathbf{A}_{ij}t}\mathbf{J}_j(0) + \int_0^t e^{-\mathbf{A}_{ij}(t-t')}\frac{\rho_j}{m_j} \mathbf{f}^{\rm static}_j(t') dt',
\end{align}
where the species component of the tensor $\mathbf{A}$ is defined as in Eq.~\eqref{eq:A_time_matrix}:
\begin{equation}
\label{eq:A_time_matrix_2}
    \mathbf{A}_{ij} \equiv \boldsymbol{\mathcal{R}}_{ij}\rho_i/m_i.
\end{equation}
\end{subequations}
In the discussion below, we assume diagonalizability of $\mathbf{A}$ for notational convenience in the discussion for this section. 
If $\mathbf{A}$ is defective, the same argument can be made, but for $\mathbf{A}$ transformed into Jordan normal form appropriately.

From here, we can identify the salient relaxation timescales as the inverse eigenvalues of $\mathbf{A}$. 
By expanding both the initial flux and the static force contributions in the eigenbasis of $\mathbf{A}$, we can rewrite the explicit solution of the species and spatial flux vector:
\begin{multline} 
J_{iA}(t) = \sum_{k}^{n_c} \sum_{B}^{d} \sum_{m}^{n_c} \sum_{C}^{d} \Bigg[ Y_{iAkB} e^{-\lambda_{kB} t} J_{mC}(0) \\+ \int_0^t Y_{iAkB} e^{-\lambda_{kB} (t - t')} f^{\rm static}_{mC}(t') dt' \Bigg] (Y^{-1})_{kBmC}, 
\end{multline}
where $Y_{iAkB}$ denotes the $i$-th species and $A$-th spatial component of the $k$-th species and $B$-th spatial component right eigenvector of $\mathbf{A}$, $\lambda_{kB}$ are the corresponding eigenvalues, and $\mathbf{Y}^{-1}$ contains the left eigenvectors, such that ${\mathbf{Y}^{-1} \mathbf{Y} = \mathbf{I}}$.
We explicitly include spatial indices $A, B, C, \ldots$ to make the summation over spatial dimensions clear.
With this form of the flux vector, we can identify the key conditions for the constitutive relation in Eq.~\eqref{eq:species_momentum_balance_linear} to hold. 
The first condition is that $\mathbf{A}$ must have purely positive eigenvalues. 
This corresponds to the stability of the system, specifically, forces must be dissipative such that relative motion between particles of different species will produce forces that dampen the motion, ensuring species flux stays in the linear regime.
This is consistent with the expectation that we are interested in transport close to stable steady states.
The second condition is related to the timescales of interest. 
As we are interested in the memory-less regime, or the regime for which measured fluxes are independent of the flux history, we can identify that the eigenvalues $\lambda_{kB}$ provide an inverse time-scale for relaxation times of different coordinated species movements.
We therefore require that forces vary and measurements are made on timescales $t >> \tau$, for:
\begin{align}
    \frac{1}{\tau} = \inf_\mathbf{x} \{ \Re(\lambda_{\rm min}(\mathbf{A}(\mathbf{x}))\}.
\end{align}
We can gain some intuition for this by looking at the form of ${\mathbf{A}_{ij} = -\frac{\rho_i}{m_i}\frac{\partial\mathbf{f}_i^{\rm eff}}{\partial\mathbf{J}_j}}\Big|_{\mathbf{J} = \mathbf{0}}$.
The smallest magnitude of eigenvalue here corresponds to collections of species for which there is the \textit{least} change in mass-normalized force from an applied velocity. This means the timescales of the overdamped limit are controlled by the combination of species that can flow and generate the weakest system response.
Therefore, we must measure flux response to forces which vary on timescales much longer than that given by this weakest system response.
Since the relaxation times may have a spatial dependence, to truly reach the overdamped limit of our system of interest, we must consider forces varying on timescales much longer than the supremum of all relevant timescales.

This timescale $\tau$ describes timescales over which $\mathbf{J}$ has transient behavior in response to some change in $\mathbf{f}^{\rm static}$. Similarly, $\tau^{\rm NL}$ (introduced in Appendix~\ref{sec:app_locality}) describes timescales over which $\mathbf{J}$ must be slowly varying such that $\mathbf{f}^{\rm eff}$ depends only on the current value of $\mathbf{J}$. While the exact values of these timescales are not the same, the maximum of the two tells us when we can treat $\mathbf{J}$ as an instantaneous local function of $\mathbf{f}^{\rm static}$.

\section{Overdamped Transport in Galilean Invariant Systems}
\label{sec:app_markovian_transport_gi_systems}
In systems without a clear velocity reference frame, if there exists a frame for which $\mathbf{f}^{\rm eff}=\mathbf{0}$, there necessarily exists a linear subspace of fluxes that generate no corresponding force. 
Specifically, as shown in Eq.~\eqref{eq:null_space_of_interactions}, we expect:
\begin{align}
    \sum_{j}^{n_c} \boldsymbol{\mathcal{R}}_{ij}\cdot\mathbf{u}\rho_j  = \mathbf{0},
\end{align}
to hold for any constant velocity vector $\mathbf{u}$ and for all species $i$. 
This condition reveals the presence of a null space of the resistance tensor $\boldsymbol{\mathcal{R}}$ and consequently a corresponding null space of $\mathbf{A}$, defined in Eq.~\eqref{eq:A_time_matrix}.
We will find that while Galilean invariance yields a condition on the right null space, the left nullspace of $\boldsymbol{\mathcal{R}}$ will be the one that defines the ``inertial'' flux contributions.
Then, out of equilibrium when $\boldsymbol{\mathcal{R}}$ is not symmetric we know Galilean invariance will lead to a nontrivial left null space of $\boldsymbol{\mathcal{R}}$, but only on a system-by-system basis will we be able to identify inertial modes of the flux.
However, in equilibrium, the symmetric nature of $\boldsymbol{\mathcal{R}}$ (as demonstrated in Sec.~\ref{sec:passive_transport}) allows us to explicitly characterize the null space of $\mathbf{A}$. 
In particular, the left kernel of $\mathbf{A}$ in equilibrium is spanned by vectors whose species components are given by $m_i\mathbf{u}$ for an arbitrary uniform velocity $\mathbf{u}$.
We can then identify the portion of $\mathbf{J}$ which we hope to model in the form of Eq.~\eqref{eq:flux_onsager_mat} by projecting it onto the row space of $\mathbf{A}$. 
We then find that the inertial component of the flux that is not overdamped, denoted as $\mathbf{J}^{\rm inertial}$, is aligned with the center-of-mass velocity $\mathbf{v}^{\rm com}$, and can be expressed as:
\begin{subequations}
\begin{align}
    &\mathbf{J}_i^{\rm inertial} = \rho_i\mathbf{v}^{\rm com},\\
    &\mathbf{v}^{\rm com} = \frac{\sum_{j}^{n_c} m_j\mathbf{J}_j}{\sum_{k}^{n_c} m_k \rho_k}.
\end{align}
\end{subequations}
The overdamped component of each species flux, defined relative to the center-of-mass motion, is then:
\begin{align}
    \mathbf{J}_i^{\rm com} =\rho_i(\mathbf{v}_i - \mathbf{v}^{\rm com}).
\end{align} 
Therefore, for Galilean invariant systems, only $\mathbf{J}^{\rm com}$ and its linear transformations admit a constitutive relation of the form ${\mathbf{J} = \mathbf{L} \cdot \mathbf{f}}$, and thus are well-described by our linear mechanical transport framework.

We project Eq.~\eqref{eq:species_momentum_balance_linear} onto the overdamped subspace associated with nonzero positive eigenvalues.
At sufficiently long times (determined by the eigenvalue with the smallest real part within this subspace), we recover the following relation:
\begin{align}
    \sum_j^{n_c} \boldsymbol{\mathcal{R}}_{ij}\cdot \mathbf{J}_j = \mathbf{f}^{\rm static}_i - m_i\sum_{j}^{n_c}\frac{\mathbf{f}^{\rm static}_j\rho_j}{\sum_k^{n_c} \rho_k m_k}.\label{eq:diffusive_flux_force}
\end{align}
Although $\boldsymbol{\mathcal{R}}$ is not strictly invertible due to the presence of inertial modes, by construction, the RHS of Eq.~\eqref{eq:diffusive_flux_force} lies entirely within the row and column space of $\boldsymbol{\mathcal{R}}$. 
Thus, we can invert $\boldsymbol{\mathcal{R}}$ within this restricted subspace to find:
\begin{align}
    \label{eq:flux_star_onsager_relation}
    \mathbf{J}_i^* = \boldsymbol{\mathcal{R}}^{-1}_{ij}\cdot \left(\mathbf{f}^{\rm static}_i - m_i\sum_{j}^{n_c}\frac{\mathbf{f}^{\rm static}_j\rho_j}{\sum_k^{n_c} \rho_k m_k}\right).
\end{align}
Because $\mathbf{J}^*$ must lie in the column-space of $\boldsymbol{\mathcal{R}}$, it satisfies:
\begin{subequations}
\begin{align}
    \sum_i^{n_c} \rho_i \mathbf{J}_i^* = \mathbf{0},
\end{align}
allowing us to express the flux explicitly in terms of the velocity vector $\mathbf{v}^*$:
\begin{align}
&\mathbf{J}^*_i = \rho_i(\mathbf{v}_i - \mathbf{v}^*), \\
&\mathbf{v}^* = \frac{\sum_i^{n_c} \rho_i^2 \mathbf{v}_i}{\sum_i^{n_c}\rho_i^2}.
\end{align}
\end{subequations}

If the symmetry of the Onsager tensor $\mathbf{L}$ were not essential, we could simply apply a linear transformation to both sides of Eq.~\eqref{eq:flux_star_onsager_relation} to obtain a linear relationship between $\mathbf{J}^{\rm com}$ and $\mathbf{f}^{\rm static}$. 
However, since $\boldsymbol{\mathcal{R}}$ is symmetric, it is desirable to preserve this symmetry when transforming the transport equation to retain the physical interpretation of $\mathbf{L}$ as an Onsager tensor. 
In Appendix \ref{sec:app_alternative_relative_flux_definitions}, we demonstrate that such a transformation exists, allowing us to express the equation in terms of the flux relative to the center-of-mass velocity:
\begin{subequations}
\begin{align}
    &\mathbf{J}_i^{\rm com} = \sum_{j,k,r}^{n_c}\mathbf{T}_{ij}\cdot\boldsymbol{\mathcal{R}}^{-1}_{jk}\cdot\mathbf{T}^{\intercal}_{kr} \cdot \mathbf{f}^{\rm static}_r,\label{eq:expression_for_flux_full_force} \\
    &\mathbf{T}_{ij} = \mathbf{I}_d\left(\delta_{ij} - \frac{\rho_i m_j}{\sum_k^{n_c} \rho_k m_k} \right).
\end{align}
\end{subequations}
This leads to the final expression:
\begin{align}
    &\mathbf{J}^{\rm com} = \mathbf{L}^{\rm com}\cdot\mathbf{f}^{\rm static}, \\
    &\mathbf{L}^{\rm com} = \mathbf{T}\cdot\boldsymbol{\mathcal{R}}^{-1}\cdot\mathbf{T}^{\intercal},
\end{align}
where the superscript $\intercal$ applied to the tensor with $(dn_c)^2$ components denotes a transpose taken over both the spatial and species dimensions.
We note that the center-of-mass contribution to the force, appearing in Eq.~\eqref{eq:flux_star_onsager_relation}, lies in the null space of $\mathbf{L}^{\rm com}$. 
In this way, we recover a well-defined and symmetric transport coefficient $\mathbf{L}^{\rm com}$ for diffusive fluxes measured in the center-of-mass frame. Further transformations to define flux relative to other species’ velocities are discussed in Appendix~\ref{sec:app_alternative_relative_flux_definitions}.

\section{Alternative Relative Flux Definitions}
\label{sec:app_alternative_relative_flux_definitions}
In molecular systems, it is often more convenient to express fluxes relative to a selected frame of reference, such as the system center-of-mass velocity $\mathbf{v}^{\rm com}$.
In this section, we demonstrate how relative fluxes can be transformed between reference frames while preserving the key equilibrium symmetry of the Onsager transport tensor $\mathbf{L}$ and that changing the reference frame does not affect the applicability of the mechanical theory.
Specifically, we show the example of transforming fluxes $\mathbf{J}^{*}$ measured relative to velocity $\mathbf{v}^{*}$ (defined in Appendix~\ref{sec:app_markovian_transport_gi_systems}) to fluxes $\mathbf{J}^{\rm com}$ measured relative to the center-of-mass velocity $\mathbf{v}^{\rm com}$ with preserved symmetry.
Parts of our derivation follow the mathematical framework outlined by DeGroot and Mazur~\cite{DeGroot2013}.

We begin by expressing the species flux relative to an arbitrary reference velocity, denoted by $\tilde{\mathbf{v}}$:
\begin{align}
 \tilde{\mathbf{J}}_i = a_i(\mathbf{v}_i - \tilde{\mathbf{v}}),
\end{align}
where $a_i$ is a per-species weighting which could be, for example, the species number density, mass density, or volume fraction. 
We then consider a general linear constitutive relation of the form:
\begin{align}
    \tilde{\mathbf{J}} = \tilde{\mathbf{L}}\cdot \tilde{\mathbf{f}},
\end{align}
where $\tilde{\mathbf{f}}$ is defined in this frame and need not be identical to $\mathbf{f}^{\rm static}$. 
We consider a transformation of the species flux to a new reference velocity frame given by:
\begin{align}
\label{eq:flux_frame_transform}
    &\mathbf{J}_i' = a_i(\mathbf{v}_i - \mathbf{v}'), \\
    &\mathbf{v}' = \frac{\sum_{j}^{n_c} a_jb_j \mathbf{v}_j}{\sum_{k}^{n_c} a_k b_k},
\end{align}
for some new velocity weighting scheme $a_ib_i$, where we expect $a_i$ to be spatially varying like a density field, and $b_i$ to be a constant species property such as mass, particle volume, or other constant.
Specifically, the transformation from number density flux (e.g., $\mathbf{J}^*$) into the center-of-mass velocity frame corresponds to choosing $a_i = \rho_i$ and $b_i = m_i$. 
We can express the transformation of the flux using the linear map:
\begin{align}
    \mathbf{T}_{ij} = \mathbf{I}_d \left(\delta_{ij} - \frac{a_i b_j}{\sum_{k}^{n_c} a_k b_k}\right),
\end{align}
which acts as an idempotent oblique projection satisfying $\mathbf{T}^2 = \mathbf{T}$.
Its left null space is spanned by the vector $b_i$, while its right null space is spanned by the vector $a_i$.
Applying the transformation operator to both sides of the linear constitutive relation, we obtain:
\begin{align}
    \mathbf{T} \cdot \tilde{\mathbf{J}} = \mathbf{T}\cdot \tilde{\mathbf{L}}\cdot \tilde{\mathbf{f}}.
\end{align}
This constitutes a valid linear relationship in the transformed reference frame. 
By defining ${\mathbf{J}' = \mathbf{T}\cdot\tilde{\mathbf{J}}}$ and ${\mathbf{L}' = \mathbf{T}\cdot\tilde{\mathbf{L}}}$, we immediately recover a linear relation between $\mathbf{J}'$ and the original force vector ${\mathbf{f}'=
\tilde{\mathbf{f}}}$, without requiring additional transformations.
However, in equilibrium, symmetry of the Onsager transport tensor $\tilde{\mathbf{L}}$ typically depends on the chosen reference frame. Thus, to preserve symmetry, the transformation to the new reference frame must be performed carefully.

We note that the operator $\mathbf{T}$ has a non-trivial null space and therefore is not invertible. 
However, since the original transport tensor $\tilde{\mathbf{L}}$ is rank-deficient -- with a null space corresponding to collective motion -- there exists a set of constants ${c_1, c_2, \dots, c_{n_c}}$ such that ${\sum_i^{n_c} c_i\tilde{\mathbf{J}}_i=\mathbf{0}}$.
In the example from Appendix~\ref{sec:app_markovian_transport_gi_systems}, this corresponds specifically to ${\sum_i^{n_c}\rho_i\mathbf{J}^*_i=0}$.
Consequently, adding any tensor of the form $d_i c_j\mathbf{I}_d$ (for an arbitrary vector $d_i$) to the transformation operator $\mathbf{T}_{ij}$ leaves both $\mathbf{T}\cdot\tilde{\mathbf{J}}$ and $\mathbf{T}\cdot\tilde{\mathbf{L}}$ unchanged. 
Thus, we can write an new invertible transformation tensor $\boldsymbol{\mathcal{T}}$, defined in component form:
\begin{align}
    \boldsymbol{\mathcal{T}}_{ij} = \mathbf{I}_d \left(\delta_{ij} - \frac{a_i b_j}{\sum_k^{n_c} a_k b_k} - d_i c_j\right),
\end{align}
where we are free to choose $d_i$. 
We can now preserve the symmetry of $\tilde{\mathbf{L}}$ under transformation with the following operations:
\begin{subequations}
\begin{equation}
\mathbf{J}' = \mathbf{L}'\cdot \mathbf{f}',
\end{equation}
where the new transformation of each component of the equation is defined as:
\begin{align}
&\mathbf{J}' = \boldsymbol{\mathcal{T}}\cdot \tilde{\mathbf{J}},\\
&\mathbf{L}' = \boldsymbol{\mathcal{T}}\cdot \tilde{\mathbf{L}}\cdot \boldsymbol{\mathcal{T}}^{\intercal}, \\
&\mathbf{f}' = (\boldsymbol{\mathcal{T}}^{\intercal})^{-1}\cdot \tilde{\mathbf{f}}.
\end{align}
\end{subequations}
As a result of the newly defined $\mathbf{L}'$, the equilibrium symmetry property is now preserved. 
We can explicitly compute the inverse $(\boldsymbol{\mathcal{T}}^{\intercal})^{-1}$ using the Woodbury matrix identity, which reveals that $\boldsymbol{\mathcal{T}}^{\intercal}$ is invertible provided the following conditions hold:
\begin{align}
    \label{eq:woodbury_condition}\sum_{i,j}^{n_c} a_ic_i d_j b_j \neq 0, \quad \sum_{k}^{n_c} a_k b_k \neq 0.
\end{align}
Additionally, the formula recovers that we can decompose $(\boldsymbol{\mathcal{T}}^{\intercal})^{-1}$ as follows:
\begin{align}
    (\boldsymbol{\mathcal{T}}^{\intercal})^{-1}_{ij} = \mathbf{I}_d\left(\delta_{ij} -  q_i a_j - r_i d_j\right),
\end{align}
for suitably defined coefficients $q_i$ and $r_i$.

Now we are ready to explicitly define the transformation from $\mathbf{J}^*$ to the center-of-mass reference frame flux $\mathbf{J}^{\rm com}$. 
In this particular transformation, we have $a_i = \rho_i$, $b_i = m_i$, and $c_i = \rho_i$, and therefore we can choose $d_i = \rho_i$ to satisfy the invertibility conditions specified by Eq.~\eqref{eq:woodbury_condition}. 
With $\mathbf{f}^{*}$ defined previously in Eq.~\eqref{eq:flux_star_onsager_relation}, we have:
\begin{align}
    \mathbf{f}^{*}_i = \left(\mathbf{f}^{\rm static}_i - m_i\sum_{j}^{n_c}\frac{\mathbf{f}^{\rm static}_j\rho_j}{\sum_k^{n_c} \rho_k m_k}\right).
\end{align}
We can now apply the transformation to the force while preserving the symmetry of $\mathbf{L}$. 
First, we observe the following identity:
\begin{align}
    \sum_i^{n_c} \rho_i \mathbf{f}^*_i = \mathbf{0},
\end{align}
which implies $\mathbf{f}^*$ will be invariant under $\boldsymbol{\mathcal{T}}^{\intercal-1}$, so that:
\begin{align}
(\boldsymbol{\mathcal{T}}^{\intercal})^{-1}\cdot \mathbf{f}^* = \mathbf{f}^*.
\end{align}
Recognizing that the term ${m_i\sum_{j}^{n_c}\frac{\mathbf{f}^{\rm static}_j\rho_j}{\sum_k^{n_c} \rho_k m_k}}$ lies in the null space of  $\mathbf{L}^{\rm com}$, we arrive at the simple linear relationship we expected:
\begin{align}
    \mathbf{J}^{\rm com} = \mathbf{L}^{\rm com}\cdot \mathbf{f}^{\rm static}.
\end{align}

\section{The Local Equilibrium Hypothesis}
\label{app:local_equilibrium_hypothesis}

The local equilibrium hypothesis is often stated as the assumption in irreversible processes that local thermodynamic functions are well-defined~\cite{DeGroot2013}.
This implies that there exists some separation of time and length scales in the system such that macroscopic fields have variations relevant to dynamics, but some neighborhoods around each point behave as a system in thermal equilibrium.
Fluxes are postulated to be driven by the product of gradients in thermodynamic variables and a transport matrix $\mathbf{L}$.
The symmetry of $\mathbf{L}$ depends on the time-reversal symmetry of observables, and the fact that $\mathbf{L}$ is positive semi-definite is derived from a local second law.
Further implications and use cases for the local equilibrium hypothesis are well explored in several references discussing irreversible thermodynamics~\cite{Prigogine1968, DeGroot2013}.

Here, we offer a complementary interpretation of the local equilibrium hypothesis based on our derivation of Eq.~\eqref{eq:flux_onsager_mat}.
For passive systems, we expect that $\mathbf{f}^{\rm static}$ must be related to an equilibrium thermodynamic force.
The standard linear irreversible thermodynamics treatment identifies this force as ${\mathbf{f}^{\rm static} = -T\boldsymbol{\nabla}\left(\boldsymbol{\mu}/T\right)}$.
We can now take a statistical perspective on this thermodynamic force.
Let us consider a system at steady state with ${\mathbf{J} = \mathbf{0}}$.
We can define the phase space of our system as the set of all degrees of freedom needed to describe the fields of interest.
While the distribution of these degrees of freedom, $f(\boldsymbol{\Gamma})$, is time-independent it may correspond to a \textit{nonequilibrium steady state}.
A hallmark of such steady states is the presence of finite probability currents which reflect the presence of forces that do not directly alter the distribution.
Moreover, these currents are divergence-free in phase space and thus necessarily cannot be express as a gradient of the distribution: these are not conservative thermodynamic forces.

Let us now consider a system at steady state that does not exhibit finite probability currents. 
We can marginalize this distribution to obtain the distribution of the fields of interest.
For simplicity, we focus on the coarse-grained number density fields and find:
\begin{subequations}
\begin{align}
   & f^{\rm cg}[\boldsymbol{\rho}(\mathbf{x})] = \langle \delta[\boldsymbol{\rho} - \hat{\boldsymbol{\rho}}^{\rm cg}]\rangle, \\
    &\delta[g - \hat{g}^{\rm cg}] = \prod_{\mathbf{x}\in V}\delta(g(\mathbf{x}) - \hat{g}^{\rm cg}(\boldsymbol{\Gamma}, \mathbf{x})),\\
    &\hat{\rho}^{\rm cg}_i(\boldsymbol{\Gamma}, \mathbf{x}) = \sum_\alpha^{N_i} \delta(\mathbf{x} - \mathbf{r}^\alpha_i),
\end{align}
\end{subequations}
where the expectation here is taken over $f$. 
The distribution of these density fields, $f^{\rm cg}$, will also be a steady state current-free distribution.
This distribution is expected to be sharply peaked about specific configurations (e.g., spatially uniform densities for stable systems in the absence of external fields).
Of course, finite fluctuations about these states are possible and we can identify a statistical force that will restore the system back to its typical state as:
\begin{align}
\label{eq:f_static_detailed_balance}
    \mathbf{f}_i^{\rm static}(\mathbf{x}) = k_BT \boldsymbol{\nabla} \frac{\delta}{\delta \rho_i(\mathbf{x})}\ln f^{\rm cg}[\boldsymbol{\rho}, T].
\end{align}
If we further define a free energy functional (up to an additive constant) ${-\mathcal{A}[\boldsymbol{\rho}(\mathbf{x})]/k_BT = \ln f^{\rm cg}}$, we can appreciate configurations that minimize $\mathcal{A}$ will be stable ``force-free'' states.
We can finally use Eq.~\eqref{eq:f_static_detailed_balance} with this free energy to recover:
\begin{subequations}
\begin{align}
    &\mathbf{f}^{\rm static} = -T\boldsymbol{\nabla}\left(\frac{\boldsymbol{\mu}}{T}\right), \\
    &\boldsymbol{\mu} = \frac{\delta \mathcal{A}}{\delta \boldsymbol{\rho}(\mathbf{x})}.
\end{align}
\end{subequations}
We interpret the local equilibrium hypothesis as an ansatz that the same statistical force is generated regardless of whether a configuration represents a spontaneous equilibrium fluctuation or is in response to nonequilibrium conditions.
This ansatz shares similarities to Onsager's regression hypothesis, which was invoked to connect transport coefficients to microscopic fluctuations. 
Some passive systems (systems with steady states free of probability currents) may not satisfy this condition, especially if the nonequilibrium driving results in strong departures from the preferred equilibrium configurations. 
However, the success of the local equilibrium hypothesis in predicting and understanding transport phenomena suggests that this picture of coupling between dynamical forces and equilibrium probability gradients holds quite generally in macroscopic systems. 

\section{Linear Response and Onsager Reciprocal Relations for Passive Systems}
\label{sec:app_linear_response}
Using the framework of linear response theory as outlined in Ref.~\cite{Evans2008}, we show that for canonically distributed passive systems, $\boldsymbol{\mathcal{R}}$ is symmetric. 
We can use this framework to calculate $\frac{\partial \mathbf{f}^{\rm int}}{\partial \mathbf{J}} \Big|_{\mathbf{J} = \mathbf{0}}$, and therefore determine $\boldsymbol{\mathcal{R}}$, by considering an ensemble which is a perturbation of a passive system at equilibrium.
We denote the \emph{equilibrium} distribution over the system phase space $f_0(\boldsymbol{\Gamma})$, where $\boldsymbol{\Gamma}$ is once again a vector of system degrees of freedom.
For the following derivation, we take:
\begin{align}
    f_0 \propto \exp\{-\beta \mathcal{U}(\{\mathbf{r}_\alpha\})\},
\end{align}
and the only position dependence of the equilibrium distribution is through this potential energy. 
We will need to introduce some further notation in this discussion.
We consider the expectation $\langle \cdot \rangle_0$ to be an expectation over the equilibrium distribution, and $\langle \cdot \rangle_1$ an expectation over a perturbed distribution where the velocities of a given species are uniformly shifted by some small velocity starting at time $0$.
This perturbed ensemble then represents a finite $\mathbf{J}$ ensemble that we can use to calculate system response, and therefore $\boldsymbol{\mathcal{R}}$.

In order to complete this calculation we will take advantage of the translational symmetry of the equilibrium distribution and our applied perturbation, and observe that we can write:
\begin{align}
    &\mathbf{f}^{\rm int}_i = \frac{1}{V\rho_i} \langle \mathbf{F}_i\rangle, \\
    &\mathbf{F}_i = \sum_{\alpha}^{N_i} \mathbf{F}_i^\alpha, \\
&\mathbf{F}_i^\alpha = -\frac{\partial \mathcal{U}}{\partial \mathbf{r_i^\alpha}}.
\end{align}
where $\mathbf{F}_i$ is an observable giving the total force on species $i$.

We can calculate the required derivative $\frac{\partial \mathbf{f}^{\rm int}}{\partial \mathbf{J}}$ by understanding $\langle \mathbf{F}_i\rangle_1 - \langle\mathbf{F}_i\rangle_0$. 
In the framework of linear response theory, we consider a small perturbation to the distribution evolution operator $\mathcal{L}$:
\begin{subequations}
\begin{align}
    \frac{\partial f}{\partial t} = -\mathcal{L} f, \\
    \mathcal{L} = \mathcal{L}_0 + \Delta \mathcal{L}, \\
    f = f_0 + \Delta f,
\end{align}
\end{subequations}
where $\mathcal{L}_0$ is the original (unperturbed) evolution operator, $f_0$ is the corresponding equilibrium distribution, and $\Delta \mathcal{L}$, $\Delta f$ the perturbed evolution operator and perturbed phase space distribution, respectively.
A manifestation of the assumption that $\mathbf{f}^{\rm eff}$ varies smoothly with $\mathbf{J}$ is that since $\mathbf{f}^{\rm eff}$ depends on the distribution function $f$ we also assume that $f$ also varies smoothly with $\mathbf{J}$ such that we can safely discard $\Delta \mathcal{L}\Delta f$~\cite{Evans2008}.
Since the perturbation takes the form of a small velocity shift applied uniformly to all particles of species $j$, we can write the corresponding perturbed distribution explicitly as follows:
\begin{subequations}
\begin{align}
    &\frac{\partial \Delta f}{\partial t} + \mathcal{L}_0 \Delta f = -\Delta \mathcal{L}f_0 + O(\Delta^2), \\
    &\Delta \mathcal{L} = \sum_{\beta \in j}^{N_j} \mathbf{v}^{\rm applied} \cdot \frac{\partial}{\partial \mathbf{r}_\beta}.
    \end{align}
\end{subequations}
We arrive at the explicit solution for the perturbed distribution $\Delta f(t)$:
\begin{subequations}
\begin{align}
    \Delta f(t) &= -\int_0^t \exp\{-\mathcal{L}_0 t'\}\Delta \mathcal{L}f_0 dt', \\
    &= -\beta \int_0^t \exp\{-\mathcal{L}_0 t'\}f_0 \sum_{\beta}^{N_j} \mathbf{F}_\beta \cdot \mathbf{v}^{\rm applied}dt'.
\end{align}
\end{subequations}
We can calculate an expectation under this perturbation using this expression for the perturbed distribution:
\begin{subequations}
\begin{align}
    &\langle \mathbf{F}_i(t)\rangle_1 - \langle\mathbf{F}_i(t)\rangle_0 = \int_\gamma d\Gamma  \mathbf{F}_i(\Gamma) \Delta f(\Gamma, t),\\
    &= -\beta \int_\gamma d\Gamma \int_0^t \mathbf{F}_i \exp\{-\mathcal{L}_0 t'\}f_0 \sum_{\beta}^{N_j} \mathbf{F}_\beta \cdot \mathbf{v}^{\rm applied}dt' ,\label{eq:app_substitute_perturbed_dist}\\
    &= -\beta \int_0^t \left\langle \mathbf{F}_i(t')  \mathbf{F}_j(0)\right\rangle_0\cdot \mathbf{v}^{\rm applied} dt'\label{eq:app_adjoint_to_time_evolution}.
\end{align}
\end{subequations}
The step from Eq.~\eqref{eq:app_substitute_perturbed_dist} to Eq.~\eqref{eq:app_adjoint_to_time_evolution} uses the fact that the adjoint of the operator governing the evolution of the distribution is the same as the microscopic evolution operator for an observable. 
Finally, we arrive at the key linear response result:
\begin{align}
\label{eq:app_linear_response_r}
    \frac{\partial \mathbf{f}^{\rm int}_i}{\partial \mathbf{J}_j}\Bigg|_{\mathbf{J} = \mathbf{0}} = -\frac{\beta}{V\rho_i\rho_j} \int_0^\infty \left\langle \mathbf{F}_i(t')  \mathbf{F}_j(0)\right\rangle_0 dt', 
\end{align}
where the integral is evaluated in the long-time limit ($t \rightarrow \infty$) to ensure the coefficient correctly captures the appropriate long-time behavior, and in particular must be much longer than $\tau^{\rm NL}$. 
Equation~\eqref{eq:app_linear_response_r} is used in the main text to establish the symmetry of the resistance tensor $\boldsymbol{\mathcal{R}}$ for passive equilibrium systems.

\section{Simulation and Calculation Details}
\label{sec:simulation_detail}
We require that the underlying dynamics of the simulations recover an equilibrium distribution in the absence of external fluxes.
To ensure this, we adopt overdamped Langevin dynamics, which not only satisfy detailed balance in the flux-free limit but also provide a practical advantage: the presence of dissipative drag forces explicitly breaks Galilean invariance. 
This feature allows us to define and measure absolute species fluxes directly, making Langevin dynamics particularly suitable for calculating multicomponent transport coefficients.
The particle dynamics are taken to follow the overdamped Langevin equation:
\begin{equation}
\label{eq:langevin_eom}
    \dot{\mathbf{x}}^\alpha_{i} = \frac{1}{\zeta_i} \left[\sum_j^{n_c} \sum_{\beta\neq \alpha}^{N_j}\mathbf{F}_{ij}(\mathbf{x}^\alpha_i - \mathbf{x}^\beta_j;t) + \mathbf{F}_{s}^\alpha(t) \right],
\end{equation}
where $\dot{\mathbf{x}}^\alpha_{i}$ is the velocity of the $\alpha$th particle of species $i$, $\mathbf{F}_{s}^\alpha$ is a stochastic force with a mean of ${\langle\mathbf{F}_{s}^\alpha(t) \rangle = \mathbf{0}}$ and variance of ${\langle \mathbf{F}_{s}^\alpha(t)\mathbf{F}_{s}^\beta(t')\rangle = 2k_BT \zeta_i \delta^{\alpha\beta}\delta(t-t')\mathbf{I}_d}$, where $\delta (t-t')$ is the Dirac delta function.
We set the drag coefficients $\boldsymbol{\zeta}_i = \zeta_i\mathbf{I}_d$ to be isotropic and identical for all species, such that ${{\zeta}_i  = \zeta}$.
We consider particles interacting via a Lennard-Jones (LJ) potential with a cutoff at $2.5\sigma$, where $\sigma$ is the LJ diameter (identical for all species). 
We define the overall volume fraction of the system to be $\phi \equiv N\pi(2^{1/6}\sigma)^3/6V$ where $V$ is the system volume and we take $2^{1/6}\sigma$ to be the physically relevant particle diameter.
Here we introduce our system timescale as the self-diffusion time of noninteracting particles, $\tau^{\rm self} = \zeta \sigma^2/k_BT$.
All simulations were performed using the \texttt{HOOMD-Blue} simulation software~\cite{Anderson2020} and consisted of at least 49999 particles. 

\subsection{One Component Passive Systems}
\begin{figure}
    \centering
    \includegraphics[width=1.0\linewidth]{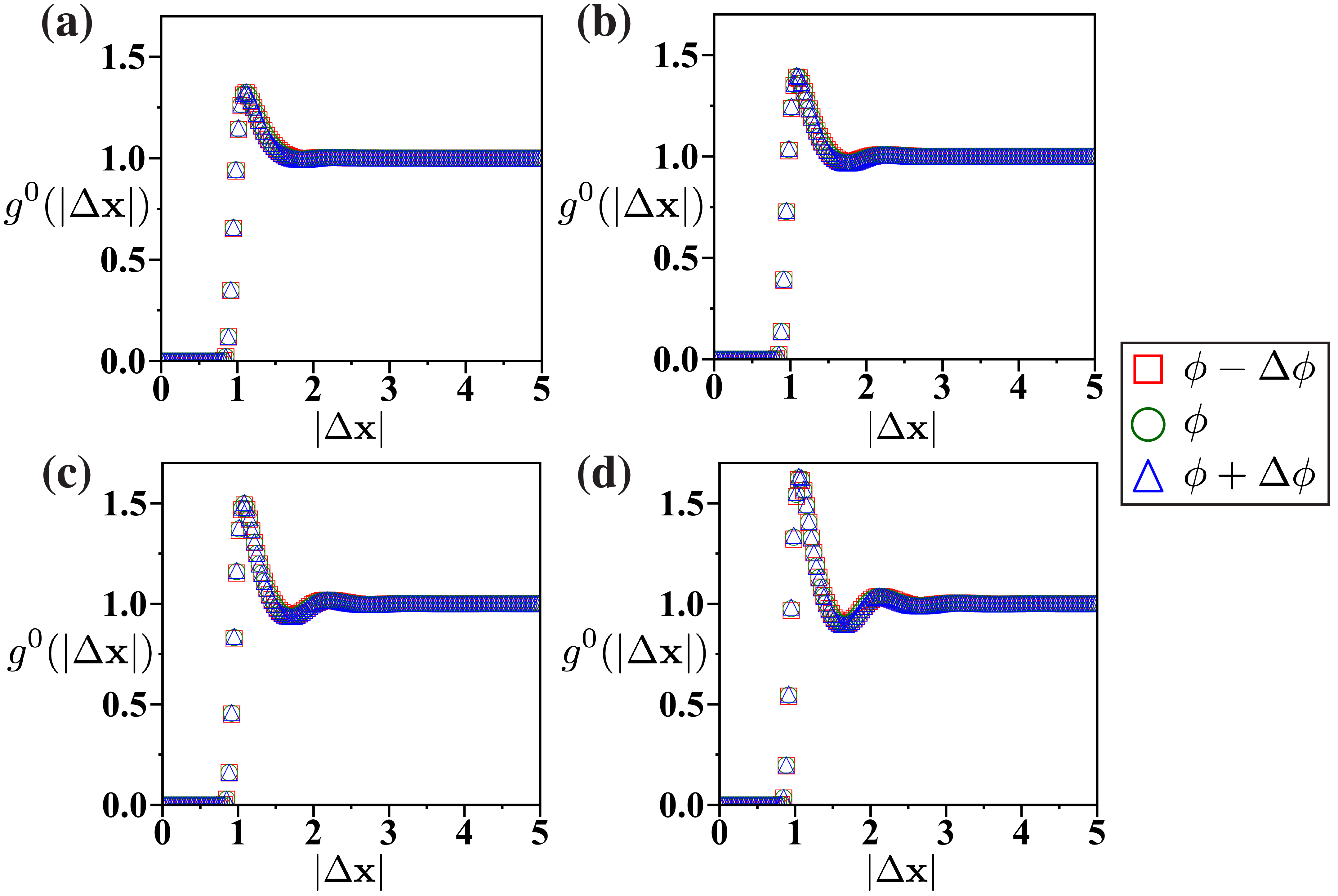}
    \caption{Homogeneous pair distribution function $g^0$ computed over the entire trajectories at (a) $\phi = 0.1$, (b) $\phi = 0.2$, (c) $\phi = 0.3$ and (d) $\phi = 0.4$, along with values at neighboring volume fractions $\phi \pm \Delta\phi$ where $\Delta\phi = 0.005$.
    Here $\Delta \mathbf{x}$ is measured in units of $\sigma$. 
    }
    \label{fig:g_allphi}
\end{figure}
The particle dynamics follow the overdamped Langevin equations as given in Eq.~\eqref{eq:langevin_eom}, but reduced to a single-component system (i.e.,~${n_c = 1}$) with $N$ total particles.
The LJ interaction energy is set to $\varepsilon/k_BT=0.25$.
For each simulation, for every $\phi$ considered, we discard the initial $250\tau^{\rm self}$ of the simulation time to allow the system to reach a steady state and subsequently run each simulation for a minimum duration of $5000\tau^{\rm self}$.
For statistical averaging, each trajectory is divided into segments of length $1000\tau^{\rm self}$, and all quantities described below are computed independently within each segment.

To compare the mechanical and Green-Kubo formulations of collective diffusion, we independently compute the force Jacobian and Onsager transport coefficient for a one-component system. 
From Eq.~\eqref{eq:l_one_component_passive}, the Onsager coefficient is known analytically within the mechanical framework, so only the force Jacobian $\mathcal{F}$ must be measured to determine $D$ from our mechanical theory. 
Below, we detail the numerical procedures used to compute each quantity.

\textit{Numerical Determination of $\mathcal{F}$.--} To compute the force Jacobian using the mechanical formalism [Eq.~\eqref{eq:force_jacobian_one_component_passive}], we calculate the steady-state pair distribution function $g^0(|\Delta \mathbf{x}|)$ at various volume fractions.
Due to the spatial isotropy of the system, $g^0$ depends only on the magnitude of the pair separation, $|\Delta \mathbf{x}|$, rather than its vector form.
The partial derivatives of $g^0$ with respect to density were then obtained using the finite difference method, based on simulations performed at neighboring densities with a spacing of $\Delta\phi = 0.005$ (corresponding to $\Delta\rho \approx 0.00675$).
\begin{figure}
    \centering
    \includegraphics[width=0.9\linewidth]{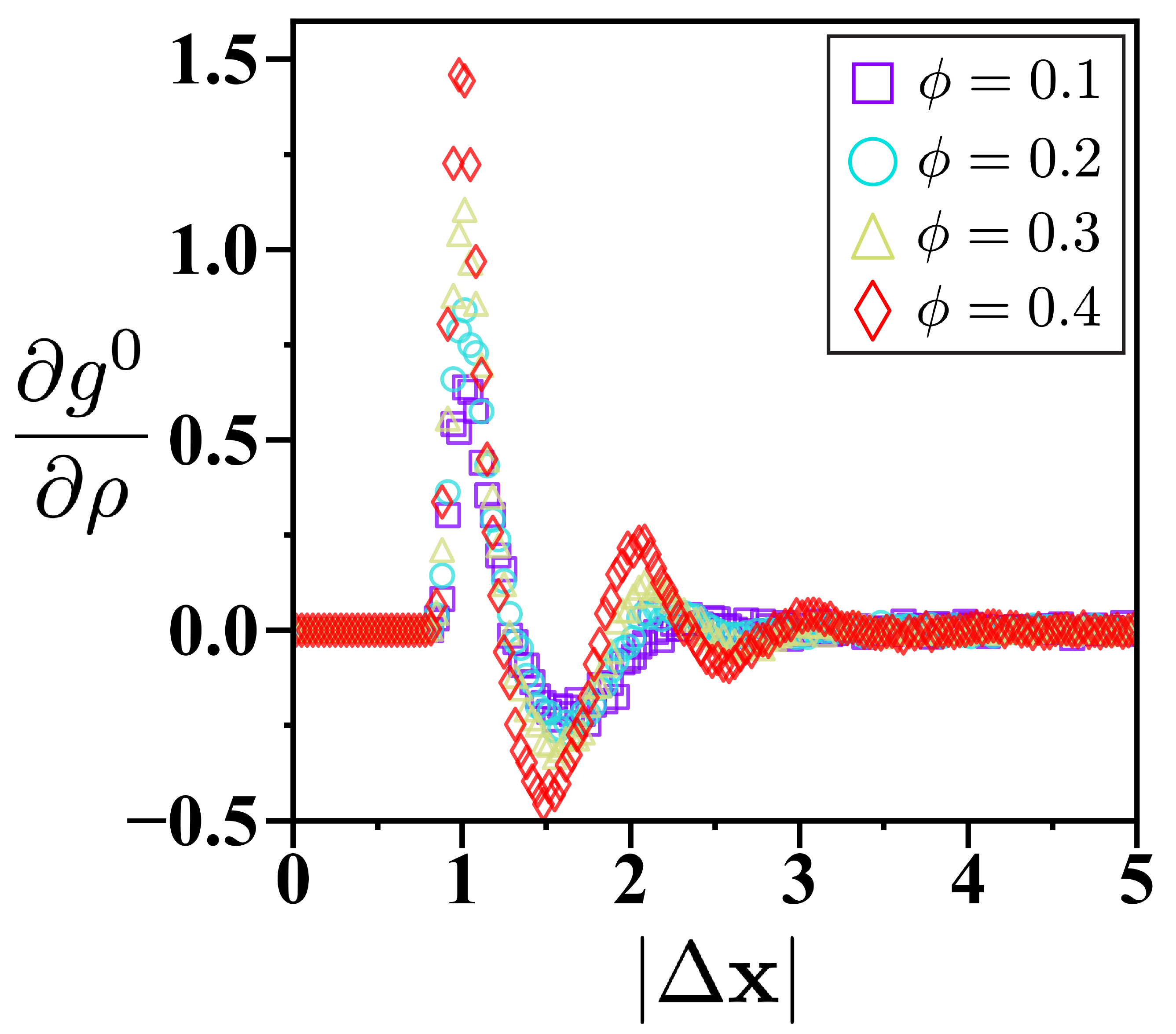}
    \caption{The partial derivatives of homogeneous pair distribution function $g^0$ with respect to density $\rho$ calculated using finite difference method.
    Here $\frac{\partial g^o}{\partial \rho}$ and $\Delta \mathbf{x}$ are measured in units of $\sigma^3$ and $\sigma$ respectively.
    }
    \label{fig:dg_drho}
\end{figure}
Figure~\ref{fig:g_allphi} presents the computed $g^0$ for all volume fractions $\phi$ of interest, including those at $\phi \pm \Delta\phi$.
The corresponding partial derivatives of $g^0$ with respect to density $\rho$ are shown in Fig.~\ref{fig:dg_drho}.

\textit{Numerical Determination of $\mathcal{F}^{GK}$.--}
The force Jacobian $\boldsymbol{\mathcal{F}}^{\rm GK}$ for one-component systems using the Green-Kubo formalism is obtained from the inverse of the large wavelength limit of the structure factor [see Eq.~\eqref{eq:F_GK}]:
\begin{equation}
    \boldsymbol{\mathcal{F}}^{\rm GK} = \frac{1}{\rho S} \mathbf{I}_d.
\end{equation}
For a one-component system, $S$ takes the form:
\begin{equation}
    S = \frac{1}{N} \lim_{\mathbf{k} \rightarrow \mathbf{0}} \langle \hat{\rho}(\mathbf{k},0) \hat{\rho}(-\mathbf{k},0)\rangle.
\end{equation}
The Fourier transform of the microscopic density is straightforwardly $\hat{\rho}(\mathbf{k}) = \sum_{\alpha = 1}^N \exp(- \mathrm{i}\mathbf{k} \cdot \mathbf{x}^\alpha)$, allowing us to express:
\begin{equation}
\label{eq:one_comp_structure_factor_micro}
    S = \lim_{\mathbf{k} \rightarrow\mathbf{0}} \frac{1}{N}  \left\langle \left| \sum_{\alpha = 1}^N \exp(- \mathrm{i}\mathbf{k} \cdot \mathbf{x}^\alpha) \right|^2\right\rangle.
\end{equation}
Due to the spatial isotropy of the system, we can consider the expression in Eq.~\eqref{eq:one_comp_structure_factor_micro} before taking the limit as a function of the scalar wavenumber $|\mathbf{k}|$ rather than the full vector $\mathbf{k}$.
\begin{figure}
    \centering
    \includegraphics[width=0.9\linewidth]{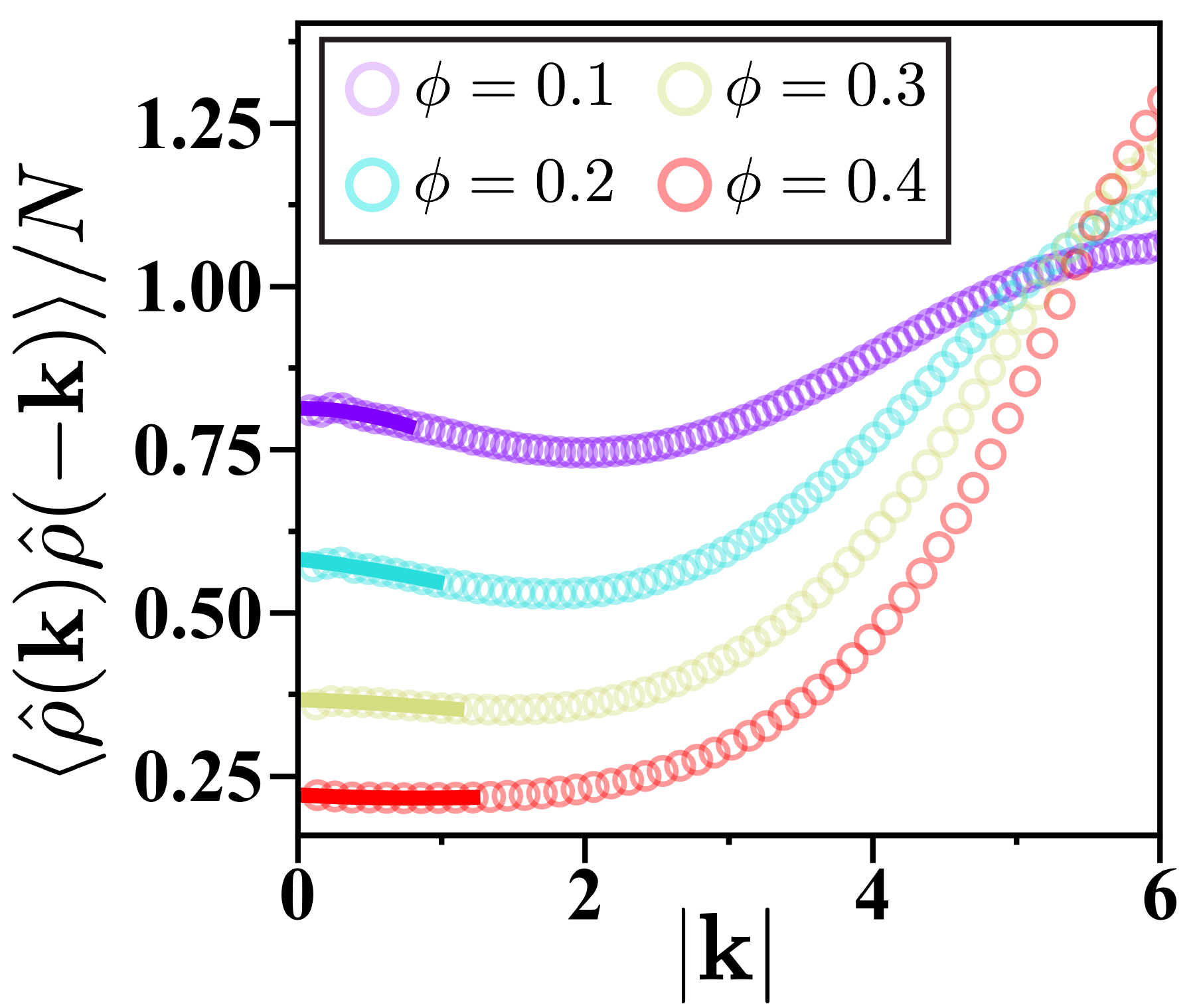}
    \caption{The equilibrium structure factor computed over the entire trajectories.
    Circle markers indicate the measured data, while the solid line denotes the fit used to find the large wavelength limit $S$ of the structure factor.
    Here $\mathbf{k}$ is measured in units of $\sigma^{-1}$.
    }
    \label{fig:structure_factor_fit}
\end{figure}
We estimate the structure factor $S$ by fitting the lowest 10 bins of $|\mathbf{k}|$ of the form to a second-degree polynomial~\cite{Cheng2022}, as illustrated in Fig.~\ref{fig:structure_factor_fit}.

\textit{Determination of $L^{\rm GK}$.--} The Onsager transport coefficient $L^{\rm GK}$ for one-component systems is computed using the Green-Kubo relation:
\begin{equation}
\label{eq:1compLGK}
    \mathbf{L}^{\rm GK} = \frac{1}{V} \lim_{\mathbf{k} \rightarrow \mathbf{0}} \int_0^t dt' \langle \hat{\mathbf{J}}(\mathbf{k}, t') \hat{\mathbf{J}}(-\mathbf{k}, 0) \rangle,
\end{equation}
which reduces from the general multicomponent form given in Eq.~\eqref{eq:L_GK}.
We omit the static flux-density correlation term $\mathbf{E}$ in this expression, as it vanishes for underdamped systems and is expected to be negligible in our case.
Exploiting the spatial isotropy and parity symmetry of the system and substituting the microscopic definition of the Fourier-transformed flux, we obtain the microscopic form of $L^{\rm GK}$:
\begin{multline}
    L^{\rm GK} = \frac{1}{dV}\times\\
    \lim_{|\mathbf{k}| \rightarrow\mathbf{0}}\int dt' \left\langle \sum_{\alpha = 1}^N \sum_{\beta = 1}^N \dot{\mathbf{x}}^\alpha (t')\cdot\dot{\mathbf{x}}^\beta (0) e^{- \mathrm{i}\mathbf{k} \cdot (\mathbf{x}^\alpha(t') - \mathbf{x}^\beta(0))} \right\rangle.
\end{multline}
We can understand the characteristic timescale of the integrand appearing in Eq.~\eqref{eq:1compLGK} through a simple analysis. 
If we set the wavenumber identically to zero, $\hat{\mathbf{J}}(\mathbf{0}, t)$ is related to the overall system velocity. 
The equation-of-motion for the system volume only contains contributions from the Langevin bath as all interaction forces sum to zero. 
It is straightforward to show that the flux correlations then take the form:
\begin{align}
    \langle\hat{\mathbf{J}}(\mathbf{0}, t')\hat{\mathbf{J}}(\mathbf{0},0)\rangle = 2\frac{k_BT}{\zeta} N\delta(t')\mathbf{I}_d.
\end{align}
Substitution of these correlations into Eq.~\eqref{eq:1compLGK} results in $L^{\rm GK} = \rho k_BT/\zeta$, precisely what we anticipate in equilibrium.

\subsection{Multicomponent Passive Systems}
\label{sec:simulation_detail_passive}
The particle dynamics follow the overdamped Langevin equations as given in Eq.~\eqref{eq:langevin_eom}.
The LJ interaction energies are set as follows: $\varepsilon_{AA}/k_BT = 0.5$, $\varepsilon_{BB}/k_BT = 0.6$, and $\varepsilon_{AB}/k_BT = 0.4$ for dissimilar particles. These values are chosen to prevent phase separation while ensuring that species $A$ and $B$ remain distinguishable.
The number ratio of type $A$ to type $B$ particles was fixed at 1:3 across all volume fractions considered.

\begin{figure}
    \centering
    \includegraphics[width=1.\linewidth]{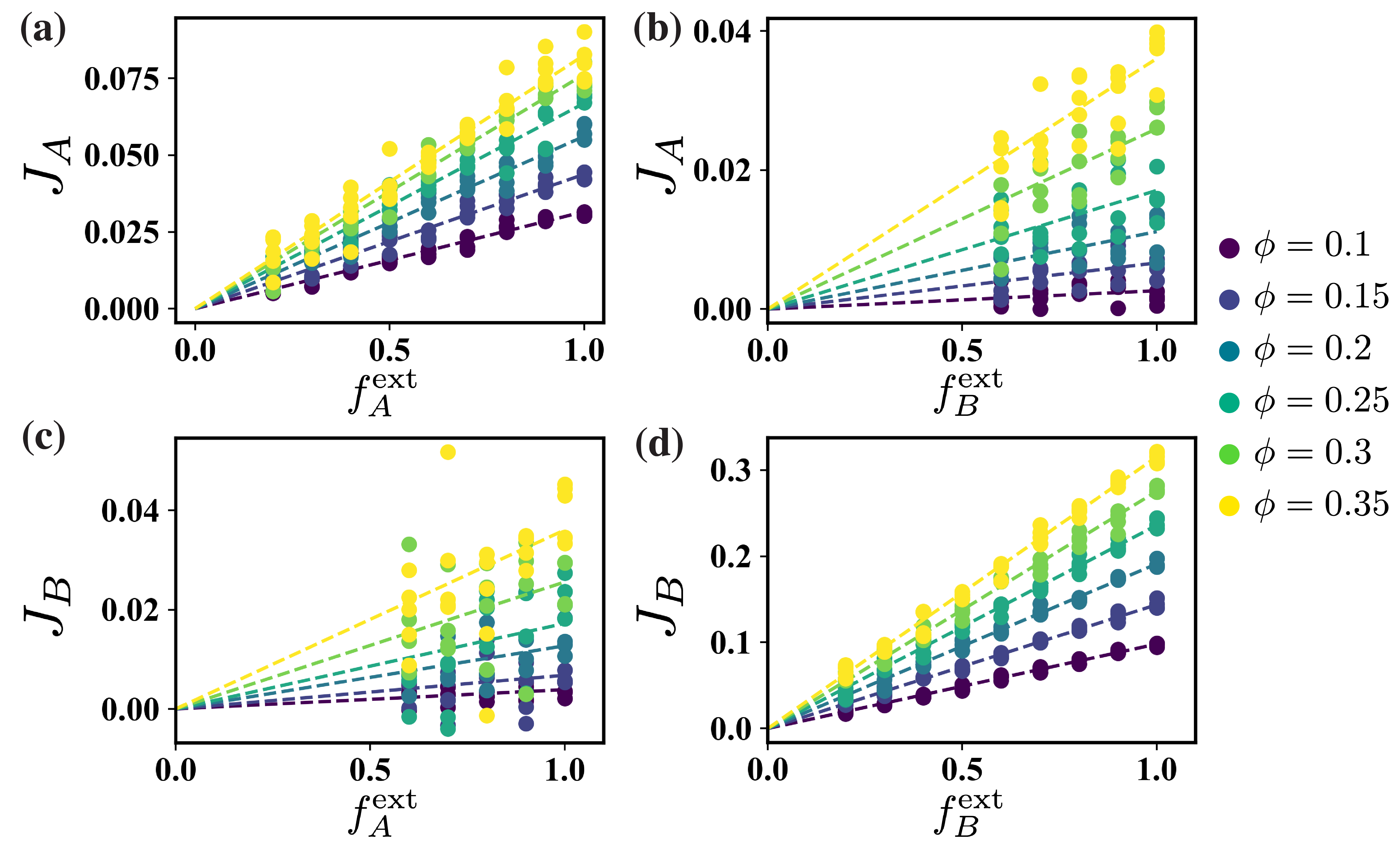}
    \caption{Species flux response to an applied external force. 
    (a) and (b) show the flux of species $A$ under external forces applied to $A$ and $B$, respectively, while (c) and (d) show the flux of species $B$ under the same external force conditions. 
    Different colors indicate the volume fractions of the system considered, and dashed lines represent linear fits.
    For every $\phi$ and $f^{\rm ext}$ considered, the plot includes multiple data points, each representing the species flux measured from individual simulation segments. 
    Here $J_i$ and $f_i^{\rm ext}$ are measured in units of $(\tau^{\rm self}\sigma^2)^{-1}$ and $k_BT\sigma^{-1}$ respectively.
    }
    \label{fig:flux_force_response}
\end{figure}

To measure the Onsager transport tensor using color field approach, we applied a species-specific external force, $\mathbf{f}^{\rm ext}_i$, to species $A$ and $B$ separately and measured the resulting flux.
The spatial isotropy and parity symmetry of the system allows us to express $\mathbf{L}_{ij} = L_{ij}\mathbf{I}_d$.
The external force was applied along a single axis, and the corresponding flux response was measured along the same axis.
To capture the full Onsager transport tensor in a single simulation, we applied a constant external force to species $A$ along the $x$-direction and to species $B$ along the $y$-direction, then measured the resulting fluxes along each respective axis.
This protocol relies on the assumption of spatial isotropy and parity symmetry—that is, the system’s response along one axis is independent of the force applied in the orthogonal direction, allowing us to decouple the flux responses and extract the full transport tensor from orthogonal force components.
With a system containing $d$ or fewer species, spatial isotropy and parity symmetry allows us to determine all components of the transport tensor $L_{ij}$ within a single simulation by applying orthogonal forces and decoupling the resulting flux responses.
For all $\phi$ reported, $f_j$ was varied from 0.1 to 1.0 (in units of $k_BT\sigma^{-1}$) for each species, and the corresponding species flux $J_i (\forall i \in \{A, B\})$ was recorded.
Species flux is measured as $J_i = \frac{1}{V}\sum_\alpha^{N_i} \dot{x}_i^\alpha$.
The Onsager transport tensor coefficients were extracted by determining the slope of $J_i$ as a function of $f_j$.
To ensure statistical significance when measuring the flux response of one species to an external force applied to another (e.g., $J_A$  in response to $f^{\rm ext}_B$), we only included data where $f^{\rm ext}_i \geq 0.5$.
In each simulation, for every $\phi$ and $f^{\rm ext}$ considered, we discard the initial $250\tau^{\rm self}$ of simulation time to allow the system to reach a steady state and subsequently run each simulation for a minimum duration of $8000\tau^{\rm self}$.
Each trajectory was segmented into intervals of $2000\tau^{\rm self}$, and $L_{ij}$ was obtained for each segment. 
The statistical average and standard deviation of $L_{ij}$ were then computed across all segments.
The measured species flux response to an applied species external force, obtained from simulations, is presented in Fig.~\ref{fig:flux_force_response}.

We also use the Green-Kubo relation to measure the Onsager transport coefficient.
We use the displacement form of the Green-Kubo relation, which uses the particle position instead of velocity~\cite{Wheeler2004_1,Evans2008,Allen2017, Fong2020} (which comes from setting $\mathbf{E}_{ij} = \mathbf{0}$ and $\mathbf{k} = \mathbf{0}$ in Eq.~\eqref{eq:L_GK}):
\begin{multline}
\label{eq:GK_msd}
    L_{ij}^{\rm GK} = \frac{V\rho_i\rho_j}{6} \times\\
    \lim_{t \rightarrow \infty} \frac{d}{dt} \left\langle  \sum_{\alpha}^{N_i} \sum_{\beta}^{N_j} [\mathbf{x}_i^\alpha(t) - \mathbf{x}_i^\alpha(0)]\cdot [\mathbf{x}_j^\beta(t) - \mathbf{x}_j^\beta(0)]
    \right\rangle.
\end{multline}
The presence of the dissipative Langevin force explicitly breaks Galilean invariance, allowing us to describe the absolute flux of each species directly.
This broken invariance ensures that the Onsager transport tensor $\mathbf{L}$ is full rank and enables a well-defined linear relationship between the absolute flux and the applied force. 
We choose the smallest practical time $t$ while ensuring it remains larger than the velocity correlation time, as the relative error in the mean square displacement increases with the number of steps~\cite{Tripathi2024}.
For each $\phi$ considered, we conducted simulations with a minimum duration of $15000\tau^{\rm self}$. Each trajectory was divided into segments of $3000\tau^{\rm self}$, and $L_{ij}$ was measured for each segment to compute the statistical average and standard deviation. 

\subsection{Multicomponent Nonreciprocal Systems}
\label{sec:simulation_detail_nonreciprocal}
The particle dynamics follow the overdamped Langevin equations as given in Eq.~\eqref{eq:langevin_eom}, but with interparticle interaction forces modified according to Eq.~\eqref{eq:modelforces}.
The conservative force is derived from the Lennard-Jones potential, with interaction energies identical to those used in the passive case, as detailed in Appendix~\ref{sec:simulation_detail_passive}.
The interaction nonreciprocity in our simulations is adjusted with a single scalar parameter, $\Delta$.
We set the reciprocity diameter $d_{\rm rec} = 2^{1/2}\sigma$ such that all particle pairs experience reciprocal \textit{repulsion} within separation distances of $d_{\rm rec}$, and generally experience nonreciprocal attraction for interparticle separations greater than $d_{\rm rec}$ for interspecies pairs.

To measure the Onsager transport tensor in nonreciprocal systems using color field theory, we applied a species-specific external force, $\mathbf{f}^{\rm ext}_i$, to species $A$ and $B$ separately and measured the resulting flux. 
We again assume spatial parity in our simulation, where the system’s response along one axis is unaffected by forces applied in the orthogonal direction.
As a result, and just as in the passive case, we then expect ${\mathbf{L}_{ij} = L_{ij}\mathbf{I}_d}$ for our isotropic system. 
The absence of spatially odd mobility allows us to extract the full Onsager transport tensor from a single simulation by applying a constant external force to species $A$ along the a direction and an external force to species $B$ in an orthogonal direction, and measuring the resulting fluxes in each direction.
For all reported $\Delta$, the external force was varied from 0.1 to 1.0 in units of $k_BT\sigma^{-1}$, and the corresponding species flux $J_i (\forall i \in \{A, B\})$ was recorded.
The measurement procedure for species flux and Onsager transport tensor coefficients follows the same method as in the passive case (detailed in Appendix~\ref{sec:simulation_detail_passive}).

\section{Trajectory Ensembles for Nonequilibrium Mechanical Transport}
\label{app:onsager_machlup}
In the main text, we make arguments for why nonreciprocal forces lead to an asymmetry in $\mathbf{L}$, but the analysis is limited to dilute systems, and we rely on our expectations for the influence of a flux on the pair distribution function to determine the sign of the asymmetry.
In order to push this understanding further, we will use ideas of trajectory ensembles for Langevin systems.
We will make use of the expression for $\mathbf{L}$ in the absence of other static forces used in Sec.~\ref{sec:color_field} and Sec.~\ref{sec:nonlinear_transport}:
\begin{align}
    \mathbf{L} = \frac{\partial \langle \mathbf{J}\rangle}{\partial \mathbf{f}^{\rm ext}}.
\end{align}
Here, we take the expectation over a translationally-invariant steady state of interest. To see why we might be able to make use of trajectory ensembles, we first make the observation that at steady state for translationally invariant ergodic systems, we have:
\begin{align}
    &\mathbf\langle\mathbf{J}\rangle = \lim_{t\rightarrow\infty}\boldsymbol{\mathcal{J}}(t), \\
    &\boldsymbol{\mathcal{J}}_i(t) = \frac{1}{tV}\int_0^t \sum_\alpha^{N_i}\mathbf{v}_i^\alpha(t')dt',
\end{align}
for total trajectory flux $\boldsymbol{\mathcal{J}}$.
If we understand long-time trajectory behavior, this is equivalent to sampling the ensemble of interest. 
For deterministic systems, the only source of ``randomness'' is through the initial condition, and while we might expect a system to be ergodic such that we can obtain steady state observables from simulation, in order to understand the system analytically we need access to the steady state distribution.
However, for stochastic systems we can divide the probability distribution into the initial condition distribution and a conditional distribution over trajectories.
We can break up the probability of observing a given trajectory through phase space:
\begin{align}
    P[\mathbf{X}] = f^{\rm ss}(\mathbf{X}(0))P[\mathbf{X} | \mathbf{X}(0)],
\end{align}
where $\mathbf{X}(t)$ is the system configuration at time $t$ ($\mathbf{X}: \mathbb{R} \rightarrow \gamma$), and $f^{\rm ss}$ is the steady-state distribution of interest.
A particularly useful framework is the case of Gaussian noise in Langevin systems, as $P[\mathbf{X}|\mathbf{X}(0)]$ provides a formally well-defined non-zero weight on continuous trajectories through phase space.
By using the formalism of the Onsager-Machlup action, we can then understand a nonequilibrium steady state by sampling modified observables from equilibrium dynamics~\cite{Cugliandolo2019, Maes2020, Lesnicki2021, Poggioli2023}.
This opens up a perturbative perspective on the nonequilibrium dynamics with no explicit reference to the nonequilibrium steady state.
In the long-time limit, many stochastic systems of interest become independent of their initial condition, and so these trajectory ensembles truly encode the steady-state distribution.

We can construct the explicit form of the trajectory probability distribution for a nonequilibrium system of interacting particles by considering the dynamics of particle $\alpha$ of species $i$ given by the Langevin equation:
\begin{subequations}
\begin{align}
&m_i\dot{\mathbf{v}}_i^\alpha = -\zeta_i \mathbf{v}_i^\alpha + \mathbf{F}_{i}^{\alpha\rm E} + \epsilon \mathbf{F}_{i}^{\alpha\rm NE} + \mathbf{f}^{\rm ext}_{i} + \boldsymbol{\xi}_i^\alpha, \\
    &\langle \boldsymbol{\xi}_i^\alpha\rangle = \mathbf{0},\\
    &\langle \boldsymbol{\xi}_i^\alpha(t)\boldsymbol{\xi}^\beta_j(t')\rangle = 2k_BT\zeta_i\delta_{ij}\delta_{\alpha\beta}\delta(t-t')\mathbf{I}_d,
\end{align}
\end{subequations}
where we differentiate between forces that can be derived from a potential $\mathbf{F}_i^{\alpha\rm E}$ and nonequilibrium forces $\mathbf{F}_i^{\alpha\rm NE}$ whose strength is controlled by the parameter $\epsilon$, and $\boldsymbol{\xi}_i^\alpha$ is delta-correlated Gaussian noise. 
Using the Onsager-Machlup formalism, we can express the probability of observing a trajectory $\mathbf{X}_{t_n}$ up to time $t_n$ under the given dynamics as~\cite{Cugliandolo2019}:
\begin{subequations}
\begin{align}
    &P[\mathbf{X}_{t_n} | \mathbf{X}(0)] \propto\exp\left(-\beta \sum_{i=1}^{n_c} \sum_{\alpha =1}^{N_i}\int_0^{t_n}\frac{S_i^\alpha(t)}{4\zeta_i}dt\right),\\
    &S_i^\alpha(t) = \left(\boldsymbol{\eta}_i^{\alpha}(t) - \epsilon \mathbf{F}_{i}^{\alpha\rm NE} - \mathbf{f}^{\rm ext}_{i}\right)^2,\\
    &\boldsymbol{\eta}_i^{\alpha} = m_i\dot{\mathbf{v}}_i^\alpha + \zeta_i \mathbf{v}_i^\alpha - \mathbf{F}_{i}^{\alpha\rm E},
\end{align}
\end{subequations}
where $S^\alpha_i(t)$ is the trajectory action for particle $\alpha$ of species $i$, and $\boldsymbol{\eta}^{\alpha}_i$ corresponds to an equilibrium noise, or the noise when $\epsilon = 0$ and $\mathbf{f}^{\rm ext}=\mathbf{0}$.
Often, to identify the particular influence of different nonequilibrium processes in system response, $\boldsymbol{\eta}_i^{\alpha}$ is broken up into a frenesy contribution and a flux contribution~\cite{Baiesi2009, Lesnicki2021}.
However, for the purposes of this work, we retain $\boldsymbol{\eta}_i^{\alpha}$ as a single quantity representing the ``random kick'' imparted by the bath, as this perspective provides clearer physical intuition for our analysis.
We observe that this formulation allows us to express $P[\mathbf{X}]$ in terms of an expectation under an equilibrium trajectory distribution. Specifically, we obtain:
\begin{subequations}
\label{eq:relative_path_probabilities}
\begin{align}
    &\frac{P[\mathbf{X}_{t_n} | \mathbf{X}(0)]}{P^{\rm eq}[\mathbf{X}_{t_n} | \mathbf{X}(0)]} = \exp(\beta U'), \\
    &U'(\mathbf{X}_{t_n}) = U^{\rm ext}(\mathbf{X}_{t_n}) + U^{\rm NE}(\mathbf{X}_{t_n}) + U^{\rm cross}(\mathbf{X}_{t_n}),\\
    &U^{\rm ext}(\mathbf{X}_{t_n}) = \sum_{i}^{n_c}\sum_{\alpha}^{N_i}\int_0^{t_n} \frac{\mathbf{f}^{\rm ext}_{i}}{2\zeta_i} \cdot \left(\boldsymbol{\eta}_i^{\alpha} - \frac{1}{2}\mathbf{f}^{\rm ext}_{i}\right)dt, \\
    &U^{\rm NE}(\mathbf{X}_{t_n}) = \sum_{i}^{n_c}\sum_{\alpha }^{N_i}\int_0^{t_n} \frac{\epsilon\mathbf{F}^{\alpha\rm NE}_{i}}{2\zeta_i} \cdot \left(\boldsymbol{\eta}_i^{\alpha} - \frac{\epsilon}{2}\mathbf{F}^{\alpha\rm NE}_{i}\right)dt, \\
    &U^{\rm cross}(\mathbf{X}_{t_n}) = -\sum_{i}^{n_c}\sum_{\alpha}^{N_i}\int_0^{t_n} \frac{\epsilon \mathbf{F}^{\alpha\rm NE}_{i}\cdot\mathbf{f}^{\rm ext}_{i}}{2\zeta_{i}} dt, \\
    &P^{\rm eq}[\mathbf{X}_{t_n}|\mathbf{X}(0)] \propto \exp\left(-\beta \sum_{i}^{n_c}\sum_\alpha^{n_c}\int_0^{t_n}\frac{\|\boldsymbol{\eta}_i^{\alpha}\|^2}{4\zeta_i} dt \right),
\end{align}
\end{subequations}
where the equilibrium distribution $P^{\rm eq}$ weights trajectories based on dynamics in the absence of $\mathbf{F}^{\alpha \rm NE}$ and $\mathbf{f}^{\rm ext}$. 
We can now express an expectation of $\mathbf{J}$ under the steady-state dynamics:
\begin{subequations}
\begin{align}
    \langle\mathbf{J}\rangle &= \int \mathcal{D}\mathbf{X} f^{\rm ss}(\mathbf{X}(0))P^{\rm eq}[\mathbf{X}|\mathbf{X}(0)]\exp(\beta U')\boldsymbol{\mathcal{J}} \\
    &= \langle\exp(\beta U')\boldsymbol{\mathcal{J}} \rangle_{\rm eq},
\end{align}
\end{subequations}
where $\int \mathcal{D}\mathbf{X}$ is a path-integral over continuous trajectories in phase space with measure dependent on the discretization scheme~\cite{Cugliandolo2019}, we suppress the time-dependence of $\boldsymbol{\mathcal{J}}$, and the second equality comes in the long-time limit as the observable becomes independent of the initial condition for equilibrium dynamics.

Now, we aim to compute $\mathbf{L}$ in an ensemble where $\mathbf{f}^{\rm ext} = \mathbf{0}$ and analyze the impact of a small, nonzero $\epsilon$ on the symmetry of $\mathbf{L}$.
We can then expand $\mathbf{L}$ for a small nonzero $\epsilon$ as:
\begin{align}
\label{eq:L_expansion}
    \mathbf{L}(\epsilon) = \left.\frac{\partial \langle \mathbf{J}\rangle}{\partial \mathbf{f}^{\rm ext}}\right|_{\mathbf{f}^{\rm ext}=\mathbf{0}, \epsilon=0} + \left.\frac{\partial^2 \langle \mathbf{J}\rangle}{\partial\epsilon\partial \mathbf{f}^{\rm ext}}\right|_{\mathbf{f}^{\rm ext}=\mathbf{0}, \epsilon=0}\epsilon + \mathcal{O}(\epsilon^2).
\end{align} 
With Eq.~\eqref{eq:relative_path_probabilities}, we can express $\mathbf{L}$ in terms of equilibrium ensemble averages of trajectory observables.
We are now ready to evaluate the first order nonequilibrium correction in Eq.~\eqref{eq:L_expansion}:
\begin{subequations}
\begin{align}
    &\Delta\mathbf{L}_{ij} \equiv \mathbf{L}_{ij} - \mathbf{L}_{ij}^{\rm eq}\approx \epsilon\left.\frac{\partial^2 \langle \boldsymbol{\mathcal{J}}_i\exp(\beta U')\rangle_{\rm eq}}{\partial\epsilon\partial \mathbf{f}^{\rm ext}_j}\right|_{\mathbf{f}^{\rm ext}=\mathbf{0}, \epsilon=0}, \\
    &\mathbf{L}^{\rm eq}_{ij} = \left.\frac{\partial \langle \mathbf{J}\rangle}{\partial \mathbf{f}_j^{\rm ext}}\right|_{\mathbf{f}^{\rm ext}=\mathbf{0}, \epsilon=0},
\end{align}
\end{subequations}
where $\Delta\mathbf{L}_{ij}$ is the correction of interest. Evaluating these derivatives, we find the following expression for the correction:
\begin{widetext}
\begin{subequations}
\begin{align}
&\Delta\mathbf{L}_{ij} = \frac{\beta^2}{4\zeta_j}\left\langle\boldsymbol{\mathcal{J}}_i\int_0^{t_n}dt\int_0^{t_n}dt'\sum_{k=1}^{ n_c}\sum_{\beta=1}^{N_j}\sum_{\gamma=1}^{N_k}\right. \left.\frac{\mathbf{F}^{\gamma \rm NE}_k(t)}{\zeta_k}\cdot 
\hat{\boldsymbol{\Sigma}}_{kj}^{\gamma\beta}(t,t')\right\rangle_{\rm eq},\\
&\hat{\boldsymbol{\Sigma}}_{kj}^{\gamma\beta}(t,t') =\boldsymbol{\eta}_k^\gamma(t)\boldsymbol{\eta}_j^\beta(t') - \left\langle\boldsymbol{\eta}_k^\gamma(t)\boldsymbol{\eta}_j^\beta(t')\right\rangle.
\end{align}
\end{subequations}
\end{widetext}
We see then that modifications to $\mathbf{L}$ due to the introduction of a nonequilibrium force arise from changes to the correlations between the flux of species $i$ and the random force on species $j$ when fluctuations on other particles are aligned with the direction of the applied nonequilibrium force.

We can make sense of this for the addition of pairwise nonreciprocal forces as $\mathbf{F}^{\rm NE}$ for a two-component mixture of otherwise identical particles. 
We are interested then in identifying:
\begin{align}
    \mathbf{L}_{AB} - \mathbf{L}_{BA} = \Delta\mathbf{L}_{AB} - \Delta\mathbf{L}_{BA}.
\end{align}
We take advantage of the symmetry of our system to note that under the equilibrium distribution, we can swap the labels of species A and B particles and obtain an equally probable system configuration.
From this, we can observe that if the perturbative force arise from a pair potential such that $\mathbf{F}^{\rm NE}$ is a reciprocal pairwise force, swapping the labels leaves the forces on each particle invariant, implying ${\Delta\mathbf{L}_{AB} = \Delta\mathbf{L}_{BA}}$ and the symmetry of $\mathbf{L}$ is preserved.
However, if $\mathbf{F}^{\rm NE}$ is a nonreciprocal pairwise force, swapping particle labels reverses the direction of the total nonequilibrium force on each particle.
In this case, ${\Delta\mathbf{L}_{AB} = -\Delta\mathbf{L}_{BA}}$, and since this correlation generally does not vanish, we find that $\mathbf{L}$ no longer satisfies the reciprocal relations, with the antisymmetric component given by ${\mathbf{L}_{AB} - \mathbf{L}_{BA} = 2\Delta\mathbf{L}_{AB}}$.

We can make qualitative arguments about the sign of $\Delta\mathbf{L}_{AB}$ based on its definition.
The term ${\mathbf{F}^{\rm NE} \cdot \boldsymbol{\eta}}$ captures the extent to which stochastic forces act in the same direction as the nonequilibrium force. 
We can understand the sign of $\Delta\mathbf{L}_{ij}$ by considering whether the flux of species $i$ is more or less correlated with an applied force on species $j$ compared to the case with no nonreciprocal force.
Suppose the nonreciprocal force is such that species $A$ is chased by species $B$ [e.g., see Fig~\ref{fig:nonreciprocal_L}(a)]. 
In that case, we anticipate that a random force on species $B$ acting in the direction of the nonreciprocal force will ``enhance'' the nonreciprocal interaction by decreasing the interparticle separation, leading to a larger nonreciprocal force (and therefore flux) on species $A$, and as result  $\Delta\mathbf{L}_{AB}>0$.
However, a random force on species $A$ will lead to a greater average distance between species $A$ and species $B$, a weaker nonreciprocal interaction, and therefore less flux, so $\Delta\mathbf{L}_{BA}<0$.

\section{Extension of the Mechanical Framework to Thermal Transport}
\label{sec:heat_flux}
Thus far, we have explicitly demonstrated how our mechanical transport framework describes density fluxes in response to general driving forces. 
To demonstrate how we can extend our framework to describe transport phenomena beyond particle flux, we now consider the case of thermal conductivity. 
This requires two key considerations: first, the dynamics of the temperature field are governed in part by the heat flux; second, temperature gradients can drive not only heat fluxes but also particle fluxes, a phenomenon known as the Soret effect~\cite{DeGroot2013}.
Conversely, just as temperature gradients can influence particle motion, density gradients can, through the reciprocal Dufour effect, induce heat fluxes~\cite{DeGroot2013}.
Accurately capturing these coupled transport processes requires determining the Onsager transport tensor that also captures the cross-coupling between mass and energy transport.

We begin by assuming that we have an expression for heat flux and its evolution as a function of the microscopic dynamics, as one would recover from an Irving-Kirkwood procedure like the one performed in Appendix~\ref{sec:appendix_species_momentum_balance}. 
We anticipate that the evolution equation of the heat flux equation will have the form:
\begin{subequations}
\begin{align}
    &m_q\frac{\partial \bar{\mathbf{J}}_q}{\partial t} +m_q \boldsymbol{\nabla} \cdot (\bar{\mathbf{J}}_q \mathbf{J}^{\rm ref}/\rho^{\rm ref}) = \rho_q\bar{\mathbf{f}}_q^{\rm eff},\\
    &m_q = \sum_i^{n_c} m_i, \\
    &\rho_q = \sum_i^{n_c} \rho_i,
\end{align}
\end{subequations}
where $\bar{\mathbf{J}}_q$ is the heat flux, $\bar{\mathbf{f}}^{\rm eff}_q$ is an effective force governing the time dependence of $\bar{\mathbf{J}}_q$, and $\mathbf{J}^{\rm ref}/\rho^{\rm ref}$ is a reference velocity with respect to which $\bar{\mathbf{f}}^{\rm eff}_q$ is calculated.
We also define $m_q$ as the total species mass and $\rho_q$ as the total species mass density, such that $\bar{\mathbf{J}}_q$ has the same structural form as the species fluxes [see Eq.~\eqref{eq:species_momentum_balance}]. 
With these definitions and the exact form of the heat flux dynamics (e.g., from an Irving-Kirkwood procedure) $\mathbf{f}_q^{\rm eff}$ will have a well-defined microscopic expression that depends on the microscopic equations of motion and the precise definition of heat flux.

Before proceeding, we normalize the heat flux and effective force by the local thermal energy $k_B T$, where $T(\mathbf{x})$ denotes the spatially varying temperature field, such that the resulting heat flux and force have the same units as the species fluxes and forces:
\begin{subequations}
\begin{align}
    \mathbf{J}_q &= \bar{\mathbf{J}}_q/ k_BT, \\
    \mathbf{f}_q^{\rm eff} &= \bar{\mathbf{f}}_q^{\rm eff} / k_BT.
\end{align}
\end{subequations}
This will allow us the define the Onsager tensor $\mathbf{L}$ such that all elements have the same units. 
In the regime of small gradients in heat and particle fluxes, we may again expand the effective force driving heat transport.
Now, we are interested in the linear dynamics of the heat flux in addition to the species fluxes. 
We therefore expand $\mathbf{f}_q^{\rm eff}$ with respect to all of these fluxes:
\begin{align}
    \mathbf{f}_q^{\rm eff} = \mathbf{f}_q^{\rm static}+\left.\frac{\partial \mathbf{f}_q^{\rm eff}}{\partial \mathbf{J}} \right|_{\mathbf{J}=\mathbf{0}}\cdot\mathbf{J},
\end{align}
where $\mathbf{J}$ is a vector of all species fluxes $\mathbf{J}_i$ \emph{and the heat flux} $\mathbf{J}_q$.
We observe then that we can express exactly the same vector equation as for the species alone, including heat flux terms:
\begin{align}
m_i\frac{\partial \mathbf{J}_i}{\partial t} = \rho_i\left(\mathbf{f}_i^{\rm static} + \sum_j^{n_c}\frac{\partial \mathbf{f}_i^{\rm eff}}{\partial \mathbf{J}_j}\cdot\mathbf{J}_j\right) + \mathcal{O}(\|\mathbf{J}\|^2 + \|\nabla\cdot\mathbf{JJ}\|),
\end{align}
where $i$ now indexes over all $n_c$ species and $q$, and $\mathbf{f}^{\rm eff}$ is now a vector of effective forces of species \emph{and includes the effective force associated with heat flux} $\mathbf{f}^{\rm eff}_q$.
This leads to a set of $(n_c+1)$ flux evolution equations —one for each of the  $n_c$ particle species and one for the heat flux:
\begin{subequations}
\begin{align}
    &\frac{\partial \mathbf{J}}{\partial t} = \mathbf{B}\cdot\mathbf{f}^{\rm static} - \mathbf{A}\cdot\mathbf{J},\\   
    & \mathbf{A}_{ij} =  -\frac{\rho_i}{m_i}\left.\frac{\partial\mathbf{f}^{\rm eff}_i}{\partial \mathbf{J}_j}\right|_{\mathbf{J}=\mathbf{0}},\\
    & \mathbf{B}_{ij} = \frac{\rho_i}{m_i}\delta_{ij}\mathbf{I}_d.
\end{align}
\end{subequations}
As in the case of particle species, we find that an overdamped limit exists if the tensor $\mathbf{A}$ has only positive eigenvalues. 
In this limit, we can again recover the linear transport relation derived in Sec.~\ref{sec:mechanics_of_multicomponent_transport}, with
the generalized resistance tensor is again recovered as ${\boldsymbol{\mathcal{R}} =-\frac{\partial \mathbf{f}^{\rm eff}}{\partial \mathbf{J}}}$.

For systems with a non-zero heat flux, we expect there could be gradients in temperature that drive heat flux or that there may exist cross-effects which couple temperature gradients to species fluxes.
To further investigate transport driven by small temperature gradients, we may simply expand $\mathbf{f}^{\rm static}$ in terms of temperature gradients $\boldsymbol{\nabla} T$:
\begin{subequations}
\begin{align}
    &\mathbf{f}^{\rm static} = -\boldsymbol{\mathcal{F}}^{q}\cdot \boldsymbol{\nabla}T + \mathcal{O}\big((\boldsymbol{\nabla} T)^2\big), \\
    &\boldsymbol{\mathcal{F}}^q = -\left.\frac{\partial\mathbf{f}^{\rm static}_k}{\partial\boldsymbol{\nabla}T}\right|_{\boldsymbol{\nabla}T=\mathbf{0}}.
\end{align}
\end{subequations}
By substituting the static force expansion into the linear transport relation in Eq.~\eqref{eq:flux_beff}, we find:
\begin{subequations}
\begin{align}
&\mathbf{J} = -\mathbf{L}\cdot\boldsymbol{\mathcal{F}}^{q}\cdot\boldsymbol{\nabla} T, \\
&\mathbf{L} = \boldsymbol{\mathcal{R}}^{-1}.
\end{align}
\end{subequations}

We can now define the thermal conductivity, $\boldsymbol{\kappa}$, using our mechanical expressions:
\begin{subequations}
\label{eq:thermal_conductivity}
\begin{align}
    &\mathbf{J}_q = -\boldsymbol{\kappa}\cdot \boldsymbol{\nabla} T ,\\
    &\boldsymbol{\kappa} = \sum_j^{n_c+1} \mathbf{L}_{qj}\cdot\boldsymbol{\mathcal{F}}^q_j,
\end{align}
\end{subequations}
where we see that this framework allows for the case that temperature gradients generate effective forces that may drive heat and species flux.
The thermal conductivity defined in Eq.~\eqref{eq:thermal_conductivity} therefore includes contributions not only from direct energy transport but also from the cross-coupling effect with species motion.

\begin{acknowledgments}
We thank Daniel Evans for insightful discussions on the structure of the mutual diffusion tensor and acknowledge discussions with Kranthi Mandadapu. 
This research was supported by the U.S. Department of Energy (DOE), Office of Science, Basic Energy Sciences (BES), under Award No. DESC0024900.
\end{acknowledgments}

\end{document}